\documentclass[%
superscriptaddress,
nofootinbib,
amsmath,amssymb,
aps,
pra,
showkeys]{revtex4-1}
\setcitestyle{authoryear,round}
\usepackage[dvipsnames]{xcolor}
\usepackage{amsmath,amsfonts,amssymb,amsthm, tabu}
\usepackage{xspace}
\usepackage{hyperref}
\usepackage{dsfont}
\usepackage{tikz}
\usetikzlibrary{arrows, decorations.markings}
\usepackage{mathrsfs}
\usepackage{lipsum}
\usepackage{braket}
\usepackage{enumitem}
\usepackage{pgfplots}
\pgfplotsset{compat=1.11}
\usepgfplotslibrary{fillbetween}
\usetikzlibrary{patterns}
\usepackage{amsbsy}
\usepackage{caption}
\usepackage{subcaption}
\usepackage{algorithm}
\usepackage[noend]{algpseudocode}

\newcommand{\com}[1]{\ensuremath{{\rm Com}\,\left(#1\right)}\xspace}

\newcommand{\ps}[2]{\ensuremath{\left\langle #1\mid#2\right\rangle}\xspace}

\newcommand{\diag}[1]{\boldsymbol{\Delta}\left(#1\right)}
\newcommand{\bigcom}[1]{}

\newcommand{\tr}[1]{{\rm Tr}\left(#1\right)}

\renewcommand{\Re}{\mathfrak{Re}}

\newcommand{\peq}[1]{p_{\text{eq},#1}}

\newcommand{\preq}[1]{\gamma_{\text{eq},#1}}
\newcommand{\peqv}{\mathbf{p}_{\text{eq}}}
\newcommand{\preqv}{\boldsymbol{\gamma}_{\text{eq}}}
\newcommand{\circslash}{\ensuremath{/\kern-0.5em{\circ}}}
\newcommand{\alphap}{\ensuremath{\alpha^\prime}}
\newcommand{\betap}{\ensuremath{\beta^\prime}}
\newcommand{\omegap}{\ensuremath{\omega^\prime}}
\newcommand{\R}{\ensuremath{\mathbb{R}}}
\newcommand{\mn}[1]{\ensuremath{\mathcal{M}_{#1}\left(\R\right)}}
\newcommand{\mnm}[2]{\ensuremath{\mathcal{M}_{#1,#2}\left(\R\right)}}
\newcommand{\gsim}{\raisebox{-0.13cm}{~\shortstack{$>$ \\[-0.07cm]
      $\sim$}}~}

\DeclareMathAlphabet{\mathscrbf}{OMS}{mdugm}{b}{n}
\tikzstyle{vecArrow} = [thick, decoration={markings,mark=at position
   1 with {\arrow[semithick]{open triangle 60}}},
   double distance=1.4pt, shorten >= 5.5pt,
   preaction = {decorate},
   postaction = {draw,line width=1.4pt, white,shorten >= 4.5pt}]
\tikzstyle{innerWhite} = [semithick, white,line width=1.4pt, shorten >= 4.5pt]

\newcommand\Algphase[1]{%
\vspace*{.1\baselineskip}\Statex\hspace*{\dimexpr-\algorithmicindent-2pt\relax}\rule{\textwidth}{0.4pt}%
\Statex\hspace*{-\algorithmicindent}\textbf{#1}%
\vspace*{-.3\baselineskip}\Statex\hspace*{\dimexpr-\algorithmicindent-2pt\relax}\rule{\textwidth}{0.4pt}%
}
  
 \algdef{SE}% flags used internally to indicate we're defining a new block statement
[CLASS]% new block type, not to be confused with loops or if-statements
{Class}% "\Struct{name}" will indicate the start of the struct declaration
{EndClass}% "\EndStruct" ends the block indent
[1]% There is one argument, which is the name of the data structure
{\textbf{class} \textsc{#1}}% typesetting of the start of a struct
{\textbf{end class}}% typesetting the end of the struct

\algdef{SE}% flags used internally to indicate we're defining a new block statement
[ATTRIBUTES]% new block type, not to be confused with loops or if-statements
{Atr}% "\Struct{name}" will indicate the start of the struct declaration
{EndAtr}% "\EndStruct" ends the block indent
[1]% There is one argument, which is the name of the data structure
{\textbf{attributes} \textsc{#1}}% typesetting of the start of a struct
{}% typesetting the end of the struct

\algdef{SE}% flags used internally to indicate we're defining a new block statement
[METHODS]% new block type, not to be confused with loops or if-statements
{Methods}% "\Struct{name}" will indicate the start of the struct declaration
{EndMethods}% "\EndStruct" ends the block indent
[1]% There is one argument, which is the name of the data structure
{\textbf{methods} \textsc{#1}}% typesetting of the start of a struct
{}% typesetting the end of the struct

\makeatother
\numberwithin{equation}{section}

\begin{document}

\makeatletter
\floatname{algorithm}{Procedure}
\algnewcommand{\algorithmicinput}{\textbf{Input:}}
\algnewcommand{\algorithmicoutput}{\textbf{Output:}}
\algnewcommand\Input{\item[\algorithmicinput]}%
\algnewcommand\Output{\item[\algorithmicoutput]}%
\newtheorem{theo}{Theorem}[section]
\newtheorem{lem}{Lemma}[section]
\newtheorem{mydef}{Definition}[section]
\newtheorem{prop}{Proposition}[section]
\newtheorem{princ}{Principle}[section]
\newtheorem{conj}{Conjecture}[section]

\title{Out-of-Equilibrium Dynamics and Excess Volatility in Firm Networks}

\author{Th\'{e}o Dessertaine}
\email{theo.dessertaine@polytechnique.edu}
\affiliation{LadHyX UMR CNRS 7646, Ecole polytechnique, 91128 Palaiseau Cedex, France}
\affiliation{Chair of Econophysics \& Complex Systems, Ecole polytechnique, 91128 Palaiseau Cedex, France}

\author{Jos\'{e} Moran}
%\email{jose.moran@polytechnique.org}
\affiliation{Mathematical Institute and Institute for New Economic Thinking at the Oxford Martin School, University of Oxford, Oxford, United Kingdom}
\affiliation{Chair of Econophysics \& Complex Systems, Ecole polytechnique, 91128 Palaiseau Cedex, France}
\affiliation{Complexity Science Hub Vienna, Josefst\"adter Stra{\ss}e 39, A-1080, Austria}
% \affiliation{Centre d'Analyse et de Math\'{e}matique Sociales, EHESS, 54 Boulevard Raspail, 75006 Paris}

\author{Michael Benzaquen}%
\affiliation{LadHyX UMR CNRS 7646, Ecole polytechnique, 91128 Palaiseau Cedex, France}
\affiliation{Chair of Econophysics \& Complex Systems, Ecole polytechnique, 91128 Palaiseau Cedex, France}
\affiliation{Capital Fund Management, 23 Rue de l'Universit\'{e}, 75007 Paris, France\medskip}

\author{Jean-Philippe Bouchaud}%
\affiliation{Chair of Econophysics \& Complex Systems, Ecole polytechnique, 91128 Palaiseau Cedex, France}
\affiliation{Capital Fund Management, 23 Rue de l'Universit\'{e}, 75007 Paris, France\medskip}

\date{\today}
\begin{abstract}
We study the conditions under which input-output networks can dynamically attain a competitive equilibrium, where markets clear and profits are zero. We endow a classical firm network model with minimal dynamical rules that reduce supply/demand imbalances and excess profits. We show that the time needed to reach equilibrium diverges to infinity as the system approaches an instability point beyond which the Hawkins-Simons condition is violated and competitive equilibrium is no longer admissible. We argue that such slow dynamics is a source of excess volatility, through accumulation and amplification of exogenous shocks. Factoring in essential physical constraints absent in our minimal model, such as causality or inventory management, we then propose a dynamically consistent model that displays a rich variety of phenomena. Competitive equilibrium can only be reached after some time and within some restricted region of parameter space, outside of which one observes spontaneous periodic and chaotic dynamics, reminiscent of real business cycles. This suggests an alternative explanation of excess volatility in terms of purely endogenous fluctuations. Diminishing return to scale and increased perishability of goods are found to ease convergence towards equilibrium. 
%Other regimes include deflationary equilibria and intermittent crises characterised by bursts of inflation. 
%Our model can in principle be calibrated using highly disaggregated data on individual firms and prices, and may provide a powerful tool to describe out-of-equilibrium economies. 
\end{abstract}
\maketitle

\tableofcontents
%\newpage

\setlength{\parskip}{\medskipamount}

\section{Introduction}

\subsection{Motivation and Economic Relevance}

What is the origin of macroeconomic fluctuations? Textbook macroeconomic models picture the world as a succession of equilibria where markets clear perfectly and firms maximise their profits.  Each equilibrium is characterised by a different level of productivity or household preferences, themselves driven by exogenous ``shocks'', which are the primary cause of fluctuations. Drawing an analogy from physics, one may call such an approach ``adiabatic'', in the sense that the time needed for the system to reach equilibrium is much shorter than the time over which the environment changes, so out-of-equilibrium effects can be neglected. The time evolution of the economy is then {\it slaved} to the time evolution of the exogenous parameters. This assumption is at the core of DSGE models (see \cite{gali_book}), but also central to the analysis of~\cite{acemoglu2012network} in their now classic paper on the network origins of aggregate fluctuations.  

The central proposition of the present work is that the standard economic equilibrium may actually be dynamically unattainable. Correspondingly, the ``small shock, large business cycle'' paradox (i.e. aggregate fluctuations much too large to be explained by exogenous shocks alone, see e.g. \cite{cochrane1994shocks} and \cite{bernanke1994financial}) would be chiefly explained by {\it out-of-equilibrium} effects. Indeed, in such out-of-equilibrium situations, the dynamics is mostly of endogenous origin and cannot be accounted for by traditional equilibrium arguments, like those of e.g. \cite{long1983real} and \cite{acemoglu2012network, carvalho_review, farhibaqaee2019}.

From a conceptual point of view, our point is the following: economic equilibrium requires so much cooperation between rational, forward looking agents, that the only way such equilibrium can plausibly be achieved is through some kind of adjustment process, that inevitably takes some time to complete.\footnote{This is actually even the case for financial markets where transactions take place at the second time scale. In reality, a large amount of the supply/demand volume is latent and is only slowly revealed, see~\cite{slowly2009}.} We will argue that even in cases where equilibrium is eventually reached, this time can be much longer than the evolution time of technology or of any other type of shocks (political, social, geopolitical, sanitary, etc.) that do affect the economy, in which case the adiabatic hypothesis is doomed to fail. 

Such a situation requires a richer modelling framework where out-of-equilibrium dynamics is an integral part of the description: We do not only need to describe the final equilibrium state, but also the path to equilibrium, which may in fact never converge.\footnote{In fact, if one delves into the history of the notion of economic equilibrium from Walras up to the Arrow-Debreu general equilibrium theory, it is striking to see that the focus has been mainly on the existence and the properties of an economic equilibrium. It is assumed that a mechanism exists that leads the economy towards that point, but it is not made explicit, as shown by~\cite{ingrao_equilibrium_2004} and \cite{ingrao1990the}.} 
%We might then realize that in some cases it is in fact never reached -- opening the possibility of purely endogenous macro-economic fluctuations.  

\subsection{Literature Review}

Past literature in macroeconomics (see e.g.~\cite{grandmont, grandmont_paper, fisher, benassy, chiarella,gintis} and \cite{beaudry_putting_2020}) has been mainly concerned with ``disequilibrium'' effects, which in that context means studying the impact of price or wage frictions and rigidities that prevent the economy from reaching full equilibrium. In a sense, these models postulate economies that are not able to reach an idealised state of equilibrium because of certain imperfections, but they still mostly deal with the \textit{static} properties of such economies with no particular focus on their dynamics or on how said state is reached. 

Effectively, the dynamics that are studied through these models are those in which the agents in the economy are all capable of optimising their behaviour, exchanging goods and coordinating between themselves instantly. The main source of fluctuations is therefore given by \textit{external} shocks to the economy. This is, for example, the case of \cite{long1983real} and \cite{Kydland1982}, despite their common acknowledgment of the need to take into account the ``time to build'' in the economy, which we choose to interpret too as the time required for all the agents to coordinate and exchange enough information to reach equilibrium. 

Another strand of the literature considers ``reduced-form'' differential equations that describe the coupled evolution of a set of aggregate variables (for example employment, wage and output in the original model by~\cite{Goodwin1982} and revived in \cite{Flaschel2008TheMO}). These low-dimensional dynamical equations can generate various types of dynamics, such as business cycles in the Goodwin model which is, {\it mutatis mutandis}, equivalent to the classic Lotka-Volterra (or predator-prey) model of \cite{Lotka1920, Volterra1926}. Note also that \cite{Liu2020} establish a system of coupled dynamical equations whose dynamics are determined by certain matrices that describe the production network. This is, in a way, a similar approach to ours, but our model considers a fully non-linear model with a \textit{linearised} evolution governed by a production-network determined matrix that is only valid close to equilibrium.

Yet another direction is explored by Agent Based Models (ABMs), where individual agents/firms make decisions based on plausible heuristic rules. ABMs are explicitly dynamical models, in the sense described by~\cite{ballot_agent-based_2015}: decision rules lead to actions (buy/sell, produce, update prices and wages, etc.) that move the economy one step forward in time (see the work of \cite{delligatti2005,DelliGatti2008, dawid, raberto, roventini, gualdi2015tipping} and \cite{poledna} for recent examples). Note that some of these heuristics actually correspond to the agents acting rationally but with limited information or by only being able to \textit{forecast} the other agents' behaviour, an approach taken by \cite{bonart2014instabilities} which we also follow in this work. The approach of ABMs, although very prolific, has been heavily criticised by those who argue in favour of ``micro-founded'' models where agents are forward-looking and optimise inter-temporal utility functions.

In the present paper, we revisit these ideas within the framework of network economies, where firms interact through a supply/demand (or input/output) network. As mentioned above, such models have recently become popular as a way to generate excess aggregate volatility, as shocks may possibly propagate through the input-output network. However, the seminal papers of \cite{long1983real}, and of \cite{acemoglu2012network} are studied within the ``adiabatic'' framework in which the system instantaneously adapts to productivity shocks (for a recent enlightening review of these models, see \cite{carvalho_review}). 

Furthermore, most of these papers assume a Cobb-Douglas production function, which ensures that an equilibrium always exists, whatever the input-output network and independently of the productivities of the firms. More recently, \cite{baqaee2018} and \cite{farhibaqaee2019} extended such work using Constant Elasticity of Substitution (CES) production functions, showing in particular that they induce non-linear effects that can cause the amplification of small shocks, all while remaining at economic equilibrium. Another similar line was recently exploited by~\cite{Pichler2020ProductionNA,Pichler2021InAO,Pichler2021ModelingSS}, where both CES-like production functions and non-equilibrium dynamics were exploited to forecast the economic shock due to the Covid-19 pandemic lockdowns. See in particular the Introduction of ~\cite{Pichler2020ProductionNA} for a comprehensive discussion about substitution effects in production functions and their macroeconomic consequences.

But as argued long ago by  \cite{hawkins1948some} and \cite{hs_condition} for Leontief economies, and more recently by two of us, \cite{moranb}, for more general CES production functions,  equilibrium may cease to exist when the average connectivity of the network is too large, firm productivities are too low, or markups are too large. In these cases, the description of a time evolving economy as a succession of static equilibria just does not make sense. In fact, \cite{moranb} argue, in the spirit of a conjecture by \cite{bak1993aggregate}, that real economies could generically be sitting close to a point where general equilibrium disappears.

In the present paper, we endow such a network model with a class of plausible dynamical rules, which aims at describing the fate of the economy outside of the adiabatic regime, and identify cases where equilibrium does exist mathematically but can never be reached dynamically. A step in this direction was proposed by \cite{MANDEL2015257}
and, independently, by \cite{bonart2014instabilities}, where a dynamical Cobb-Douglas economy was considered, with plausible update rules for production and prices. Interestingly, the model considered by \cite{bonart2014instabilities} leads to a phase transition between a region where equilibrium is reached (when firms slowly adapt to shocks) and a region where coordination breaks down (when firms adapt too aggressively) and where equilibrium is no longer dynamically accessible.\footnote{A similar phenomenology is reported by \cite{MANDEL2015257}, where it is stated that {\it depending on the stringency of the financial constraints the model can
settle in two very different regimes: one characterised by equilibrium, the
other by disequilibrium and financial fragility}.} In the latter phase, endogenous volatility becomes dominant. But this model only goes half-way towards a full-fledged dynamical description, since market-clearing was imposed by {\it fiat} in \cite{bonart2014instabilities}, with no excess production or excess demand -- leading to conceptual inconsistencies and, in fact, spurious instabilities. 

\subsection{Main Results and Outline}

In the present work, we propose a consistent framework to describe dynamical out-of-equilibrium effects in network economies. Our approach is a hybrid between standard economics thinking (where firms attempt to optimise profits in a competitive environment, and households optimise their utility function to balance consumption and labour) and Agent Based Models, where simplified behavioural assumptions allow one to specify the decision-making process of firms.

%{\color{red} Much to our surprise, we have found that in order to obtain well-behaved outcomes, extra care has to be devoted to treat all the decision steps in a strictly causal way (for example, goods must be produced before they are consumed) and to satisfy all inequalities (for example, consumption cannot exceed production plus inventories). Any attempt to write down ``reasonable'' dynamical equations that violate these constraints consistently lead to spurious instabilities. We in fact consider this as a blessing: physical constraints provide a discriminant straight-jacket for modelling.} 
We propose a minimal parametrisation of the heuristic rules used by firms to update production, prices and wages, which already leads to a surprisingly rich phenomenology of the resulting economy, which in some cases smoothly reaches equilibrium, and in other cases display much more complicated dynamical patterns, including cycles, chaos and crises, but also what we call ``deflationary'' equilibria. We argue that such generic scenarios could naturally explain the ``small shocks, large business cycle'' conundrum in terms of out-of-equilibrium effects. In a nutshell, when firms tend to over-react and adjust prices/productions too quickly in the face of imbalances, the economy enters an oscillatory or chaotic regime, where volatility is purely of endogenous origin. 

In a sense, our model can be seen as a multidimensional, discrete time version of the reduced form differential equations {\it \`a la}~\cite{Goodwin1982} and followers, that also lead to oscillatory dynamics. The main difference is that we describe the dynamics of the economy at a highly disaggregated level (that of firms), which is an important aspect in view of the amount of micro-data now available to calibrate such models. Given the diversity of phenomena that can take place within our framework, we are quite confident that the model is flexible enough to account for many empirical facts. However, in the current era of ``big data'', some extensions of the model may be worth investigating -- with each extension bringing one or several new parameters that need to be calibrated. In particular, some of our behavioural assumptions may appear too primitive and could be enriched later, as we discuss in section \ref{sec:extensions}. 

The most important generalisation, in our opinion, will be to explicitly include competition, debt, interest rates and bankruptcies in the model. In particular, the way the network ``rewires'' after the birth of a new competitive firm or after the removal of a bankrupt firm, with the possibility of cascading defaults, is clearly one of the most interesting aspects of firm network models when it comes to understanding business cycles and economic crises. These cascading bankruptcies are, in fact, the very motivation for studying network models, but they will not be directly addressed in the present article. 

The article is organised as follows. In section \ref{sec:static_framework}, we set up the stage for a firm network model and propose a definition of competitive equilibrium suited for our purposes. Section \ref{sec:naive} presents a simple heuristic for an out-of-equilibrium dynamical model of interacting firms. We show that reaching equilibrium might take an infinite amount of time (therefore jeopardising the adiabatic hypothesis) and that the dynamics displays excess volatility when the economy sits close to an instability. In section \ref{sec:fullyconsist}, we present a fully consistent extension of the model of section \ref{sec:naive} which incorporates natural constraints which were overlooked such as causality or shortages. We propose a numerical study of this extension in section \ref{sec:numerical_study} where we highlight and discuss the existence of other interesting dynamical regimes besides competitive equilibrium. We also provide several technical appendices for completeness. In Appendix \ref{ap:eq}, we detail the derivation of competitive equilibrium equations in the most general setting of production function. Appendix \ref{ap:relax_time} shows the computation of the relaxation time of the naive model which relies on Appendix \ref{ap:block_stability_matrix} that compiles necessary intermediate results on the stability matrix. In Appendix \ref{ap:fluctuation_sqrt_eps} we show that a marginally stable linear stochastic system creates excess volatility and we apply this result to a generic case of the naive model of \ref{sec:naive}. In Appendix \ref{ap:real_net} we provide time-series of the dynamics of our model on realistic networks. Finally in Appendix \ref{ap:pseudo_code}, we provide a pseudo-code for simulation of the fully consistent approach of section \ref{sec:fullyconsist}. The code itself is made available at:  \url{https://yakari.polytechnique.fr/dash}.

\section{Firm Networks at Competitive Equilibrium} \label{sec:static_framework}
\subsection{Network and Production Function}

Following the descriptions of \cite{long1983real, acemoglu2012network,bonart2014instabilities} and \cite{carvalho_review}, we model the economy as consisting of $N$ firms that interact with one another and with a single representative household which provides labour and consumes goods. The economy is described by a ``technology network'', namely a directed graph where each node $i=1,\ldots, N$ represents a firm  and where the link $j\rightarrow i$ exists if $i$ uses the good produced by $j$ for its own production. The node labelled $i=0$ conventionally represents households. Each edge in the graph $j\to i$ carries a ``weight'' that is a measure of the number of $j$ goods needed to make an unit of $i$. The production function gives the quantity of goods $y_i$ produced by $i$ as a function of input goods and labour (no capital at this stage) and the intrinsic, possibly time dependent, productivity of the firm $z_i$ (i.e. its efficiency in converting a given amount of inputs into outputs). We generalise the standard CES production function~\cite{Arrow1961CES} as:\footnote{The standard CES function corresponds to all $J_{ij}$ set to unity.}
\begin{equation}
    y_{i}=z_{i}\gamma_i, \qquad \gamma_i:= \left(a_{i0}\left(\frac{\ell_{i}}{J_{i0}}\right)^{-1/q}+\sum_{j=1}^{N}a_{ij}\left(\frac{x_{ij}}{J_{ij}}\right)^{-1/q}\right)^{-bq},
    \label{eq:production_function}
\end{equation}
where $x_{ij}$ is the amount of good $j$ (or labour $x_{i0}:=\ell_i$ if $j=0$) available to $i$, $J_{ij} \geq 0$ and $a_{ij} \geq 0$ link variables that measure the importance of good $j$ in the production of $i$\footnote{The $a_{ij}$ are normalised such that $\sum_{j=1}^N a_{ij}+a_{i0} = 1$, $\forall i$.}, and where we define $\gamma_i$ as the level of production of firm $i$. Note that although for all values of $q  \in ]0,\infty[$, the $J_{ij}$s can be absorbed into the $a_{ij}$, our specification allows for consistent limits when $q=0$ (Leontief) and $q = \infty$ (Cobb-Douglas), see below.

The parameter $b$ sets the return to scale: if all inputs and work hours are multiplied by a factor $\lambda$, then total output is multiplied by $\lambda^b$. 

The parameter $q$ measures the substitutability of inputs. For example, when $q \to 0^+$ we get the Leontief production function, corresponding to the case where production falls to zero if a single input is missing:
\[
y_i=z_i\left(\min\left[\frac{\ell_i}{J_{i0}},\min_{j}\left(\frac{x_{ij}}{J_{ij}}\right)\right]\right)^{b},
\]
where $J_{ij}$ is the amount of good $j$ that $i$ would need to achieve a level of production equal to $z_i$. The Leontief production function
corresponds to an economy where firms only keep a small, very optimised portfolio of suppliers that does not allow for redundancy. 

If $q\rightarrow+\infty$, we get the  Cobb-Douglas production function
\[
y_i=z_{i}\left(\left(\frac{\ell_i}{J_{i0}}\right)^{a_{i0}}\prod_{j=1}^{N}\left(\frac{x_{ij}}{J_{ij}}\right)^{a_{ij}}\right)^{b},
\]
for which some amount of substitutability is present. Indeed, halving the quantity $x_{ik}$ of input $k$ can be compensated by multiplying the input of $\ell$ by $2^{a_{ik}/a_{i\ell}}$, where the $a_{ij}$ describes the amount of substitutability between the goods in the production of $i$.

Although our dynamical model applies to any production function, and is not restricted to the CES family specified above, we will for the sake of simplicity illustrate our general arguments using the special case of a Leontief production function with constant return to scale ($b=1$), as equilibrium conditions can be solved explicitly. However, the general phenomenology of the model does not depend on this specific choice and applies to a broad family of production functions.   

\subsection{Competitive Equilibrium Conditions on Prices and Productions}

{
Given the prices $p_i$ of the goods and wage $p_0$, the profit $\pi_{i}$ of firm $i$ can be written as 
\begin{equation}
\pi_{i}=\sum_{j=0}^{N}x_{ji} p_i -\sum_{j=0}^{N}x_{ij}p_j\equiv\mathscr{G}_i -\sum_{j=0}^{N}x_{ij}p_j,
    \label{eq:profit_equilibrium}
\end{equation}
where $\mathscr{G}_i$ denotes the total proceeds of the future sales (``gains''), $x_{i0}:=\ell_i$ the working hours provided by the household and $x_{0i}:= C_i$ is the consumption of good $i$ by the households. Now, the textbook protocol at this stage is to impose that firms maximise their profit {\it assuming} that markets will clear, so that all that is produced will be sold, hence 
\begin{equation}
\mathscr{G}_i \equiv y_i p_i.
    \label{eq:gain_equilibrium}
\end{equation}
Using the production function \eqref{eq:production_function}, profit maximisation by firm $i$ then leads to the optimal quantities of input goods $x_{ij}$ and optimal production $z_i \gamma_i$. Note that when $b=1$, the corresponding solution leads to profits that are zero in equilibrium, but they are strictly positive when $b < 1$, corresponding to imperfect competition in that case.

Following the logic of our paper, we take another stance and depart from the standard definition of equilibrium in two ways:
\begin{enumerate}
    \item Since we do not assume that markets clear at each time step, firms can only compute the optimal input quantities  $\widehat{x}_{ij}$ required to reach a certain production target $\widehat{y}_i:=z_i\widehat{\gamma}_i$. Since firms do not know in advance how much of their production they will be able to sell (and consequently how much they will earn), the only lever on which they can act is the cost term that they attempt to minimise. The optimal value of $\widehat{\gamma}_i$ will then be obtained through a dynamical adjustment process, see below.
    \item We assume that our firm network is competitive in the sense that enough firms sell similar goods to drive profits down to zero at equilibrium. (An explicit description of such a competitive process would entail introducing a time dependent network where firms rewire towards cheaper suppliers. This is however beyond the scope of the present work, which {\it assumes} that this process has already taken place).  
\end{enumerate} 

\subsubsection{Step 1: Cost-minimizing Inputs}

More explicitly, the cost-minimizing input quantities are such that
\begin{equation}
    \widehat x_{ij} = \left\{\begin{matrix}
        \underset{x_{ij}}{\arg\min}\sum_{j=0}^{N}x_{ij}p_j\\
        \textrm{subject to}\;\left(\sum_{j=0}^{N}a_{ij}\left(\frac{x_{ij}}{J_{ij}}\right)^{-1/q}\right)^{-bq}=\widehat{\gamma}_i.
    \end{matrix}\right.
\end{equation}
Within the CES framework, this leads to:
\begin{equation}
\widehat x_{ik} =a_{ik}^{q\zeta}J_{ik}^{\zeta}\left(\sum_{j}a_{ij}^{q\zeta}J_{ij}^{\zeta}\left(\frac{p_{j}}{p_{k}}\right)^{\zeta}\right)^{q} \, \widehat{\gamma}_{i}^{1/b},
    \label{eq:optimal_quantities}
\end{equation}
with $\zeta=(1+q)^{-1}$. In the Leontief case with $b=1$, this boils down to 
\begin{equation}
    \widehat x_{ik} = J_{ik} \widehat{\gamma}_{i},
\end{equation}
which amounts to buying no more than the minimum amount needed to reach the target.}

\subsubsection{Step 2: Market Clearing \& Competitive Prices}

We then obtain prices and productions by assuming perfect competition, i.e. $\pi_i = 0$ for all firms (step 2), and perfect market clearing. This last condition can be written as
\begin{equation}
    y_{\text{eq},i} = C_{\text{eq},i} + \sum_{j} x_{\text{eq},ji},
\end{equation}
where $C_i$ is the households' demand for good $i$. Except when return-to-scales are constant ($b=1$), the above two-step procedure is {\it not} equivalent to the standard profit maximization where market clearing is assumed from the start, which allows firms to know their gains in advance and include them in the optimisation program.  

\subsubsection{The Leontief Case}

For Leontief production functions with $b=1$, the resulting equations are linear and identical to those of the classical equilibrium for which profits are naturally zero when markets clear. They read:
\begin{subequations}
\label{eq:eq_CRS}
\begin{eqnarray}
     \boldsymbol{\mathcal{M}} \, \peqv &=&\mathbf{V}
    \label{eq:price_eq_CRS}\\
    \boldsymbol{\mathcal{M}}^{\top} \,  \preqv&=&\frac{\boldsymbol{\kappa}}{\peqv},
    \label{eq:prod_eq_CRS}
\end{eqnarray}
\end{subequations}
where $\boldsymbol{\mathcal{M}}$ is a matrix defined as $\boldsymbol{\mathcal{M}}_{ij}= z_i \delta_{ij} - J_{ij}$ (where $\delta_{ij}$ is the Kronecker symbol $\delta_{ij}=1$ for $i=j$, $0$ otherwise), ${V}_i := p_0J_{i0}$ is the workforce need of firm $i$ and $\boldsymbol{\kappa}$ a positive vector describing final demand.\footnote{Vector division in Eq. \eqref{eq:prod_eq_CRS} is understood as component-wise division.}\textsuperscript{,}\footnote{The term $\boldsymbol{\kappa}/\peqv$ is the household's equilibrium consumption obtained through utility maximization, see below.}  For more general production functions, the equations can be written down as well -- see Appendix~\ref{ap:eq} -- but we will not consider them further in the present paper. The important features are:
\begin{itemize}
    \item For Eqs. (\ref{eq:price_eq_CRS},~\ref{eq:prod_eq_CRS}) to have non-negative solutions for prices and productions, $\boldsymbol{\mathcal{M}}$ must be a so-called $M$-matrix, as shown by \cite{hs_condition,Fiedler1962} and, in the context of production networks, \cite{moranb}. Owing to its particular shape, with non-negative terms on the diagonal and negative terms on the off-diagonal, this is equivalent to the spectrum of $\boldsymbol{\mathcal{M}}$ having a non-negative real part. For a given set of input-output coefficients $J_{ij}$, this imposes that firms productivities must be large enough, otherwise no admissible equilibrium exists where all $N$ firms are alive.\footnote{For an admissible equilibrium to reappear with the same level of productivity, one must necessarily remove some firms from the network and from the production function of the surviving firms.}
    \item For all finite values of $q < + \infty$ in the CES production function, some analogous conditions must be fulfilled for an admissible equilibrium to exist, see \cite{moranb}.
    \item When $q = +\infty$ (i.e. in the Cobb-Douglas case), positive solutions to the equilibrium equations {\it always exist}, independently of productivities or network coefficients (see Appendix \ref{ap:eq}), as is the case in the article by \cite{acemoglu2012network}.
\end{itemize}

The possible non-existence of static solutions for generic production functions and network topologies urges us to go beyond equilibrium and formulate dynamical equations that would still make sense in such cases. But even in situations where an admissible equilibrium exists, it is by no means automatic that the economy is able to reach it on its own device. And even if it does, the description of non-adiabatic situations, i.e. those for which technologies and productivities evolve on a time shorter than the time needed to reach equilibrium, also require consistent dynamical equations. 

Interestingly, when the economy is close to an instability, e.g. when the smallest eigenvalue of $\boldsymbol{\mathcal{M}}$ tends to zero in the Leontief case, the time needed to reach equilibrium will turn out to be infinitely large. This not only makes the adiabatic assumption moot but, as we shall see, compels the modeller to handle dynamical effects with special care.  

\section{A First ``Naive'' Approach}
\label{sec:naive} 

In this section, we introduce the simplest version of a dynamical model aimed as describing out-of-equilibrium effects (transient or permanent) in a network economy. The equations we will postulate are based on reasonable ``rules of thumb'' that firm decision makers are likely to use in real life conditions, see~\cite{kahneman_psychology_1973,tversky_judgment_1974, gigerenzer}.
%There is obviously still a  demarcation line between purists, who insist that these decision rules must be based on rational, forward looking optimisation programs, and a growing cohort of pragmatists who believe that modellers should embrace radical uncertainty and adopt behavioural rules closer to reality, with enough flexibility to avoid absurd paradoxes and accommodate, at least to some extent, Lucas' critique~\cite{LUCAS197619}.

In looking for such reduced form dynamical equations, we draw inspiration from what physicists call ``phenomenological approaches'', based on symmetry, plausibility and dimensional arguments. Such arguments avoid getting lost in the ``wilderness'' of possible models -- to paraphrase Sims -- once the straight-jacket of rationality is jettisoned.

%\footnote{As we have learnt from physics, general arguments can often be used to write down correct equations before the underlying foundations have been worked out. For example, the Navier-Stokes equations for fluid motion have been postulated in the XIXth century based on general arguments, 50 years before Boltzmann's statistical theory of molecular motion gave a solid, first principle justification of these equations.} 

\subsection{Forces Restoring Equilibrium}

Whereas in the economic equilibrium as defined in the previous section profits are zero and markets clear, out-of-equilibrium situations tautologically imply non zero profits and/or excess supply or demand. So we naturally introduce, for each firm, two indicators that measure the distance from equilibrium: $\mathscr{E}_i(t)$ is the excess production at time $t$ (interpreted as unsatisfied demand if $\mathscr{E}_i(t) < 0$), and $\pi_i(t)$ the instantaneous profit or losses of the firm at time~$t$.

Prices and productions must then adapt through some kind of adjustment process to reduce these imbalances:
\begin{itemize}
    \item Faced with excess production, firms will lower prices to prop up demand, and/or reduce production to limit losses.
    \item Faced with excess demand, on the other hand, firms can consider increasing prices and/or increase production.
    \item Similarly, when profits are positive, firms may be tempted to increase production but at the same time competition, attracted by the prospect of a profit, should put pressure on prices.
    \item If profits are negative, firms will try to adapt by lowering production and increase prices, with the hope of better compensating production costs. 
\end{itemize}   

All these rules are common sense and it is hard to argue that they do not play a crucial role in the real economy with boundedly rational agents. What is more debatable, however, is how to model them quantitatively. In this work, we further assume that restoring forces are all {\it linear} in $\mathscr{E}_i(t),\pi_i(t)$, at least when these imbalances are small enough. If only for dimensional reasons, all quantities determining price and production relative changes must appear as relative, non-dimensional quantities, i.e. ratios of $\mathscr{E}_i(t)$ to total production $y_i(t)$ and 
$\pi_i(t)$ to total sales $y_i(t) p_i(t)$. 

Hence we posit the following adjustment rules for prices and production:
\begin{subequations}
\begin{eqnarray}
\log \left(\frac{p_{i}(t+\delta t)}{p_i(t)}\right) &=& \left(- \alpha \frac{\mathscr{E}_{i}(t)}{y_{i}(t)} - \alphap \frac{\pi_{i}(t)}{p_{i}(t)y_{i}(t)} \right) \, \delta t
\label{eq:price_rule_toy}\\
\log \left(\frac{y_{i}(t+\delta t)}{y_i(t)}\right) &=&  \left(\beta \frac{\pi_{i}(t)}{p_{i}(t)y_{i}(t)} - \beta^{\prime} \, \frac{\mathscr{E}_{i}(t)}{y_{i}(t)}  \right) \, \delta t
\label{eq:prod_rule_toy},
\end{eqnarray}
\label{eq:rule_toy}
\end{subequations}
where $\delta t$ is an elementary time step, and the parameters $\alpha, \alphap, \beta, \beta^{\prime}$ characterise the speed of adjustment in the face of imbalances. From our general arguments above, we expect that all these parameters are non-negative, i.e. that firm policies and market forces tend to dampen imbalances. Whether these will be sufficient to stabilise the whole economy around the classical equilibrium described in the previous section is the whole point of the present research.  

These parameters could depend on the firm $i$, with some firms choosing to be more aggressive than others in their adjustment policy. Throughout the present work we will stick to time-independent and firm-independent values for $\alpha, \alpha', \beta, \beta^{\prime}$.\footnote{Note that one could imagine a version of the model where firms attempt to {\it learn} optimal values of these adjustment parameters, adding an extra level of complexity in the dynamical rules.} The simple rules of Eqs. (\ref{eq:price_rule_toy}, \ref{eq:prod_rule_toy}) are very similar in spirit to those used in several well studied Agent Based Models -- see \cite{DelliGatti2008, gualdi2015tipping}. Note that $\alphap > 0$ reflects our hypothesis that competition is at play in the economy, pushing prices down when profits are positive.

Although null profits and market clearing obviously imply from Eqs. (\ref{eq:price_rule_toy}, \ref{eq:prod_rule_toy}) that prices and productions are time invariant, the converse is more subtle. Assume indeed that there exists quantities $p_i^\star$, $y_i^\star$, $\mathscr{E}_i^\star$ and $\pi_i^\star$ towards which prices, productions, and imbalances converge under the dynamics (\ref{eq:price_rule_toy}, \ref{eq:prod_rule_toy}). These values should satisfy
\begin{equation}
\begin{array}{ccc}
- \alpha \dfrac{\mathscr{E}_{i}^\star}{y_{i}^\star} - \alpha' \dfrac{\pi_{i}^\star}{p_{i}^\star y_{i}^\star} &=& 0\\
\vspace{0.03pt}\\
- \beta^{\prime}  \dfrac{\mathscr{E}_{i}^\star}{y_{i}^\star} +\beta \dfrac{\pi_{i}^\star}{p_{i}^\star y_{i}^\star}&=& 0 \end{array}\,\,\Longleftrightarrow \, \,
\begin{pmatrix}
\alpha & \alphap \\
\betap & -\beta 
\end{pmatrix}
\begin{pmatrix}
\dfrac{\mathscr{E}_{i}^\star}{y_{i}^\star} \\
\dfrac{\pi_{i}^\star}{p_{i}^\star y_{i}^\star}
\end{pmatrix}=\begin{pmatrix}
0 \\
0
\end{pmatrix}.
\end{equation}

If the matrix of parameters is non-singular, then the only solution is trivial and $\mathscr{E}_{i}^\star=\mathscr{\pi}_{i}^\star=0$ which coincides with the equilibrium defined in the previous section. The only way through which this matrix can be singular is when $\alpha\beta+\alphap\betap=0$. Since $\alpha, \alpha', \beta, \beta^{\prime}$ are chosen to be positive, this happens only when at least one of the pairs $(\alpha,\alphap)$, $(\beta,\betap)$, $(\alpha,\betap)$ or $(\alphap,\beta)$ is equal to $(0,0)$. In the first two cases, prices or productions are frozen in time, making the dynamical rules moot. In the last two cases, prices and productions are driven either only by profits or only be production surplus. The dynamics will converge towards a partial competitive equilibrium with only one of the two conditions of null profits or market clearing fulfilled. This is again not a satisfying choice of parameters, because it implies no reaction from the firms to either supply/demand imbalances or to profits/losses. Therefore for generic cases our dynamical rules have fixed points that correspond precisely to competitive equilibria.\footnote{It is unclear at this stage if the general equations of Appendix \ref{ap:eq} could yield multiple solutions. Even if they did, stationary points of the dynamics would still coincide with these solutions.}

\subsection{Dynamical Equations}

Eqs. (\ref{eq:price_rule_toy}, \ref{eq:prod_rule_toy}) may now be closed by expressing imbalances in terms of prices $p_i$ and productions $y_i$, as:
\begin{subequations}
\begin{eqnarray}
\pi_i(t) &=&p_i(t) y_i(t) - \sum_{j=1}^N x_{ij}(t)p_j(t) - p_0(t) \ell_{i}(t) = \gamma_i(t) \left(z_i p_i(t) - \sum_{j=1}^N J_{ij}p_j(t) -  J_{i0} p_0(t)\right)
\label{eq:profit_mc2}\\
\mathscr{E}_i(t) &=& y_i(t) - \sum_{j=1}^N x_{ji}(t) - C_i(t) = z_i\gamma_i(t) - \sum_{j=1}^N J_{ji}\gamma_j(t) - C_i(t)
\label{eq:surplus_mc2},
\end{eqnarray}
\end{subequations}
where $C_i(t)$ is the consumption of households, $\ell_i(t)$ the quantity of labour, and where we have again restricted our analysis to constant returns to scale Leontief production functions.

With this, one must too model the consumption of households. For simplicity, we assume that households work full time, and denote by $L_0$ the total amount of available labour (this assumption will be relaxed below, as we will allow for unemployment, see section \ref{sec:households}). Consumption is obtained by saturating the current budget $p_0(t)L_0$ to maximise a log-consumption utility, i.e.
\begin{equation}
    \max_{C_i(t)}\;\;{\sum_{i}\theta_i\log{C_i(t)}}\qquad \text{with} \qquad 
    \sum_{i} p_i(t)C_i(t)\leq p_0(t) \sum_i{\ell_{i}(t)}=p_{0}(t)L_0
    \label{eq:utilitx_max_toy},
\end{equation}
where $\theta_i$ is the preference for good $i$. The optimal consumption is then  ${C}_i(t)={L_0\theta_i}/{\mu(t){p}_i(t)}$ with $\mu(t)=\sum_i{\theta_i}/p_0(t)$. 

Putting all these ingredients together and taking the continuous time limit $\delta t \to 0$ in \eqref{eq:rule_toy} yields the following  system of coupled non-linear ordinary differential equations (with ${V}_i = p_0 J_{i0}$):
\begin{subequations}
\label{eq:sys_tox_model}
\begin{eqnarray}
    z_i \gamma_i(t) \frac{\mathrm{{d}}p_{i}}{\mathrm{{d}}t} &=& -\alpha p_i(t)\left(\sum_j \mathcal{M}_{ji}\gamma_j(t)-\frac{L_0\theta_i}{\mu(t)p_i(t)}\right) - \alpha' \gamma_i(t)\left(\sum_j \mathcal{M}_{ij}p_j(t)-V_i\right) \\
    z_i p_i(t)\frac{\mathrm{{d}}\gamma_{i}}{\mathrm{{d}}t} &=& \beta \gamma_i(t)\left(\sum_j \mathcal{M}_{ij}p_j(t)-V_i\right) - \beta^{\prime} p_i(t)\left(\sum_j \mathcal{M}_{ji}\gamma_j(t)-\frac{L_0\theta_i}{\mu(t)p_i(t)}\right).
\end{eqnarray}
\end{subequations}

Interestingly, these equations bear a strong resemblance to generalised Lotka-Volterra models used in theoretical ecology by~\cite{Biroli_2018}, where an ecosystem self-organises into a configuration that is highly susceptible to amplify external perturbations. Newer extensions to such models, along the lines of~\cite{Roy2020}, show that they can also explain anomalous, persistent fluctuations in the populations of the different species that make up an ecosystem. The different analogies linking the study of firm networks and ecosystem have also been fruitful in linking the notion of trophic levels, namely the position of a species along the food web, to the ``upstreamness'' of a firm along the supply chain, as done by \cite{Antrs2012}, and in the work of \cite{MacKay2020} where these concepts are used to study the properties of production networks. 
Interesting parallels between these two domains could also arise when studying the impact of technological innovation or biological evolution within these models. 

The economic intuition behind the analogy with Lotka-Volterra equations is the following: when dealing with a complex assembly of interacting entities, be it an ecosystem with species having attained a certain evolutionary level or an economy with firms capable of using certain technologies, one is considering a complex system with a large amount of feedback loops. The different entities depend on one another in a way that creates feedback loops that can lead to very volatile oscillatory behaviour or even chaos, and in general to crises where certain firms or certain species must become ``extinct'' to re-stabilise the system, a point that was made by \cite{moranb} and \cite{bak1993aggregate} arguing in favour of ``self-organised criticality''. Although the analogy with ecology is chosen here because it is easy to understand, we stress that we believe this is a generic characteristic of a large class of systems with a large number of inter-dependencies, and so that this is particularly the case for firm networks.

\subsection{Perturbations Around Equilibrium} \label{sec:pert_naive} 

Equations~\eqref{eq:sys_tox_model} are our ``naive'' candidate equations for the out-of-equilibrium dynamics of the firm network model, which limitations will be discussed below. One immediately checks that the equilibrium solutions  $p_{\text{eq},i}$ and $\gamma_{\text{eq},i}$ (given by Eqs. (\ref{eq:price_eq_CRS}, \ref{eq:prod_eq_CRS})) are fixed points of these equations, as it should be. 

One can also study the linear stability of this equilibrium. Writing $p_i(t)=p_{\text{eq},i} + \delta p_i(t)$ and 
$\gamma_i(t)=\gamma_{\text{eq},i} + \delta \gamma_i(t)$ and keeping only terms of order 1 in $\delta(.)$, one finds a linear evolution equation for a $2N$ dimensional vector $\mathbf{U}=(\mathbf{\delta p},\mathbf{\delta \boldsymbol \gamma})$, of the form: 
\begin{equation}\label{eq:dyn_fluctuations}
    \frac{\mathrm{{d}}\mathbf{U}(t)}{\mathrm{{d}}t} = \mathbf{\mathbb{D}} \mathbf{U}(t).
\end{equation} 
The equilibrium stability is determined by the sign of the eigenvalues of the corresponding $2N \times 2N$  dynamical matrix~$\mathbf{\mathbb{D}}$. Such an analysis is detailed in Appendix~\ref{ap:relax_time}. 

When all eigenvalues are negative, equilibrium is locally stable. Any small perturbation away from equilibrium decays towards zero, at a rate asymptotically given by the eigenvalue closest to zero. The corresponding relaxation time $\tau_{\text{relax}}$ can be computed explicitly when the Hawkins-Simon conditions are on the verge of being violated, i.e. when the smallest eigenvalue of the network matrix $\boldsymbol{\mathcal{M}}$ is at a distance $\varepsilon \to 0$ away from $0$. We find (see Appendix~\ref{ap:relax_time}):

\begin{equation}\label{eq:relax_time_toy}
\tau_{\text{relax}} \approx 
\frac{2\max_{j} z_j}{\varepsilon} \times
\left\{
\begin{matrix}
\left(\alpha^{\prime}+\beta^{\prime}+\alpha-\sqrt{(\alpha^{\prime}+\beta^{\prime}+\alpha)^{2}-4(\alpha\beta+\alpha^{\prime}\beta^{\prime})}\right)^{-1} &\text{ if }& (\alpha^{\prime}+\beta^{\prime}+\alpha)^{2}>4(\alpha\beta+\alpha^{\prime}\beta^{\prime}) \\
 \left(\alpha^{\prime}+\beta^{\prime}+\alpha\right)^{-1} &\text{ if }& (\alpha^{\prime}+\beta^{\prime}+\alpha)^{2}\leq4(\alpha\beta+\alpha^{\prime}\beta^{\prime}).
\end{matrix}
\right.
\end{equation}
This expression allows us to draw two important conclusions:
\begin{itemize}
    \item When $\varepsilon \to 0$, the relaxation time of the system {\it diverges}, i.e. it takes an infinitely long time to reach equilibrium. As we mentioned in the introduction, this makes the adiabatic approximation unsuitable as changes in the technologies and in the network structure will happen before equilibrium can be reached. This long time scale also leads to an amplification of exogenous volatility in the system, see below.
    \item As long as $\alpha, \alpha^\prime$ or $\beta^\prime$ are strictly positive, the relaxation time is finite. (The equilibrium is still stable if some coefficients are negative provided others are positive and sufficiently large.)
\end{itemize}

\begin{figure}[t!]
    \centering
    \includegraphics{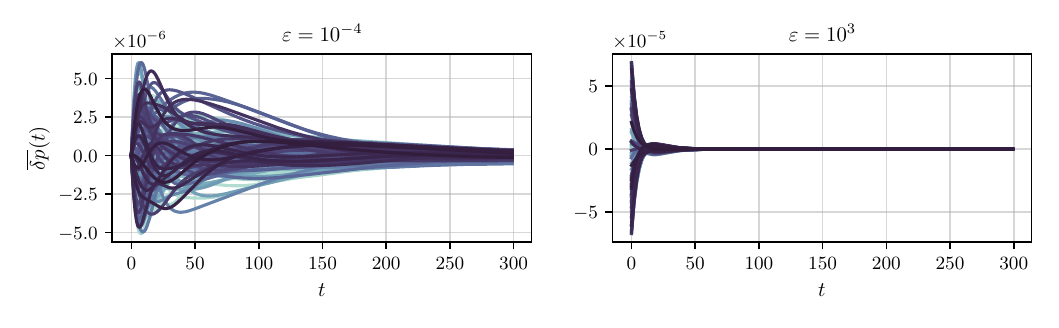}
    \caption{Relative distance to equilibrium values of prices for the non-linear discrete dynamics Eqs. \eqref{eq:sys_tox_model} for $N=100$ firms. The initial relative distance in the simulation is taken to be $\delta=10^{-3}$. The high productivity regime corresponds to a high value of $\varepsilon = 1000$ and leads to a very short relaxation time $\tau_{\text{relax}}$. On the other hand, in the low productivity regime where $\varepsilon\to 0$, the system takes longer and longer to reach equilibrium again, and the relaxation time $\tau_{\text{relax}}$ diverges.}
    \label{fig:example_tox_linear_pert}
\end{figure}

A numerical illustration of the type of weakly out-of-equilibrium dynamics predicted by the model is shown in Fig.~\ref{fig:example_tox_linear_pert}. One sees a complex interplay of spontaneous oscillations (coming from the imaginary part of the eigenvalues 
of the dynamical matrix $\mathbf{\mathbb{D}}$) with a slowly decaying envelope, $\propto \exp(-t/\tau_{\text{relax}})$. 

Note however that one must be particularly careful about spurious numerical effects when simulating \eqref{eq:sys_tox_model}. Indeed, such differential equations fall into the category of so called \emph{stiff} ordinary differential equations (ODEs). They are characterised by an evolution governed by two (or more) very different timescales. For a dynamical system of the form \ref{eq:dyn_fluctuations}, we denote by $\sigma_\nu$ the eigenvalues of the matrix $\mathbb{D}$ (as in Appendix \eqref{ap:relax_time}). We call $\bar{\sigma}$ and $\underline{\sigma}$ the two eigenvalues such that 
\[\left|\Re{\overline{\sigma}}\right|\geq\left|\Re{\sigma_\nu}\right|\geq\left|\Re{\underline{\sigma}}\right|,\;\forall\nu,\]
i.e. respectively the \emph{fast} and \emph{slow} timescales of the system. The stiffness ratio is defined as 
\[r=\frac{\left|\Re{\overline{\sigma}}\right|}{\left|\Re{\underline{\sigma}}\right|},\] and the system is said to be stiff if this ratio is large. In our case, as $\varepsilon\to 0$, $\overline{\sigma}$ will remain finite whereas $\underline{\sigma}$ is of order $\varepsilon$ making the stiffness ratio $r$ behaves as $\varepsilon^{-1}$ (see Appendix \ref{ap:relax_time}). Stiff ODEs require special care for their simulation. More precisely, one cannot use simple explicit integration routines with fixed step-size but rather implicit schemes such as Radau integration (see \cite{hairer1993solving}).

\subsection{Excess Volatility} 
\label{sec:excess_vol}

Now, suppose that the parameters describing the economic equilibrium (such as productivities or household preferences, etc.) are changing over time, the dynamical equation governing economic fluctuations, Eq.~(\ref{eq:dyn_fluctuations}), becomes:
\begin{equation}\label{eq:dyn_fluctuations2}
    \frac{\mathrm{{d}}\mathbf{U}(t)}{{\text{d}}t} = \mathbf{\mathbb{D}} \mathbf{U}(t) + \mathbf{\xi}(t),
\end{equation} 
where $\mathbf{\xi}(t)$ represents the (weak) exogenous shocks to the economy. It is then not hard to show (see Appendix \ref{ap:fluctuation_sqrt_eps}) that in the limit $\varepsilon \to 0$, the volatility of prices and output is proportional to $\varepsilon^{-1/2}$, and can thus be much larger than the variance of the exogenous shocks when the system approaches the limit of stability.
The intuitive reason is that past shocks linger a very long time (comparable to $\tau_{\text{relax}}$) in the system and aggregate with more recent shocks, leading to a much larger overall perturbation. 

\begin{figure}[t!]
    \centering
    \includegraphics{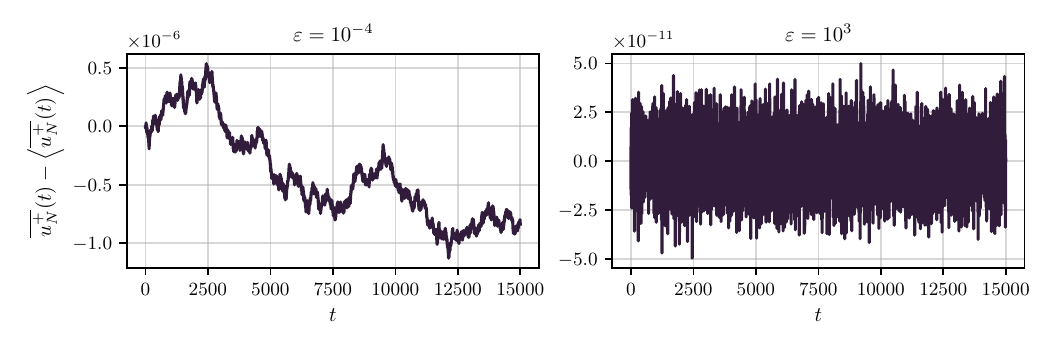}
    \caption{Evolution of the projection $u_N^+(t)$ of $\mathbf{U}(t)$ onto the eigenvector of $\mathbb{D}$ associated to the marginal eigenvalue, responsible for the volatility increase (see Appendix \ref{ap:fluctuation_sqrt_eps}) after productivity shocks with volatility $\sigma=10^{-8}$ and $\varepsilon=10^{-4}$, y-scale $10^{-6}$ (left),  $\varepsilon=10^{3}$,  y-scale $10^{-11}$ (right). For $\varepsilon=10^{-4}$, the volatility of output and prices is of the order of $10^{-6}$, i.e. 100 times larger than $\sigma$, as expected from theory.} 
    \label{fig:example_tox_linear_pert_with_shocks}
\end{figure}

Hence, the proximity to the point of instability is a natural candidate to explain the ``small shocks, large business cycle'' paradox (see \cite{bonart2014instabilities} for a related discussion). An illustration of this phenomenon for our model is given in Fig.~\ref{fig:example_tox_linear_pert_with_shocks}. However, in this scenario, fluctuations are predicted to persist over long times $\sim \varepsilon^{-1}$.  

We will discuss in section \ref{sec:oscillations} below another scenario for ``large business cycles'' based on non-linear, {\it endogenous} fluctuations rather than on long-lived {\it exogenous} fluctuations.   

\subsection{Limitations}

The above results suggest that, although ``naive'', our equations already provide an interesting generic scenario for anomalous fluctuations of output, namely the proximity of an instability. Note that 
the  dynamics we have described is directly linked to a large body of work concerned with the stability of large complex systems (see the historical precursors \cite{may1972will,Gardner_Ashby_1970}, and \cite{Fyodorov_Khoruzhenko_2016} and \cite{Bunin_2017} for recent general approaches), using random matrix techniques to represent generic interactions. These papers highlight the importance of studying of the eigenvectors and eigenvalues of large random matrices for understanding of complex systems, with other noteworthy contributions by~\cite{metz2018spectra,Tarnowski2020} and \cite{mambuca2020dynamical}.

However, the naive approach above sweeps under the rug important constraints that, while irrelevant at equilibrium, turn out to be essential out-of-equilibrium: 
\begin{itemize}
\item Causality: firms must plan production {\it before} they know how much they will manage to sell. 
\item Supply/demand imbalances (which are zero if markets clear): when supply exceeds demand, inventories accumulate, whereas when demand exceed supply (including inventories) involuntary savings increase. These extra variables should play a role in the out-of-equilibrium evolution of the economy, but are totally absent from Eqs.\eqref{eq:sys_tox_model}.
Furthermore, if some input goods is missing, Eqs. \eqref{eq:surplus_mc2} incorrectly account for imbalances.
\end{itemize}

In the next section, we will propose a minimal, fully consistent model that allows one to account for both causality and imbalances. 
Interestingly, we will see that hard constraints -- such as the impossibility to consume more than what is available -- lead to intrinsically non-linear dynamics, even for small perturbations close to equilibrium. As a consequence, limit cycles or chaotic behaviour will spontaneously emerge, when Eqs. \eqref{eq:sys_tox_model} can only lead to damped oscillations converging to equilibrium.  

Such generalised equations in fact allow one to obtain legitimate dynamics even in the region where the equilibrium is no longer defined, i.e. when $\varepsilon < 0$, whereas Eqs. \eqref{eq:sys_tox_model} cease to make sense in this case (prices and productions are are always dragged below zero).

\section{A Fully Consistent Approach}\label{sec:fullyconsist}

As we just mentioned, the naive approach of the previous section, although interesting, is at best approximate and incomplete since it overlooks some incontrovertible constraints, such as physical bounds (consumption cannot be larger than production plus inventories) and causality (or ``time to build''). In particular, when shortages are present, input goods must be allocated among customers in a specific way, which we will choose to be proportional to the posted demand. But in turn, such shortages will lead to production undershooting targets. Still, some of the results of the previous section will turn out to be useful to understand the extended model presented below. 

\subsection{Imbalances and Causality}

Accounting for the first constraint implies the following. If demand exceeds supply, all of a firm's production will be sold and exchanged, whereas if supply exceeds demand, only the quantity that was demanded will be traded, leaving a surplus that will add to the firm's inventories. Hence, the flow of goods going from $i$ to $j$ must be computed with care; instead of the single quantity $x_{ji}(t)$ considered in the previous section, we need to introduce the amount of goods $i$ {\it demanded} by firm $j$,  $x_{ji}^{\text{d}}$, that can only be smaller or equal to the quantity actually exchanged, $x_{ji}$. This can be understood as a contract that may only be fully honoured if firm $i$ produces enough to meet all demands. In a similar fashion, we distinguish the amounts $C^{\text{d}}$ demanded by households from what they will effectively be able to buy, $C$. Similarly, the work hours posted by firms $\ell_i^{\text{d}}$ may not be equal to the total amount of work $L^{\text{s}}$ households are willing to provide. To handle the situation where supply exceeds demand, we keep track of firm $i$'s inventory of good $j$, denoted by $I_{ij}(t)$ and to which we successively add the goods that the firms did not manage to sell or to use and subtract those that perished. 

%\red{[attention à nouveaux, $\ell$ est homogène à $p\times L$ non ? Donc $\ell$ et $L$ ne peuvent pas être tous deux des "amount".]}

Implementing causality in the dynamics also means dissecting the firms' decision processes. Clearly, goods can only be sold at time $t$ after they have been produced at time $t-1$, and prices may change (if only slightly) between these two times. More importantly, firms only have partial information about the amount of goods they will be able to buy and sell when they plan for the next production cycle. Likewise, the number of employees they will be able to hire is not known precisely, because it depends on the amount of work deemed acceptable by the households. It is at this stage that we will  introduce a heuristic rule that allows firms to plan for the next production round by making more or less informed guesses about these unknown quantities. In the present work, we assume that firms base their estimate on what happened in the previous time step, although more complicated and more general rules can already be imagined. 

\subsection{Time-line}
\label{ssec:time-line}
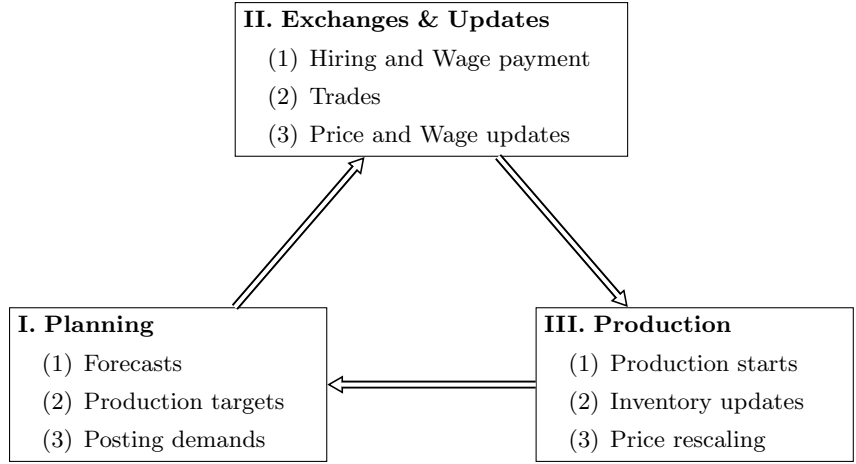
\begin{figure}[!t]
    \centering
    \begin{tikzpicture}

	\node[rectangle, draw,text width=4cm] (action_tm) at (-3.5,0) 
	{
	{\bf{I. Planning}}
	\begin{enumerate}[label=(\arabic*)]
	\item Forecasts
	\item Production targets
	\item Posting demands
	\end{enumerate}};
	
	\node[rectangle, draw,text width=5cm] (action_t) at (0,4.06)
	{
	{\bf{II. Exchanges \& Updates}}
	\begin{enumerate}[label=(\arabic*)]
	\item Hiring and Wage payment 
	\item Trades
	\item Price and Wage updates
	\end{enumerate}};
	
	\node[rectangle, draw,text width=4cm] (action_tp) at (3.5,0) 
	{
	{\bf{III. Production}}
	\begin{enumerate}[label=(\arabic*)]
	\item Production starts
	\item Inventory updates
	\item Price rescaling
	\end{enumerate}};
	\draw[vecArrow] (action_tm) to (action_t);
	\draw[vecArrow] (action_t) to (action_tp);
    \draw[vecArrow] (action_tp) to (action_tm);

\end{tikzpicture}
    \caption{Time-line of the model.}

    \label{fig:time_line}
\end{figure}

In order to keep all causal constraints satisfied, one must carefully set up a consistent chronology for the actions of firms and households. The resulting time-line of the model is schematised in Fig.~\ref{fig:time_line}. Each time step $\delta t$ ($\delta t=1$ hereafter) is conveniently sliced in three successive ``epochs'', represented as boxes in Fig.~\ref{fig:time_line}. At the end of time step $t-1$, goods have been produced and are available for consumption at $t$ in quantities $y_i(t)$ and prices $p_i(t)$. 

\subsubsection{{Planning}}

At any given time, firms must plan how much to produce for the following period. To capture this, we keep exactly the same adjustment rule as in the naive version of our model, Eq. (\ref{eq:prod_rule_toy}), but using now the {\it expected} profits $\mathbb{E}_t[\pi_i]$ and excess productions $\mathbb{E}_t[\mathscr{E}_i]$ at the end of the period, which we specify below. 
    
Thus, the {\it target} production for time $t+1$, $\widehat{y}_i(t+1)$, is set using
\begin{equation}
    \log \left(\frac{\widehat{y}_{i}(t+1)}{y_i(t)}\right) =  2\beta \frac{\mathbb{E}_t[\pi_i(t)]}{\mathbb{E}_t[\mathscr{G}_i(t)]+\mathbb{E}_t[\mathscr{L}_i(t)]} - 2\beta^{\prime}  \frac{\mathbb{E}_t[\mathscr{E}_i(t)]}{\mathbb{E}_t[\mathscr{S}_i(t)]+\mathbb{E}_t[\mathscr{D}_i(t)]},
\label{eq:final_target_update}
\end{equation}
where $\mathscr{G}_i(t)$ denotes the proceeds of the sales (``gains''), $\mathscr{L}_i(t)$ the production costs (``losses''), $\mathscr{D}_i(t)$ the overall demand for good~$i$ and $\mathscr{S}_i(t)$ the supply of good~$i$, which is already known to the firm at time $t$, hence  $\mathbb{E}_t[\mathscr{S}_i(t)]\equiv \mathscr{S}_i(t)$.

Once the target productions for $t+1$ are decided, the corresponding quantities $\widehat{x}_{ij}$ are computed according to Eq~\eqref{eq:optimal_quantities}. Firm $i$ then posts its demands for inputs $j$ for delivery at time $t$, taking into account their current stock of $I_{ij}$ of said inputs, with the rule%\footnote{A more complicated expression must be written in the general CES case with $b \neq 1$.}
\begin{equation}\label{eq:Qd_Qhat}
x^{\text{d}}_{ij}=\left\{
\begin{matrix}
    \max\left(0,\widehat{x}_{ij}-I_{ij}\right)&i=1,\ldots,N;\;j=1,\ldots,N\\
    \widehat{x}_{i0}&i=1,\ldots,N;\;j=0.
\end{matrix}
\right.
\end{equation}
Thus, if stocks are plentiful, the firm will prefer drawing from them instead of buying new inputs. In the meantime, households calculate their own consumption target for good $i$ as detailed below and they also decide, given offered wages, how much labour they are willing to supply, a quantity we call $L^{\text{s}}(t)$ that now may not correspond to full employment. 

\subsubsection{{Exchanges \& Price/Wage Updates}}
\label{subsec:exchanges}

At this point, firms start hiring workers from the job market, albeit without exceeding the total supply of work $L^{\text{s}}$, i.e.
\begin{equation}
    \ell^{ }_i(t) = \ell_i^{\text{d}}(t) \min\left(1, \frac{L^\text{s}(t)}{L^{\text{d}}(t)}\right); \qquad L^{\text{d}}(t):=\sum_i \ell_i^{\text{d}}(t),
\end{equation}
where $\ell_i^{ }$ is the real amount of work contracted by firm $i$.
Workers are paid the same wage $p_0(t)$ independently of their employer.\footnote{Extending the model to firm-dependent wages would be interesting but requires one to move beyond a representative agent description of the household sector.} Conventionally, we prescribe that wages are paid immediately upon hiring -- regardless of any technical unemployment in the future caused by shortages of inputs -- which allows the household to compute its available budget for the present period:
\begin{equation}
    B(t)=S(t)+p_0(t)\sum_i \ell^{ }_i(t),
    \label{eq:budget_real}
\end{equation}
with $S(t)$ the household's savings. The household's demands for goods $C_i^{\text{d}}(t)$ are computed in section \ref{sec:households}.
%\footnote{\label{fn:nu_parameter}Note that, since consumption demands are computed with an optimistic estimation of the future budget (full employment), the real budget may be smaller. The household may choose to adapt its consumption demands with a behavioural parameter $\nu$ 
%\begin{equation}
%    C_i^{\text{d}}(t) \leftarrow C^{\text{d}}_i(t) \left(\nu + (1-\nu)\min\left(1,\frac{B(t)}{\mathbb{E}_t[B(t)]}\right)\right),
%    \label{eq:consumption_demands}
%\end{equation}
%which interpolates between a ``wait-and-see'' ($\nu=1$) behaviour and   a ``virtuous'' anticipatory ($\nu=0$) behaviour}.

Trading can now start, whereby firms sell their production and buy the goods they need, in a way to satisfy the constraint that the total amount of goods sold cannot exceed production plus inventory, viz. 
\begin{equation}
    C_i(t) + \sum_{j} x_{ji}(t) \leq \mathscr{S}_i(t) \equiv y_i(t) + I_{ii}(t).
\end{equation}

If demand exceeds supply, buyers are satisfied proportionally to their posted demand, and so quantities $x$ that are effectively exchanged are given by
\begin{equation} 
x^{ }_{ji}(t) = x_{ji}^{\text{d}}(t) \min\left(1,\frac{\mathscr{S}_i(t)}{\mathscr{D}_i(t)}\right); \qquad \mathscr{D}_i(t):=
C^{\text{d}}_i(t) + \sum_{j} x^{\text{d}}_{ji}(t),
\end{equation} 
where $\mathscr{D}_i(t)$ is the total demand for good $i$ at time $t$. The equation for $C_i(t)$ is slightly more convoluted because we do not give households access to debt, see Eq.~\eqref{eq:real_cons} below.

At this point, firms have an exact knowledge of their earnings and expenses. Their profit at round $t$ may now be computed:
\begin{equation}
    \pi_i(t)= p_i(t) \left(\sum_j x_{ji}(t) + C_i(t) \right)
    -  \left( \sum_j p_j(t) x_{ij}(t) + p_0(t) \ell_i(t) \right):= \mathscr{G}_i(t)-\mathscr{L}_i(t),
    \label{eq:realized_profits}
\end{equation}

Firms also know how much excess supply or demand they actually registered:
\begin{equation}
    \mathscr{E}_i(t)= {\mathscr{S}_i(t)}-
    {\mathscr{D}_i(t)}.
    \label{eq:realized_balance}
\end{equation}

Realised profits and supply/demand imbalances then generate price updates. We describe them exactly as in Eq.~\eqref{eq:price_rule_toy}, which now reads:
\begin{equation}
    \log\left(\frac{p_i(t+1)}{p_i(t)}\right) = -2\alpha \, \frac{\mathscr{E}_i(t)}{\mathscr{S}_i(t)+\mathscr{D}_i(t)}- 2\alphap \, \frac{\pi_i(t)}{\mathscr{G}_i(t)+\mathscr{L}_i(t)},
    \label{eq:final_price_update}
\end{equation}
where all quantities are now known.\footnote{Since markets do not clear and profits are non zero, we choose symmetric normalisation factors involving the average of supply and demand for the first term, and the average of sales and costs for the second.}

Prices are updated due to tension between supply and demand, which is in our framework a natural channel for inflation or deflation. By the same token, tensions on the job market are bound to lead to wage updates, which we postulate to be of the same form as for price updates, namely
\begin{equation}
        \log\left(\frac{p_0(t+1)}{p_0(t)}\right) = 2\omega \, \frac{L^{\text{d}}(t) - L^{\text{s}}(t)}{L^{\text{d}}(t) + L^{\text{s}}(t)},
    \label{eq:final_wage_update}
\end{equation}
meaning that excess demand of labour increases wages, and vice-versa. This rule implements a Phillips curve at each time step (see \cite{Phillips1958} and \cite{Blanchard2016}). One could also use an asymmetric update rule, accounting for the fact that lowering nominal wages is more difficult than raising them. Finally, one could also consider adding a direct coupling between the inflation of the price of goods and wages, as an extra term in the right hand side of Eq. \eqref{eq:final_wage_update}. 

\subsubsection{{Production}}

The last epoch corresponds to the start of production. Firm $i$ uses the workforce $\ell_i$, along with available quantities $x^{\text{a}}_{ij}$ that depend on exchanges $x$, optimal inputs $\widehat x$ and inventories $I$, as
\begin{equation}
    x^{\text{a}}_{ij}(t)=x^{ }_{ij}(t)+\min\left(I_{ij},\widehat{x}_{ij}\right).
    \label{eq:quantities_prod}
\end{equation}

Indeed, if the inventory $I$ allows to provide for optimal input $\widehat{x}$, then no demand is posted (see Eq.~\eqref{eq:Qd_Qhat}): $x^{ }=0$ and $x^{\text{a}}=\widehat{x}$. Otherwise, the firm acquired a quantity $x^{ }$ that now adds to available stocks, and so $x^{\text{a}}=x^{ }+I \leq \widehat{x}$. Note that labour cannot be stored, and therefore $I_{i0} = 0$ at all times.

Now that all of the available inputs $x^{\text{a}}_{ij}$ and labour $\ell_i$ are known, the outputs are determined by the firms' production functions, which in the Leontief case with $b=1$ entails:
\begin{equation}
    y_i(t+1)=z_i(t) \min\left[\min_{j} \left(\frac{x^{\text{a}}_{ij}(t)}{J_{ij}}\right), \frac{\ell_i(t)}{J_{i0}}\right].
\end{equation}

The firms' inventories of their own production is also updated, as
\begin{equation}
    I_{ii}(t+1)= e^{-\sigma_i} \left(y_i(t) + I_{ii}(t) - \sum_j x^{ }_{ji}(t) \right),
\end{equation}
where the decay factor $\sigma_i$ measures the perishability of good $i$. For durable goods, $\sigma_i \ll 1$ and $e^{-\sigma_i}\approx 1$, whereas $\sigma_i\gg1$ and $e^{-\sigma_i} \ll 1$ for perishable goods.

Furthermore, in the Leontief framework total production is limited by the scarcest input, which is therefore depleted during production, leaving a fraction of the other inputs unused. We denote
\[
j^\star(i)=\underset{j}{\arg\min}\left(\frac{x^{\text{a}}_{ij}}{J_{ij}}\right),
\]
so that we can write the fraction of inputs $k\neq j^{\star}(i)$ effectively used as
\begin{equation}
    x^{\text{u}}_{ik}(t)=\frac{J_{ik}}{J_{ij^\star(i)}}x_{ij^\star(i)}^{\text{a}}.
    \label{eq:used_quantities}
\end{equation}

The unused inputs add to firm inventories, and their update may be written using Eq.~\eqref{eq:quantities_prod}, as 
\begin{equation}
I_{ik}(t+1)=e^{-\sigma_k} \left(x_{ik}^{\text{a}}-x_{ik}^{\text{u}}\right).
    \label{eq:inventories_off_diag}
\end{equation}
%rather tricky and finally yields
%\begin{equation}
%    I_{ik}(t+1)=e^{-\sigma_k}\left[\mathbb{I}\left[I_{ik}(t) > \widehat{x}_{ik}\right]\left(I_{ik}(t)-x_{ik}^{\text{u}}\right)+\mathbb{I}\left[I_{ik}(t) < \widehat{x}_{ik}\right]\left(x_{ik}-x_{ik}^{\text{u}}\right)\right].
%    \label{eq:inventories_off_diag}
%\end{equation}

%Let us walk through the off-diagonal update
%\begin{itemize}
%    \item If $I_{ik}>\widehat{x}_{ik}$ then stocks of good $k$ where initially enough to meet the production target. As a result $x_{ik}^d=x_{ik}^{ }=0$ and the quantity for production is $\widehat{x}_{ik}$ that firms took from our stock $I_{ik}$. The stock update amounts to substract the actually used quantity (given that $k$ may not be the scarcest input) from $I_{ik}$.
%    \item If $I_{ik}\in]0,\widehat{x}_{ik}[$, then stocks were not enough to meet the production target, and, in the end, they will be completely used. The quantity that we may need to stock is the remainder $x_{ik}^p-x_{ik}^{u}\geq0$ since $k$ might not be the scarcest input.
%\end{itemize}

Finally, for numerical purposes, it is convenient to rescale new prices $p_i(t+1)$ by the new wage $p_0(t+1)$ to avoid exponential growth (or decay) of prices induced by inflation (or deflation), effectively measuring prices in units of wages. We therefore set:\footnote{Note that profits and savings should also be appropriately rescaled, when necessary, e.g. $S(t+1) \to S(t+1)/p_0(t+1)$, etc.}
\begin{equation}
    p_i(t+1) \longrightarrow \frac{p_i(t+1)}{p_0(t+1)}; \qquad p_0(t+1) \longrightarrow 1.
\end{equation}

This concludes the third and last epoch of the time step. The process is then repeated at time $t+1$, with productions $y_i(t+1)$ and prices $p_i(t+1)$.

To close the model, we now need to specify how firms estimate their future profits/losses and excess/deficit production. The behaviour of households must also be spelled out, to allow for the determination of the demand of goods and the supply of labour. 

\subsection{Expected Profits and Imbalances}
\label{sec:extensions}

We may write the expected profit of firm $i$ as
\begin{equation}
\mathbb{E}_t[\pi_i]= p_i(t) \left(\sum_j \mathbb{E}_t[x_{ji}] + \mathbb{E}_t[C_i] \right)
-  \left( \sum_j p_j(t) \mathbb{E}_t[x_{ij}] + p_0(t) \mathbb{E}_t[\ell_i] \right),
\label{eq:expected_profits}
\end{equation}
showing that in the planning phase firms must estimate future goods and labour demand, which we will denote generically as $\mathbb{E}_t[x]$.
Similarly, the expected excess production is also  a function of $\mathbb{E}_t[x]$:
\begin{equation}
   \mathbb{E}_t[\mathscr{E}_i] = y_i(t) + I_{ii}(t) - \sum_j \mathbb{E}_t[x_{ji}] - \mathbb{E}_t[C_i]. 
   \label{eq:expected_balance}
\end{equation}

The simplest assumption we can adopt is that firms are ``sticky'', and estimate all future demands to be equal to their last observation (which follows the rationale that firms produce in order to meet total demand), i.e.
\begin{equation}
    \mathbb{E}_t[x] = x^{\text{d}}(t-1).
\end{equation}
%{Using past demands for forecasting future surpluses and profits is quite natural since firms strive to reduce imbalances and therefore to meet expressed demands.} 
However, some immediate generalisations come to mind. For example, firms may also factor in {\it realized} quantities $x^{ }(t-1)$ in their estimate, and set as a learning rule
\begin{equation} \label{eq:def_lambda}
    \mathbb{E}_t[x] = \lambda x^{\text{d}}(t-1) + (1 - \lambda) x^{ }(t-1),
\end{equation}
where $\lambda \in [0,1]$ is a parameter. Our ``sticky'' assumption that will be used henceforth thus corresponds to $\lambda=1$.

Another possible generalisation is that firms use a more sophisticated learning rule that allows them to estimate $\mathbb{E}_t[x]$ using time-series analysis, the simplest of which is ``constant gain learning'' (equivalent to computing the exponential moving average) of past realised demands. This is similar to the AR1 estimation of economic growth used by the agents of \cite{poledna} in their decision-making process. Trend-following, extrapolative rules may also be considered. All these extensions are beyond the scope of the present paper; at this stage, our ambition is to set up a minimal consistent framework, free of spurious numerical instabilities, and that can converge to competitive equilibrium in some region of parameter space.

\subsection{Household Demand and Labour} 
\label{sec:households} 

\subsubsection{Work-elastic Households}

As in standard macroeconomic models, we assume that households are represented by a single representative agent with a certain disutility for work, who seeks to maximise the following utility function\footnote{We restrict to a ``myopic'' optimisation here, that does not take into account the long-term forecasts and desires of the household. Inter-temporal effects would require to add interest rates, which we completely disregard in the present study.}
\begin{equation}
\mathcal{U}(t)=\sum_j\theta_j\log{C_j(t)}-\frac{\Gamma}{1+\varphi}\left(\frac{L(t)}{L_0}\right)^{1+\varphi},
    \label{eq:full_utility}
\end{equation}
where $L(t)=\sum_j\ell_j(t):=\sum_j x_{j0}(t)$ is the total amount of work provided by the representative household. The so called Frisch elasticity index $\varphi$, after the eponymous~\cite{Frisch1959}, gives a measure of the convexity of the disutility of work, $L_0$ is the scale of the amount of work that the household is able to provide and $\Gamma$ is a parameter that can be set to unity without loss of generality. In the limit $\varphi \to \infty$, households are indifferent to the amount of work provided $L(t) < L_0$, but refuse to work more than $L_0$.
With an utility function of this form, the household may then compute its optimal demand for good $i$, $C^{\text{d}}_i(t)$ which it will set as a consumption target for period $t$, and the optimal amount of labour $L^{\text{s}}(t)$  it is willing to provide to firms. 

\subsubsection{The Optimization Sequence}

To compute the aforementioned quantities, the household needs to know its current savings $S(t)$ and anticipate its income for the next period. The expected utility is estimated with optimistic forecasts (i.e. consumption demand will be met and offered labour will be fully utilised). Wage $p_0(t)$ and prices $p_i(t)$, on the other hand, are all known before the ``Exchange and Update'' stage, see \ref{subsec:exchanges}. Hence,
\begin{equation}
    \mathbb{E}_{t}[\mathcal{U}]=\sum_{i}\theta_{i}\log{C^{\text{d}}_i(t)}-\frac{1}{1+\varphi}\left(\frac{L^{\text{s}}(t)}{L_0}\right)^{1+\varphi},
    \label{eq:expected_utility}
\end{equation}
with an expected budget constraint that reads\footnote{In a follow-up paper, we shall introduce precautionary savings and interest rates, which lead to the appearance of inflationary equilibria. Note also that we assume that all goods are immediately consumed by the household, which does not make much sense for durable goods. This also could be reconsidered.}
\begin{equation}
    \sum_{i} {p}_{i}(t)C^{\text{d}}_i(t)=p_0(t) L^{\text{s}}(t)+S(t):=\mathbb{E}_t[B],
    \label{eq:budget_constraint}
\end{equation}
where $\mathbb{E}_t[B]$ is the expected (or in fact hoped for!) budget. For convenience, we denote as $W_0(t)=p_0(t) L_0$ the wage associated to $L_0$ work-hours.

The household optimises its expected utility while enforcing the budget constraint using a Lagrange multiplier $\mu(t)/W_0$, so that\footnote{Although not necessary for the purpose of the present paper, it is important to allow for confidence effects, which can lead to endogenous crises (see e.g. \cite{morelli_pnas}). One possibility is to couple the consumption propensity to the unemployment level, taken as a proxy of consumer confidence, i.e.: 
\begin{equation}
\log{\left(\frac{\theta_i(t)}{\theta_i^0}\right)}=2\omega^\prime \, \frac{L^{\text{d}}(t) - L^{\text{s}}(t)}{L^{\text{d}}(t) + L^{\text{s}}(t)},
    \label{eq:varying_theta}
\end{equation}
where $\theta_i^0$ are the baseline values for consumption preferences. In the following, we will fix $\omega'=\omega$.}
{
\begin{subequations}
\begin{eqnarray}
C^{\text{d}}_i(t)&=&L_0 \, \frac{\theta_i}{\mu(t)} \, \frac{p_0(t)}{p_i(t)}\label{eq:demanded_cons}\\
    L^{\text{s}}(t)&=&L_0 \, \mu(t)^{1/\varphi}\label{eq:labour_supply}.
\end{eqnarray}
\end{subequations}
}
In order to find $\mu(t)$, one must enforce \eqref{eq:budget_constraint}. We find the following equation on $\mu(t)$:
\begin{equation}
    \mu^k(t)+ \frac{S(t)}{W_0(t)} \, \mu(t)  = \bar{\theta},
    \label{eq:fixed_point_mu}
\end{equation}
with $k=1+1/\varphi$ and $\bar{\theta}=\sum_i \theta_i$. For instance, if $\varphi=\infty$ (constant work offer $L^{\text{s}}(t)=L_0$), we have 
\begin{equation}
\mu(t)=\frac{\bar{\theta}W_0(t) }{W_0(t)+S(t)}.
    \label{eq:mu_phi_inf}
\end{equation}

When $\varphi=1$ (a common value found in the literature and corresponding to a quadratic work-disutility), we have 
\begin{equation}
    \mu(t)=\frac{1}{2 W_0(t) }\left(\sqrt{S(t)^2 +4 \bar{\theta} W_0(t)^2 }-S(t)\right).
    \label{eq:mu_phi_one}
\end{equation}
Note, interestingly, that high savings lead to reduced labour supply. Also, because of possible involuntary unemployment, the household may want to consume more than it is able to spend when $L^{\text{d}}(t) < L^{\text{s}}(t)$. 

A final word on the scaling behaviour of these quantities with $N$ is in order. For large $N$ we expect that the size of the household sector will also be of order $N$. Noting that $\bar{\theta}$ is also of order $N$, one finds the following befitting scaling laws if we choose $L_0 \sim \sqrt{N}$:
\begin{equation} 
\mu \sim \sqrt{N}; \qquad 
L^{\text{s}} \sim N; \qquad 
C^{\text{d}}_i(t) \sim 1,
\end{equation}
meaning that total work-hours and total consumption are proportional to the size of the population, as it should be. 

%\subsubsection{Confidence Effects}

%In the setup above, households consume regardless of the state of the economy. Although not necessary for the purpose of the present paper, we believe it is important to introduce a notion of confidence in the economy by coupling the consumption propensity to the unemployment level, taken as a proxy of consumer confidence. Hence we allow the utility of consumption to vary as:
%\begin{equation}
%\log{\left(\frac{\theta_i(t)}{\theta_i^0}\right)}=2\omega^\prime \, \frac{L^{\text{d}}(t) - L^{\text{s}}(t)}{L^{\text{d}}(t) + L^{\text{s}}(t)},
%    \label{eq:varying_theta}
%\end{equation}
%where $\theta_i^0$ are the baseline values for consumption preferences. 

%In a booming economy where demand for workforce is high, households will tend to consume more (increased $\theta$'s); whereas it will consume less in a failing economy with high unemployment. 

%\red{[Peut-être détailler un peu pourquoi c'est important d'ajouter ça: If one omits to account for this effect... On argumente un modèle minimal donc il faut justifier pourquoi chaque feature du modèle est "essentiel", surtout les règles de pouce.]} 

\subsubsection{Savings Update}

%{\color{red}As already explained in the previous section, if $L^\text{s}$ exceeds $L^\text{d}=\sum_i \ell_i^\text{d}$, partial unemployment ensues; on the other hand, when $L^\text{s} < L^\text{d}$, firms are only allotted a fraction of their needs:
%\begin{equation}
%    \ell_i(t)=\ell^{\text{d}}_i(t)\min\left(1,\frac{L^\text{s}(t)}{L^\text{d}(t)}\right).\label{eq:labour_demand}
%\end{equation}
%This allows the household to compute their real budget, possibly corrected downwards by unemployment:  
%\begin{equation}
%    B(t)=S(t)+L(t) \leq \mathbb{E}_t[B], \qquad \text{with} \qquad L(t) = \min\left (L^\text{s}(t), L^\text{d}(t) \right).
%    \label{eq:budget2}
%\end{equation}

Because we do not allow households to borrow in the present version of the model, real consumption must be adjusted in the case of partial unemployment. In this case, the available budget is necessarily smaller than what was hoped, leading to a realised consumption:
%\footnote{An extension could be imagined, where workers borrow money to compensate the gap between the expected and realised budget, but we do not consider this in the present version of our model.}
\begin{equation}
C^{\text{r}}_i(t) = C^{ }_i(t) \min\left(1,\frac{B(t)}{\sum_j p_j(t) C^{ }_j(t)}\right); \qquad C_i(t)=C_i^{\text{d}}(t) \min\left(1,\frac{\mathscr{S}_i(t)}{\mathscr{D}_i(t)}\right),
\label{eq:real_cons}
\end{equation}
with $B(t)$ their available budget computed in \eqref{eq:budget_real}
.
The difference between $C_i(t)$ and $C^{\text{r}}_i(t)$, if positive, is added to the inventory $I_{ii}(t)$ of firm $i$. The households' savings are then updated as:
\begin{equation}
    S(t+1)=B(t)-\sum_i p_i(t) C^{\text{r}}_i(t).
\end{equation}

%Prior to being hired, households had computed their demand based on their expected budget $\mathbb{E}_t\left[B(t)\right]$, see section \ref{sec:households} below. However, once they are hired they may revise this demand, taking into account their {\it real} budget, as it could possibly be less than what they expected.

%We thus introduce an interpolation parameter $\nu$ between these two scenarios and set the actual household demand to
%\begin{equation}
%    C_i^{\text{d}}(t) = C^{\text{d}}_i(t) \left(\nu + (1-\nu)\min\left(1,\frac{B(t)}{\mathbb{E}_t[B(t)]}\right)\right),
%    \label{eq:consumption_demands}
%\end{equation}
%where $\nu=1$ corresponds to ``wait-and-see'' behaviour while $\nu=0$ accounts for ``virtuous'' anticipatory behaviour.  For definiteness, we focus on the case $\nu=1$.%footnote{Interestingly enough, the anticipatory behaviour leads to strong instabilities. An interpretation could be that the household reacts ``too quickly'' to information, leading to coordination breakdown as in \cite{bonart2014instabilities}.}

\subsection{Discussion}

The above steps look rather tedious and considerably more complex than the simple logic behind our first ``naive'' model. Nonetheless, they are quite natural when one decomposes all the stages of a real production process. But more importantly, we have found that short-circuiting any of these steps leads to inconsistent dynamics with spurious instabilities, reflecting that natural constraints are in fact violated. Furthermore, the approach of behaviour modelling as a series of actions or sequence of events is a typical feature of ABMs, where the ordering of these events is done in a coherent way as to ensure causality. 

An important difference with the naive version of section~\ref{sec:naive} is the large number of update rules that necessarily involve cusps, such as those involving taking the maximum or minimum of two expressions, see section \ref{sec:cones} below. Furthermore, the number of thumb rules used by firms and households to aid their decision has increased, and so has the number of parameters that are needed to describe a given instance of our toy economy. 

Therefore, and in spite of the fact that the naive model allows a fair  understanding of certain regions of the parameter-space of the full model, we cannot reasonably attempt an exhaustive description using analytical tools only. We  therefore resort to a numerical exploration of its properties, using computer simulations that are described in detail in the pseudo-code provided in Appendix~\ref{ap:pseudo_code}. We also provide access to an open access simulation tool that allows the reader to explore different configurations here: \url{https://yakari.polytechnique.fr/dash}.
%\red{CAN CHANGE!}.

\section{A Numerical Study\label{sec:numerical_study}}

The following section is a numerical investigation of the very rich phenomenology of the above model, supplemented with some analytical results when possible. Because of the relatively large number of parameters, we only investigate here some specific ``cuts'' in parameter space, but believe that these cuts are representative of all the possible dynamical classes that the model can generate.  

To facilitate reading this section, we will first recall the different parameters that can be adjusted. We will then explore the different types of dynamical trajectories that can be observed in our toy economy, and classify them into different ``phases''. This idea comes from physics, where the macroscopic properties of a system can be split into different parameter regions where its aggregate behaviour is qualitatively the same. These regions only depend on the values taken by a handful of parameters that describe the system; an eloquent example is that of water, which depending on the pressure or temperature can be in either the liquid, solid or gas phase. 

We will therefore present the following  ``phase diagrams'' that summarise the influence of the parameters on the broad dynamical behaviour of our model, an idea that was already advocated for economic Agent-Based Modelling in~\cite{gualdi2015tipping}. 

\subsection{Summary of Parameters\label{sec:summary_param}}

The different parameters introduced in the previous sections may be split into two categories: static parameters, describing the production network and the production function, and dynamic parameters, describing the evolution of prices, labour and outputs. We provide an overview of them and of the typical values we assign to them in our simulations below.

\paragraph{Static Parameters}

\begin{enumerate}
    \item Number of firms $N$ -- here $N=100$.
    \item Type of network -- here a random regular directed network, see \cite{MCKAY1981203, KimRegularGraph2006}, where each firm has the same number of clients and suppliers $d=15$.
    \item CES production function -- here a Leontief production function ($q=0^+$) with a return to scale parameter $b=0.95$.\footnote{\label{fn:rts_below_one}Choosing $b$ slightly below unity helps stabilising the dynamics and also prevents the relaxation time from diverging as the smallest eigenvalue of the production matrix $\varepsilon \to 0$. Arguments to this effect are detailed in Appendix \ref{ap:eq}.}
    \item The smallest eigenvalue $\varepsilon$ of the production matrix $\boldsymbol{\mathcal{M}}$, which for large values corresponds to a stable economy. 
    \item Firm inter-linkages $J_{ij}$, which we  take to be $1$ when firms $i$ and $j$ are linked and zero otherwise. 
    \item Firm productivities $z_i$, first set to 1 and then adapted to adjust $\varepsilon$ to take the required value.\footnote{Modifying the productivity factors as $z^\prime=z+\varepsilon-\min{\rm Sp}\left(\mathcal{M}\right)$ makes the minimum eigenvalue of $\mathcal{M}$ equal to $\varepsilon$.}
    \item Baseline household consumption preferences $\theta_i^0$, modelled by iid uniform random variables rescaled to have $\sum_i\theta_i^0=1$.
    \item Work disutility Frisch index, set to $\varphi=1$ (quadratic disutility of labour) and scale of workforce set to $L_0=1$.
    \item The behavioural extrapolation parameter $\lambda$, defined in Eq. \eqref{eq:def_lambda}, is set to $1$.
\end{enumerate}

Note that we shall also simulate our model on more realistic models of firm networks, including actual input-output networks constructed from the FactSet database (\cite{factset}), see Appendix \ref{ap:real_net}.

\paragraph{Dynamic Parameters}

\begin{enumerate} 
    \item Parameters describing restoring forces: $\alpha, \alphap, \beta, \betap$, (see Eqs.~\eqref{eq:final_target_update}-\eqref{eq:final_price_update}). We restrict ourselves to the case $\betap=\alphap=\beta=\alpha$ and scan for varying values of $\alpha$.
    \item Phillips curve parameter $\omega$, relating wages to tensions in the job market (see Eq.~\eqref{eq:final_wage_update}).
    \item Confidence parameter, relating consumption propensities to unemployment: $\omega^\prime$  (see Eq.~\eqref{eq:varying_theta}). For this study, we take $\omega'=\omega$. 
    \item Perishability parameters $\sigma_i$ describing the speed of decay of good $i$, all taken as $\sigma_i=\sigma$ except when otherwise indicated.
\end{enumerate}

These choices therefore reduce the number of parameters to explore to four: $\varepsilon$ (network stability), $\alpha$ (strength of restoring forces), $\omega$ (Phillips curve parameter) and~$\sigma$ (perishability). We will now show how varying them may lead to a very rich phenomenology.

\subsection{Perturbations Around Equilibrium \& Conewise-linear Dynamics}\label{sec:cones}

As stressed above, the naive model of section~\ref{sec:pert_naive} can be linearised, leading to a complete analytical estimation of the time needed to reach equilibrium. The cusps of the full model, however, imply that perturbative analysis produces at best piecewise-linear equations.\footnote{This, as the general time-line framework outlined in section~\ref{ssec:time-line}, is a feature common to other ABMs, such as Mark-0 \cite{gualdi2015tipping}, or the ABM recently developed in~\cite{del2020supply}.}

To be precise, let us attempt to linearise the different update rules by writing $\delta x(t)=x(t)-x_{\text{eq}}$ for the perturbed value of any quantity $x$ and expanding the different equations to lowest order in $\delta \cdot$. When applied to the flows $x_{ji}$ one gets:
\begin{equation}\label{eq:lin_non_lin}
    \delta x_{ji}^{ }(t)=\delta x_{ji}^{\text{d}}(t)+\frac{x_{eq,ji}}{z_i\preq{i}}\min\left(0,\delta\mathscr{S}_i(t)-\delta\mathscr{D}_i(t)\right).
\end{equation}

Depending on the sign of $\delta\mathscr{S}_i(t)-\delta\mathscr{D}_i(t)$, the flow of exchanged goods is characterised by two different linear equations, rendering the system piecewise-linear. The same feature also holds for exchanged work (replacing $\delta\mathscr{S}_i(t)-\delta\mathscr{D}_i(t)$ by $\delta L^{\text{s}}(t)-\delta L^{\text{d}}(t)$) and realised consumption (where the switch depends on $\delta\mathscr{S}_i(t)-\delta\mathscr{D}_i(t)$ as well as the budget constraint which is more cumbersome to write, see \eqref{eq:real_cons}). But this means, perhaps surprisingly,  that there does not exist a limiting case where the full model would boil down to the ``naive'' model of section \ref{sec:naive}.

Linearising around equilibrium yields piecewise-linear dynamics that can be described by the evolution of a $(N^2+4N+1)$-dimensional state vector that we denote by $\mathbf{U}(t)$, following the notation of section \ref{sec:pert_naive}. This vector encodes perturbations on stocks $\delta I_{ij}(t)$ (stacking the columns of the $N\times N$ stocks-matrix in an $N^2$-dimensional vector), current production targets $\delta\widehat{\gamma}_i(t+1)$, past targets $\delta\widehat{\gamma}_i(t)$, production levels $\delta\gamma_i(t)$, prices $\delta p_i(t)$, and finally the household's savings $\delta S(t)$. 

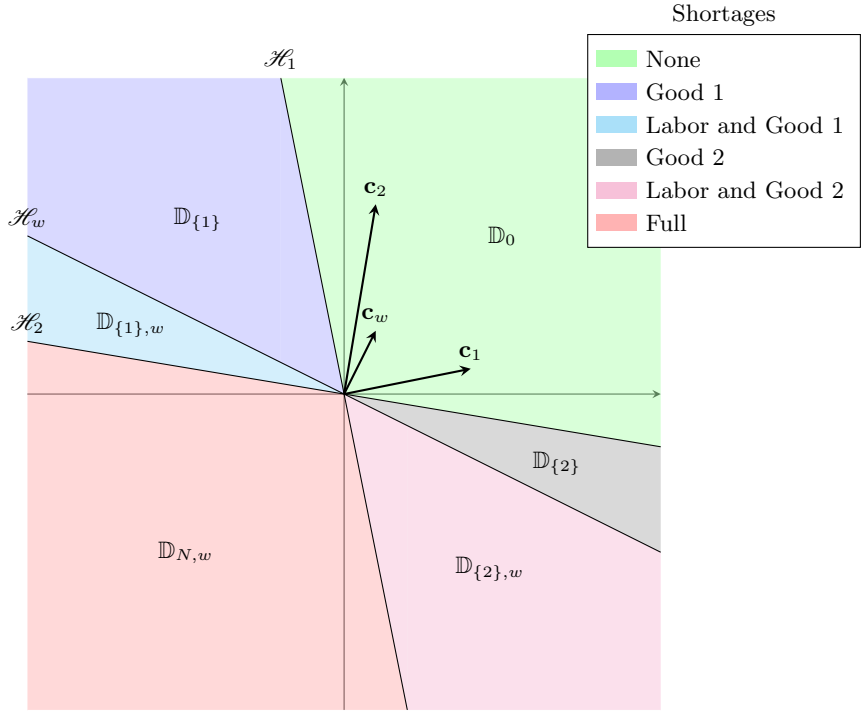
\begin{figure}
    \centering
    \begin{tikzpicture}
\begin{axis}[
width=10cm, %set bigger width
height=10cm,
xmin=-5,xmax=5,
ymin=-5,ymax=5,
axis x line = middle,
axis y line = middle,
%axis line style={->},
%axis on top,
ticks = none,clip=false,
]
\addplot[name path=H1,domain={-1:1}] {-2*x/0.4} node[pos=0, above]{$\mathscr{H}_1$};
\addplot[name path=H2,domain={-5:5}] {-0.5*x/3} node[pos=0, above]{$\mathscr{H}_2$};
\addplot[name path=Hw,domain={-5:5}] {-0.5*x/1} node[pos=0, above]{$\mathscr{H}_w$};
\addplot[name path=Up, white, domain={-5:5}] {5};
\addplot[name path=Down, white, domain={-5:5}] {-5};

% \addplot[name path=G,green,domain={-.2:6}] {-0.1*x^2+2}node[pos=.1, below]{$g$};

\addplot[color=blue!30, opacity=0.5, pattern color=brown!100]fill between[of=Up and Hw, soft clip={domain=-5:-1}];
\addplot[color=blue!30, opacity=0.5, pattern color=brown!100]fill between[of=H1 and Hw, soft clip={domain=-1:0}];

\addplot[color=magenta!30, opacity=0.5, pattern color=brown!100]fill between[of=Down and Hw, soft clip={domain=1:5}];
\addplot[color=magenta!30, opacity=0.5, pattern color=brown!100]fill between[of=H1 and Hw, soft clip={domain=0:1}];

\addplot[color=cyan!30, opacity=0.5]fill between[of=H2 and Hw, soft clip={domain=-5:0}];
\addplot[color=black!30, opacity=0.5]fill between[of=H2 and Hw, soft clip={domain=0:5}];
%\addplot[pattern=north east lines, pattern color=gray!100]fill between[of=H1 and Hw, soft clip={domain=0:5}];
%\addplot[color=blue!10]fill between[of=H2 and Hw, soft clip={domain=-5:0}];

\addplot[color=green!30, opacity=0.5]fill between[of=Up and H2, soft clip={domain=0:5}];
\addplot[color=green!30, opacity=0.5]fill between[of=Up and H1, soft clip={domain=-1:0}];
\addplot[color=red!30, opacity=0.5]fill between[of=Down and H2, soft clip={domain=-5:0}];
\addplot[color=red!30, opacity=0.5]fill between[of=Down and H1, soft clip={domain=0:1}];

\node[coordinate] (zero) at (axis cs:0,0){};
\node[coordinate,label=$\mathbf{c}_1$] (c1) at (axis cs:2,.4) {};
\node[coordinate, below, label=$\mathbf{c}_2$] (c2) at (axis cs:0.5,3) {};
\node[coordinate, below, label=$\mathbf{c}_w$] (cw) at (axis cs:0.5,1) {};
\draw[-stealth, thick] (zero) -- (c1);
\draw[-stealth, thick] (zero) -- (c2);
\draw[-stealth, thick] (zero) -- (cw);
% \node[coordinate,label=No shortage] at (axis cs:2,2){};
% \node[coordinate,label=Full shortage] at (axis cs:-2,-2){};
% \node[coordinate,label=Good 2 shortage] at (axis cs:2,-2){};
% \node[coordinate,label=Good 1 shortage] at (axis cs:-2,2){};

\node (D+) at (axis cs:2.5,2.5){$\mathbb{D}_0$};
\node (Dall) at (axis cs:-2.5,-2.5){$\mathbb{D}_{N,w}$};
\node (D1) at (axis cs:-2.304,2.764){$\mathbb{D}_{\{1\}}$};
\node (Dw1) at (axis cs:-3.362,1.092){$\mathbb{D}_{\{1\},w}$};
\node (Dw2) at (axis cs:2.304,-2.764){$\mathbb{D}_{\{2\},w}$};
\node (D2) at (axis cs:3.362,-1.092){$\mathbb{D}_{\{2\}}$};

\matrix [draw, label=above:Shortages, fill=white] at (6,4) {
  \node [rectangle, fill=green!30, label=right:\text{None}] {$\quad$}; \\
  \node [rectangle, fill=blue!30, label=right:\text{Good 1}] {$\quad$}; \\
  \node [rectangle, fill=cyan!30, label=right:\text{Labor and Good 1}] {$\quad$}; \\
  \node [rectangle, fill=black!30, label=right:\text{Good 2}] {$\quad$}; \\
  \node [rectangle, fill=magenta!30, label=right:\text{Labor and Good 2}] {$\quad$}; \\
  \node [rectangle, fill=red!30, label=right:\text{Full}] {$\quad$}; \\
};
\end{axis}

\end{tikzpicture}
    \caption{Example of the cone-separation for a schematic two dimensional space. Solid black lines represent the hyperplanes $\mathscr{H}_{1,2,w}$ separating shortage/no-shortage cones, they are orthogonal to cone vectors $\mathbf{c}_{1,2,w}$.
    Shaded areas represent the different cones of possible shortages along with the associated stability matrices. Note that for the given configuration of vectors, the matrices $\mathbb{D}_N$ and $\mathbb{D}_{0,w}$ cannot exist.}
    \label{fig:2d-conewise-lin}
\end{figure}

To study possible switches in the conditions defining our piecewise-linear dynamics, let us define two vectors $\mathbf{c}_i$ and $\mathbf{c}_w$ such that 
\begin{equation}
    \mathbf{c}_i^\top\mathbf{U}(t)=\delta\mathscr{S}_i(t)-\delta\mathscr{D}_i(t),\quad \mathbf{c}_w^\top\mathbf{U}(t)=\delta L^{\text{s}}(t)-\delta L^{\text{d}}(t).
\end{equation}
Each vector defines a hyperplane $\mathscr{H}_i=\{\mathbf{c}_i\}^{\perp}$ (resp. $\mathscr{H}_w=\{\mathbf{c}_w\}^{\perp}$) separating state space into two regions:
\begin{itemize}
    \item the no shortage region for $i$, $\mathscr{H}_i^+$ (resp. $\mathscr{H}_w^+$) where $\mathbf{c}_i^\top\mathbf{U}(t)>0$ (resp. $\mathbf{c}_w^\top\mathbf{U}(t)>0$) where $i$'s supply is enough to cope with demand (resp. work offer is enough to cope with work demands);
    \item the shortage region for $i$, $\mathscr{H}_i^-$ (resp. $\mathscr{H}_w^-$) where $\mathbf{c}_i^\top\mathbf{U}(t)<0$ (resp. $\mathbf{c}_w^\top\mathbf{U}(t)<0$) where $i$'s supply is not enough to cope with demand (resp. work offer is not enough to cope with work demands).
\end{itemize}
The intersection of these half-spaces defines regions of space called \emph{cones} in which the linearised dynamics are fully characterised by a well-defined stability matrix. Calling $S\subseteq [\![ 1,N ]\!]$, the set of firms with a shortage of goods, we define stability matrices in each cone as follows 
\begin{equation}
    \begin{aligned}
    \mathbf{U}(t)\in\bigcap_{s\in S}\mathscr{H}_s^-\,\cap\,\bigcap_{s'\in [\![ 1,N ]\!]\setminus S}\mathscr{H}_{s'}^+\,\cap\,\mathscr{H}_w^+&\Longleftrightarrow \mathbf{U}(t+1)=\mathbb{D}_S\mathbf{U}(t),\\
    \mathbf{U}(t)\in\bigcap_{s\in S}\mathscr{H}_s^-\,\cap\,\bigcap_{s'\in [\![ 1,N ]\!]\setminus S}\mathscr{H}_{s'}^+\,\cap\,\mathscr{H}_w^-&\Longleftrightarrow \mathbf{U}(t+1)=\mathbb{D}_{S,w}\mathbf{U}(t).
\end{aligned}
\end{equation}
If $S=\emptyset$ (no shortages), we call $\mathbb{D}_0$ and $\mathbb{D}_{0,w}$ the stability matrices without/with work shortage; and if $S=[\![ 1,N ]\!]$ (all firms have shortages), we call them $\mathbb{D}_N$ and $\mathbb{D}_{N,w}$. Fig.~\ref{fig:2d-conewise-lin} illustrates the previous construction in a schematic $2$-dimensional space.

{Although the dynamics inside each cone are linear, knowledge of the eigenvalues of the stability matrices is in general not sufficient to conclude on the stability of the entire system. Indeed, knowing whether a cone is preserved or not by its stability matrix is essential to understand the dynamics. If the linear dynamics corresponding to the stability matrix inside a cone preserves it, meaning that any trajectory starting in the cone will always be contained within it, the dynamics becomes trivial. On the other hand when this is not the case, a trajectory may switch back and forth between different cones, and it will therefore be described by a product of stability matrices. It is this product that one most study in order to conclude on the overall stability of the system. This can lead to quite complicated trajectories, where for example two different cones have stability matrices that are such that a trajectory starting in one inevitably ends up in the other and vice versa, leading to a pseudo-oscillation that can be stable in the long run. To our knowledge, the mathematical tools needed to account for these interesting cone-wise linear dynamics are not available in the general case.}

%Depending on the value of $\delta\mathscr{S}_i(t)-\delta\mathscr{D}_i(t)$, the system is characterized by two different linear stability matrices. As the system evolves in time, and even for very small fluctuations around equilibrium values, it may switch back and forth between these two stability matrices, explaining some of the behaviour we observe below.

%We also stress that the model is well-defined even if we choose initial conditions very far from equilibrium.  Preliminary studies show that initial conditions well above equilibrium values (up to 6 orders of magnitude) may still lead to an overall stable system. However, if the initial conditions are instead too small this can lead to divergences. 

\subsection{Phase Diagrams and Dynamical Classes}

For each set of values of the parameters ($\alpha$, $\omega$, $\sigma$, $\varepsilon$), we start from a random perturbation about equilibrium of relative magnitude $\delta=10^{-3}$, taking e.g. $p_i(t)=p_{\text{eq},i}(1+ \delta u)$ with $u$ uniform in $[-1,1]$.\footnote{When $\varepsilon<0$ and no competitive equilibrium can be defined, we start from random initial conditions between 1 and 2 for prices and productions.}  We then run the dynamics for $T=20000$ time-steps and consider only the last $2500$ to classify the trajectory into one of several classes that are detailed below.

The trajectories we will use to classify the behaviour of our model are those of  relative price differences $\overline{\delta p}(t):=p(t)/p_{\text{eq},i} - 1$.\footnote{The trajectories of produced quantities are qualitatively similar within each phase, except that, as expected, high prices correspond to production troughs, and vice versa.} In order to provide more vivid illustrations of some of these dynamical types, we have made firms slightly heterogeneous in their values of the parameters $\alpha$ and $\sigma$. In the figure captions below, the notation $\alpha, \sigma \in [A,B]$ means that these quantities are chosen uniformly in $[A,B]$, independently for each firm. Finally, for all price trajectories reported below, we highlight one firm at random to make the time-series more readable.

\begin{figure}[b]
    \centering
    \includegraphics{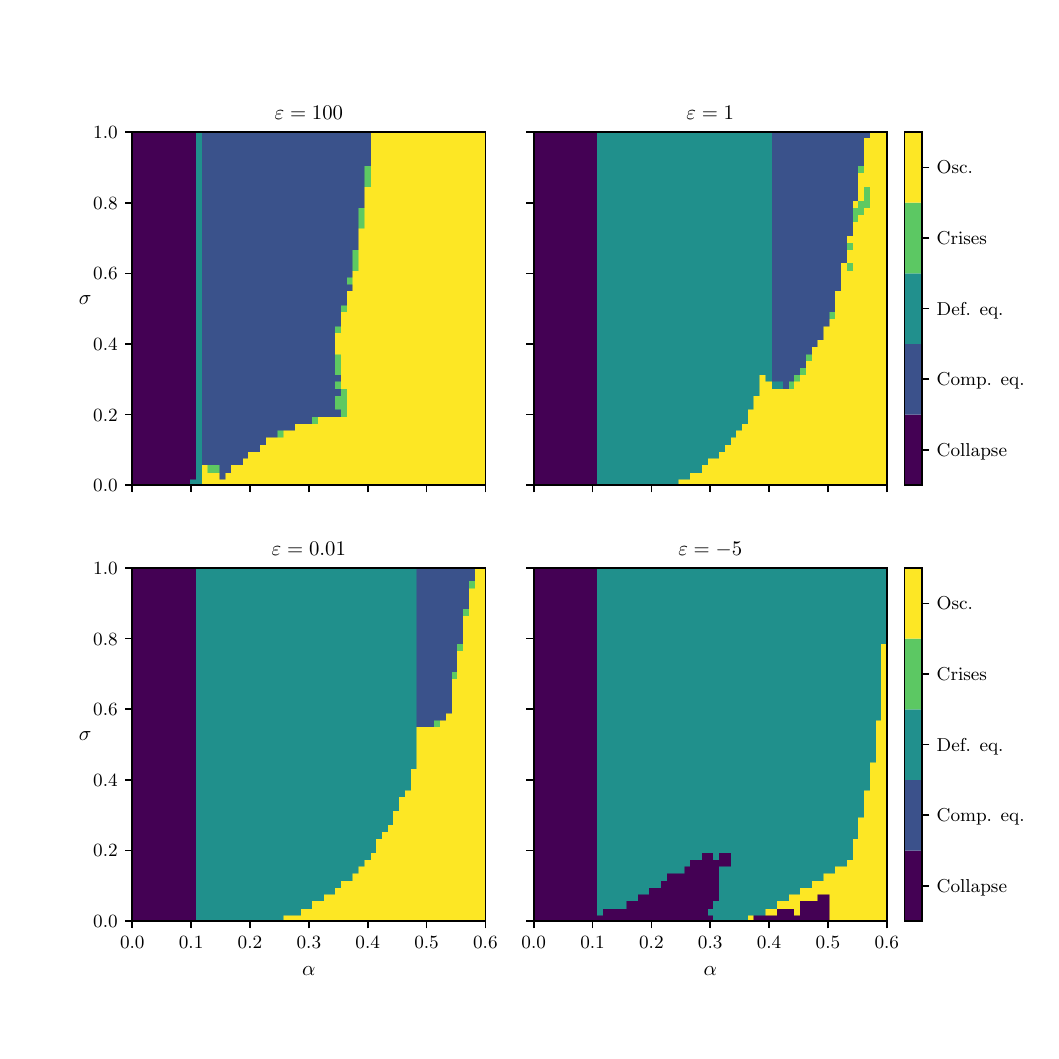}
    \caption{Phase diagrams in the plane restoring forces, perishability (i.e.  $(\alpha, \sigma)$), all for the same network economy, $\omega=0.1$ and different values of $\varepsilon$. The color code is explained in the legends. The region where the competitive equilibrium state is stable shrinks when $\varepsilon$ decreases, and disappears when $\varepsilon < 0$ as deflationary equilibria and cycles/chaos take over. One also observes regions with cycles and chaos, and crises. Finally, when restoring forces are to weak (small $\alpha$) the economy crashes.}
    \label{fig:phase_diagrams}
\end{figure}

%The behaviour detection is done automatically and can yield some mistakes. For instance, it might be difficult to distinguish between crises-like patterns (spikes and then relaxation towards competitive equilibrium) whose spikes are relatively close together from periodic patterns that resembles business cycles.
%\newpage

In general, we observe five classes of behaviour (or ``phases''): convergence towards the competitive equilibrium, convergence towards deflationary equilibria, crises, business-cycle like oscillations or chaotic oscillations and economic collapse, where the economy crashes after a finite number of time steps. Different phase diagrams corresponding to this classification can be seen in Fig.~\ref{fig:phase_diagrams} with $\varepsilon$ in $[100,1,0,01,-5]$, and the study and description of these phases is detailed in the sections below.

Note that the boundaries of the phase diagrams depend on the network of interactions, especially as $\varepsilon\to0$. If $\varepsilon\gg1$, then productivity factors are very large, and network effects can safely be neglected. However, as $\varepsilon\to0$, these network effects become more and more important and the specific type of network will play a role. The regular network chosen in this section is thus only meant to illustrate the different {classes} of dynamical trajectories that can be generated by the model. But such classes are in fact generic and appear for a broad family of networks, and are a consequence of the non-linear update rules followed by the firms. 

Before delving into the description of each of these classes, we note in particular that (see Fig. \ref{fig:phase_diagrams}):
\begin{itemize} 
\item All the different classes appear for values of parameters $\alpha$, $\omega$, $\sigma$, $\varepsilon$ of order unity, to wit: interesting dynamical behaviour do not require uncanny values of the parameters. 
\item The region where the competitive equilibrium is reached shrinks as the economy approaches the instability $\varepsilon \to 0$ from above. When $\varepsilon < 0$, there is no admissible equilibrium and only deflationary equilibria or cycles/chaos can be attained. 
\item For a fixed perishability $\sigma$ one observes the following succession of phases as the restoring parameter $\alpha$ is increased: collapse when $\alpha$ is too small, followed by deflationary equilibria, then competitive equilibrium and finally cycles and chaos for large $\alpha$, corresponding to firms that overreact to imbalances. 

\item At the boundary between competitive equilibrium and cycles and chaos, one observes intermittent crises, similar to the ones described in~\cite{gualdi2015tipping,Gualdi2016} -- see below.   
\end{itemize}

{
\subsubsection{Economic Collapse}

Starting from $\alpha=0$ and increasing its value, the model finds itself first in a collapse phase, where prices diverge exponentially and productions plummet to zero. As $\alpha$ grows we reach a critical value $\alpha_c$ corresponding to a transition from the collapse phase to one where the economy is able to stabilize (either in a deflationary of competitive equilibrium).

This transition, which can be observed in Fig. \ref{fig:phase_diagrams}, appears to be independent of both $\varepsilon$ and $\sigma$. The fact that $\alpha_c$ is independent of $\varepsilon$ means that the economy collapses when prices are too slow to adjust, regardless of the nature of the firm network. The exact value of $\alpha_c$ in Fig. \ref{fig:phase_diagrams} can be computed to be:
\begin{equation} \label{eq:alpha_c}
    \alpha_c=\omega \frac{\betap}{\beta}
\end{equation}
(i.e. $\alpha_c=\omega$ whenever $\alpha=\alphap=\beta=\betap$), where $\omega$ is the Phillips curve parameter relating wages to unemployment (see Eq.~\eqref{eq:final_wage_update}).

This value is obtained by diagonalizing the stability matrix of the system $\mathbb{D}_0$ in the no shortage cone $\delta\mathscr{S}_i(t)-\delta\mathscr{D}_i(t)>0$ for all firms $i$ (see section \ref{sec:cones}), which happen to be stable under the quasi-linear dynamics.\footnote{The precise computation is quite lengthy and beyond the scope of this paper.}
%When $\varepsilon\to\infty$ and $\sigma=\infty$, the equilibrium is no longer linearly stable whenever $\alpha<\alpha_c(\infty)$ as shown in Fig. \ref{fig:transition_dv}.  {\color{blue} JM: I'm not quite sure I follow all of this, what did you mean above with a stable economy with large epsilon? is that stable in the May's instability sense, or is that the dynamical stability of the linearised model?}

This collapse transition may also be observed along the diagonal $\omega=\alpha$ of the right plots in Fig.~\ref{fig:phase_diagrams_alpha_omega} below. (Note however that an additional wedge where the economy diverges appears for small $\omega$ which will be discussed in \ref{sec:directions_pert}.)}

%\begin{figure}[t]
%    \centering
%    \includegraphics[scale=0.9]{transition_dv.pdf}
%    \caption{\color{ForestGreen} Close up of some eigenvalues $\lambda$ of the stability matrix in the region where $\delta\mathscr{S}_i(t)-\delta\mathscr{D}_i(t)>0$ for all $i$. Whenever $\alpha<\alpha_c(\infty)$ the leading eigenvalue $\lambda_>$ has a modulus greater that $1$, rendering the equilibrium unstable in this region.\footnote{\label{fn:lin_dis_time}Since the ABM consists in a series of iterated actions, its linearisation is discrete in time. For such linear systems, the stability condition is to have every eigenvalues of the jacobian within the unit circle.} We took $\sigma=\infty$, $\varepsilon=10^7$, $\alphap=\beta=0,2$, $\betap=0.05$, $\omega=0.1$ and $\alpha=0.5\alpha_c(\infty),\,1,5\alpha_c(\infty)$.}
%    \label{fig:transition_dv}
%\end{figure}

\subsubsection{Competitive Equilibrium}

\begin{figure}[t]
    \centering
    \includegraphics{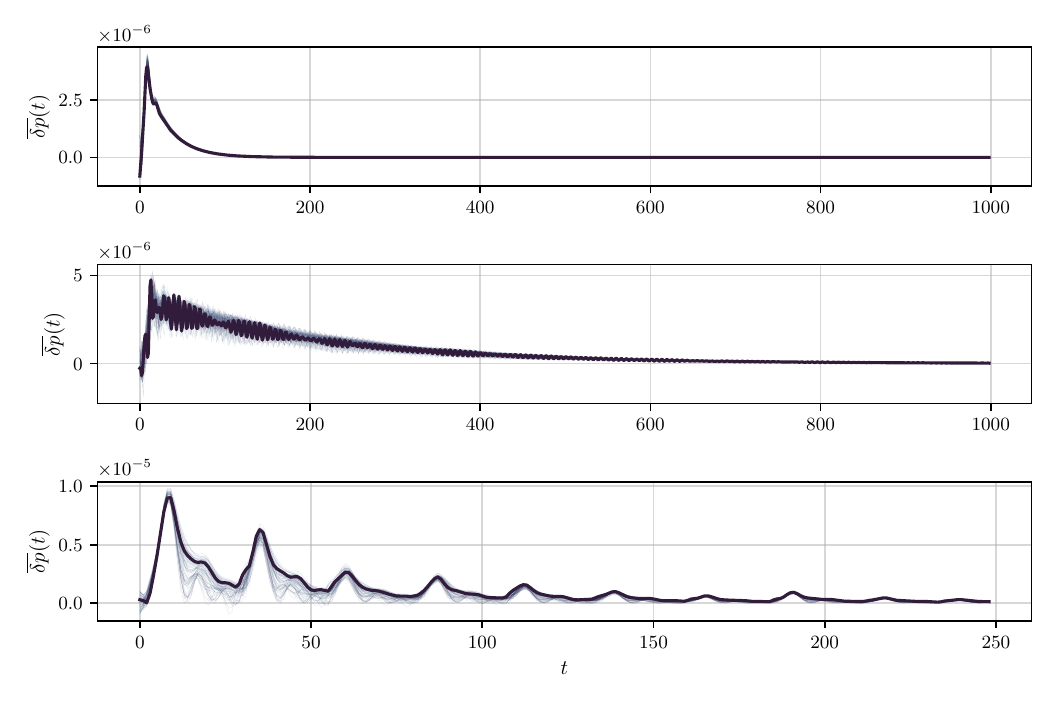}
    \caption{Relaxation towards the competitive equilibrium after a perturbation of magnitude $\delta=10^{-3}$ for a variety of firms. Top: Exponential relaxation for $\varepsilon=10$, $\omega=\omega^\prime=0.1$, $\alpha=\alphap=\beta=\betap\in[0.3,0.35]$ and $\sigma\in[0.5,0.6]$. Middle: Damped oscillations for $\varepsilon=1$, $\omega=\omega^\prime=0.1$, $\alpha=\alphap=\beta=\betap\in[0.4,0.45]$ and $\sigma\in[0.2,0.6]$. Bottom: Damped chaotic oscillations for $\varepsilon=100$, $\omega=\omega^\prime=0.1$, $\alpha=\alphap=\beta=\betap\in[0.25,0.3]$ and $\sigma\in[0.2,0.6]$. The dark lines correspond to one randomly picked firm}
    \label{fig:comp_eq_relax}
\end{figure}
\begin{figure}[t]
    \centering
    \includegraphics{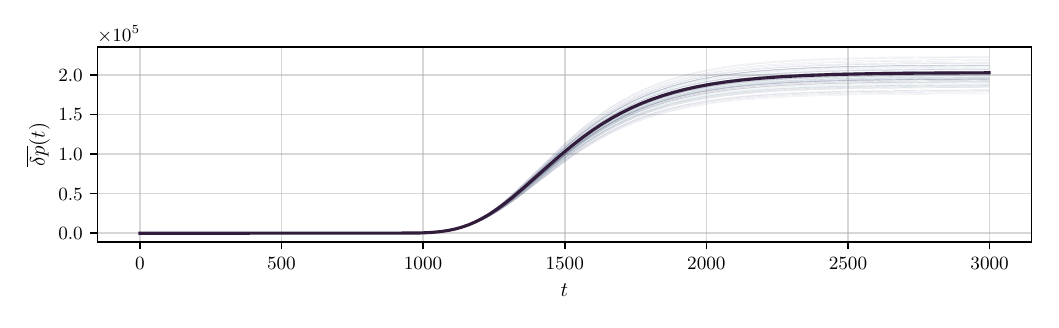}
    \caption{Example of a deflationary equilibrium with $\varepsilon=1$ and heterogeneous productivity factors. Note that we show here real prices (deflated by wages), which reach unreasonable values $10^5$ higher than at equilibrium. We choose here $\omega=\omega^\prime=0.1$, $\alpha=\alphap=\beta=\betap\in[0.25,0.3]$ and $\sigma=0.6$. The dark line corresponds to one randomly picked firm.}
    \label{fig:inf_eq}
\end{figure}

{The most natural behaviour one could expect is for the economy to converge to a competitive equilibrium, where all profits are zero and markets clear, as classically assumed in economics models.} This is indeed what happens, but, interestingly, it requires $\alpha$ to be neither too small, nor too large, i.e. when restoring forces are strong enough to stabilise the system but not too strong to avoid overshoots and the corresponding impossibility for the economy to coordinate. Perishability $\sigma$ should also be large enough, see Fig. \ref{fig:phase_diagrams}. Finally, as returns to scale diminish (i.e. as the parameter $b$ decreases), the region where competitive equilibrium can be attained becomes more extended (see Fig. \ref{fig:transition_deflat}  below).

In order to give some economic meaning to the phase where competitive equilibrium is reached, let us focus on the case $\varepsilon=1$, i.e. a firm network of moderate average productivity, relatively far from the Hawkins-Simons instability. Choose the unit time scale of the model to be a quarter (3 months), a reasonable period for firms to adjust prices and production. Competitive equilibrium cannot be reached when $\alpha < \alpha_d(\varepsilon=1) \approx 0.395$. Such a value of $\alpha$ means that a firm facing a production imbalance of say $10 \%$ will attempt to reduce it to $10 \% \times e^{-0.395} \approx 6.7 \%$ over the next quarter. Attempting to reduce it much faster leads to oscillation and chaos. For example, when $\sigma=\log 2 \approx 0.69$, corresponding to half-life of goods of a quarter, $\alpha$ should remain smaller than $\approx 0.55$ to avoid falling in the yellow region of Fig. \ref{fig:phase_diagrams}. Note that when $\sigma$ drops below $\approx 0.3$, competitive equilibrium is unattainable. 

Note further that within the competitive equilibrium phase, convergence can either be purely exponential, or correspond to damped oscillations or even damped chaos, see Fig.~\ref{fig:comp_eq_relax}. The precise nature of the relaxation depends on the relative values of $\alpha, \alphap, \beta$ and $\betap$. 

Interestingly, marginal stability and diverging relaxation times may occur even for non-zero values $\varepsilon$. Indeed, as we get closer to the transition line $\alpha_d(\epsilon)$ between competitive equilibrium and deflationary equilibrium, the time needed to converge gets larger in the same way as in the naive model. As an illustration, if we take $\varepsilon\to\infty$, one finds $\alpha_d=\alpha_c$ as given by Eq. \eqref{eq:alpha_c}. If we now set $\alpha=\alpha_c+\delta$ with $\delta\ll\alpha_c$, it is not hard to show that the largest eigenvalue of $\mathbb{D}_0$ is given by $1-\delta/(3(1+\omega))+ O(\delta^2)$. The relaxation time is then of order $\delta^{-1}$ and indeed diverges close to the destabilisation transition as in the naive model. The same scaling $\delta^{-1}$ for the relaxation time holds for generic values of $\varepsilon$, when $\alpha=\alpha_d(\varepsilon)+\delta$. Finally, note that at fixed $\sigma$, the interval for which competitive equilibrium can be reached shrinks to zero, to wit $]\alpha_d(\varepsilon), \alpha_d(\varepsilon)+f(\varepsilon)[$, with $f(\varepsilon) \to 0$ as $\varepsilon\to0$. More precisely, numerical simulations show there exists a value $\varepsilon_c$ (depending on $\sigma$, $b$ and $\omega$) such that
\begin{equation}
    f(\varepsilon)=\begin{cases}
    0,&\text{for $\varepsilon\leq\varepsilon_c$}\\
    r(\varepsilon-\varepsilon_c),&\text{for $\varepsilon\to\varepsilon_c^+$}
    \end{cases},
\end{equation}
where $r=f'(\varepsilon_c^+)$. By the same argument as above, the relaxation time will scale as $f(\varepsilon)^{-1}\sim 1/(\varepsilon-\varepsilon_c)$ and we retrieve the same behaviour as in the naive model. 

\subsubsection{Deflationary Equilibrium}\label{sec:deflat}

An interesting feature of our model is the appearance of a different kind of equilibrium, corresponding to stationary points where profits and excess demand are non-zero, but equal to a constant value. We call them ``deflationary'' equilibria  because prices synchronise with the (negative) inflation rate determined by the downward evolution of wages, induced by chronic unemployment (i.e. $L^{\text{s}} > L^{\text{d}}$).  

We denote by $\bar{\pi}^\infty_i$ and $\bar{\mathscr{E}}^\infty_i$ the rescaled values of profits and excess supply in the stationary state. These must then verify (see Eqs. (\ref{eq:final_target_update},~\ref{eq:final_price_update})):
\begin{equation}
\label{eq:sys_inf_eq}
\begin{aligned}
    \alpha\bar{\mathscr{E}}_i^\infty+\alpha^\prime\bar{\pi}_i^\infty&=\omega\frac{L^{\text{s},\infty} - L^{\text{d},\infty}}{L^{\text{s},\infty} + L^{\text{d},\infty}} \quad (> 0),\\
    \beta^\prime\mathbb{E}_\infty[\bar{\mathscr{E}}_i]&=\beta\mathbb{E}_\infty[\bar{\pi}_i].
\end{aligned}
\end{equation}
In the present case where $\lambda=1$ in \eqref{eq:def_lambda} one also has $\mathbb{E}_\infty[\bar{\mathscr{E}}_i]=\bar{\mathscr{E}}_i^\infty$ so one can simplify these equations to get
\begin{equation}
\label{eq:sys_inf_eq2}
\begin{aligned}
    \alpha\beta\mathbb{E}_\infty[\bar{\pi}_i]+\alpha^\prime\betap\bar{\pi}_i^\infty&=\omega\betap\frac{L^{\text{s},\infty} - L^{\text{d},\infty}}{L^{\text{s},\infty} + L^{\text{d},\infty}} \quad (> 0),\\
    \bar{\mathscr{E}}_i^\infty&=\frac{\beta}{\betap}\mathbb{E}_\infty[\bar{\pi}_i].
\end{aligned}
\end{equation}

In contrast with the competitive equilibrium, which is independent of the dynamical parameters $\alpha, \alphap, \beta, \betap$, deflationary equilibria are characterised by prices and production levels that depend on the parametrisation of the dynamics. Explicit expressions for the stationary prices/productions are, however, difficult to compute analytically. 

\begin{figure}[t]
    \centering
    \includegraphics[scale=0.8]{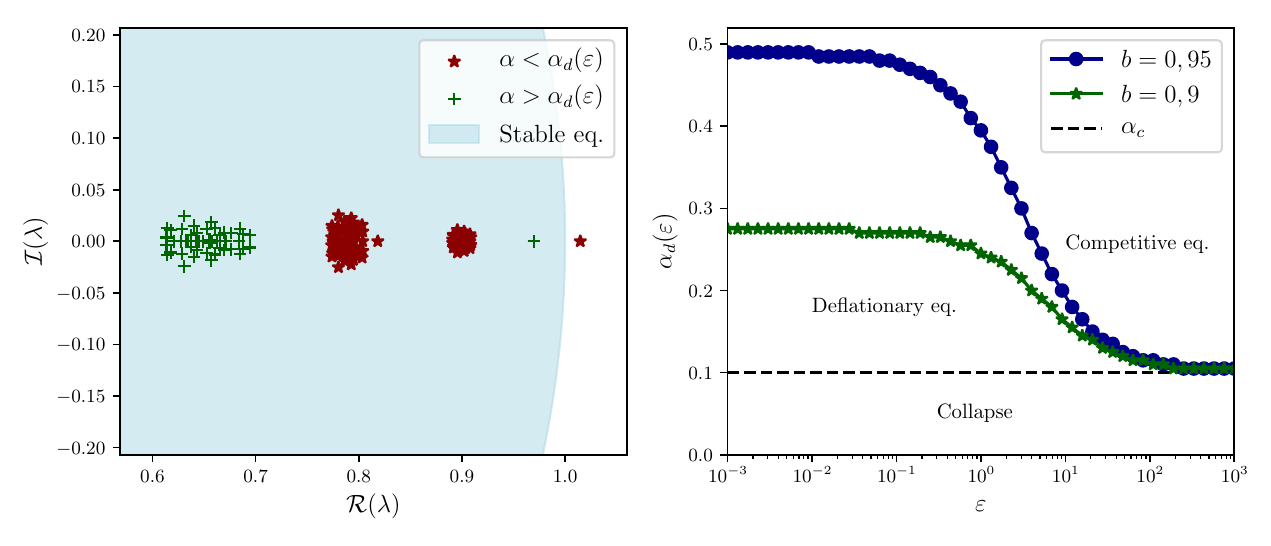}
    \caption{Left: Eigenvalues $\lambda$ of the stability matrix for $\varepsilon=10$ in the no-shortage cone  where $\delta\mathscr{S}_i(t)-\delta\mathscr{D}_i(t)>0$ for all $i$. Here, $\sigma=\infty$, $\alpha=\alphap=\beta=\betap\in\{0.12,0.25\}$ and $\omega=0.1$. For  $\alpha_c<\alpha<\alpha_d(\varepsilon)$, competitive equilibrium is not stable and the non-linear dynamics converges to a deflationary equilibrium. Right: $\alpha_d$ as a function of $\varepsilon$ (in log-scale on the x-axis) for different values of the return-to-scale parameter $b$. The horizontal dotted line corresponds to $\alpha_c = 0.1$, which is independent of $\varepsilon$. The labeled areas correspond to the different phases of Fig. \ref{fig:phase_diagrams} for $b=0.95$. As $b$ decreases, one can see that the area below $\alpha_d(\varepsilon)$ tend to decrease i.e. the region where competitive equilibrium is reached becomes larger. As mentioned in footnote \ref{fn:rts_below_one}, decreasing return-to-scale tend to stabilise the dynamics. Note that the values of $\alpha_d(\varepsilon=100)$, $\alpha_d(\varepsilon=10)$ and $\alpha_d(\varepsilon=1)$ are consistent with values observed on the phase diagrams of Fig. \ref{fig:phase_diagrams} for which $b=0.95$. Finally, as $\varepsilon\to0$, one can see in Fig. \ref{fig:phase_diagrams} that the smallest $\sigma$ over which this transition exists gets larger. As a consequence, for small $\varepsilon$, the region labelled ``competitive equilibrium'' only exists for large enough $\sigma$.}
    \label{fig:transition_deflat}
\end{figure}

Fig.~\ref{fig:inf_eq} shows an example of the convergence of inflation-adjusted prices towards their stationary values. Note that in real terms, the stationary price level is above the equilibrium value. Throughout our simulations we have found these equilibria to be rather stable. {For a fixed value of $\omega$, the transition between deflationary equilibria and competitive equilibrium occurs at a value $\alpha_d(\varepsilon)$ which is  difficult to compute analytically. We can nevertheless locate this transition numerically by simulating the stability matrix in the no-shortage cone mentioned in the previous subsection. Results are reported in Fig. \ref{fig:transition_deflat}.}

However, these deflationary equilibria make little economic sense in the long run, because (a) the stationary level of production tends to be extremely small compared to equilibrium values and (b) forecasts of consumption (for households) and profits (for firms) systematically overshoot their realized counterparts. One expects that in such situations, like in the case of economic collapse, the influence of monetary and fiscal policies cannot be neglected. Furthermore, we expect that when biases are strong and systematic, agents would soon adapt and change their forecasting rules accordingly. Such an extension is however beyond the scope of the present paper, but a natural conjecture is that firms would react more strongly to imbalances (i.e. increase the coefficients $\alpha, \alphap, \beta, \betap$), which would drive the system back in the competitive equilibrium phase or in the oscillatory phase. 

%These deflationary equilibria disappear when $\omega=0$. The corresponding phase diagrams in that case are similar, when $\varepsilon > 0$,  to those of Fig.~\ref{fig:phase_diagrams}, but with a wider region corresponding to the competitive equilibrium phase. When $\varepsilon < 0$, the only possibilities are cycles/chaos (yellow phase) or a complete crash (black phase). 

We finally point out that we have not found, within the present specification of the model, inflationary equilibria where the demand for labour exceeds the supply. However, we have found that introducing precautionary savings used to buy interest rate paying bonds leads to new phenomena, including a whole region where {\it inflationary} equilibria are now found. %We plan to report these results in a separate publication. 

\subsubsection{Oscillatory Dynamics}
\label{sec:oscillations} 

\begin{figure}[t]
    \centering
    \includegraphics{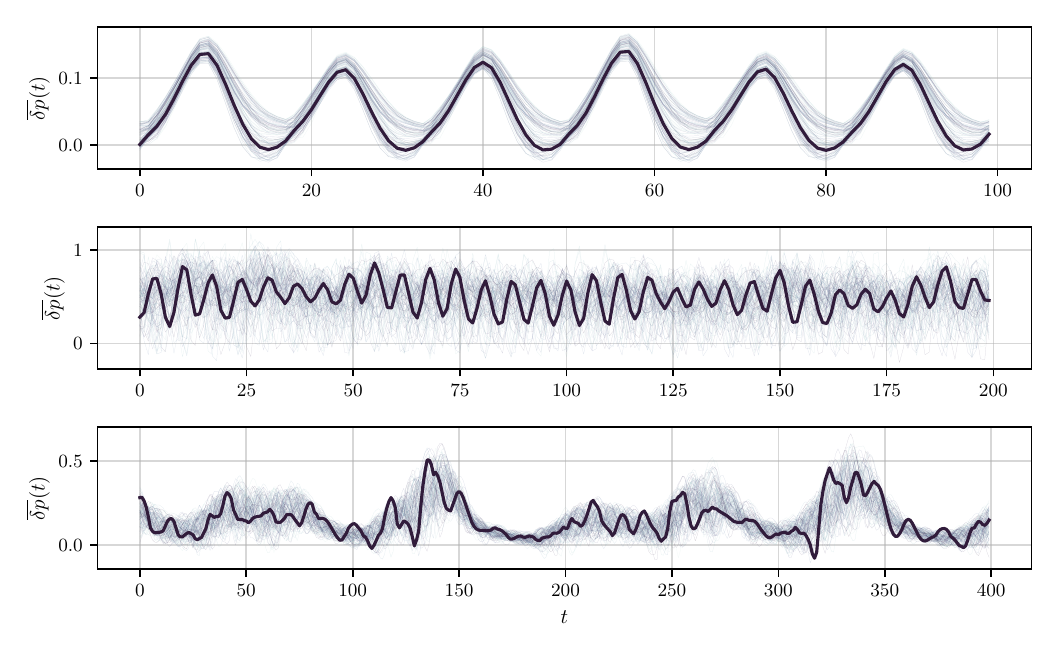}
    \caption{Different types of price (or production) oscillations around equilibrium after an initial perturbation of magnitude $\delta=10^{-3}$ from equilibrium.  Top: Synchronised business cycles for $\varepsilon=100$, $\omega=\omegap=0.05$, $\alpha=\alphap=\beta=\betap\in[0.2,0.25]$, $\sigma\in[0.1,0.4]$. Middle: unsynchronised oscillations for $\varepsilon=100$, $\omega=\omegap=0.1$, $\alpha=\alphap=\beta\in[0.25,0.4]$, $\sigma_=0.2$; $\betap=1.3\alpha$. Bottom: Chaotic oscillations for the same parameters except $\varepsilon=1$ and $\betap = 0.2\alpha$. The dark lines correspond to one randomly picked firm.}
    \label{fig:business_cycles}
\end{figure}
\begin{figure}[!b]
    \centering
    \includegraphics{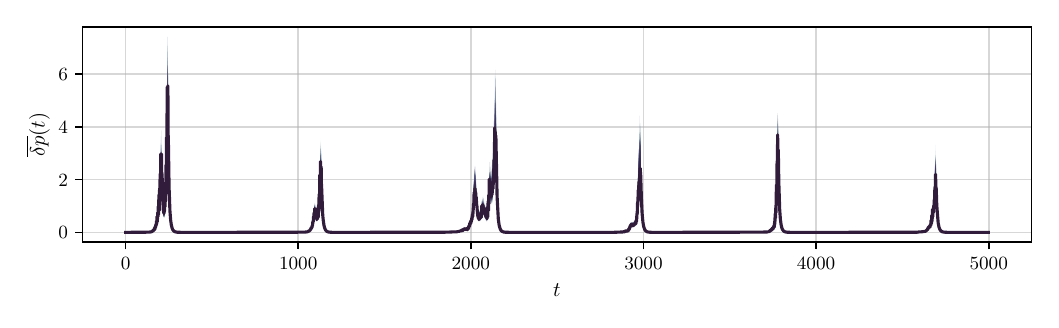}
    \caption{Crises-like price pattern for $\varepsilon=100$, $\omega=\omegap=0.1$, $\alpha=\alphap=\beta=\betap=05$, $\sigma=\infty$.}
    \label{fig:crises}
\end{figure}

Owing to the strongly non-linear dynamics defining the model, it is natural to expect that some choices of the parameters lead -- as in generic dynamical systems -- to oscillations or to chaotic dynamics, which is indeed what we observe in a whole region of parameter space -- in short, when firms tend to over-react and adjust prices/productions too quickly in the face of imbalances. 

The first interesting oscillatory behaviour is that of spontaneously emerging business cycles, as shown in Fig.~\ref{fig:business_cycles}. They can be either synchronised (Fig. \ref{fig:business_cycles}-a) or completely unsynchronised (Fig. \ref{fig:business_cycles}-b), depending on the values of $\omega$ and $\varepsilon$, and the relative values of $\alpha$ and $\betap$. Chaotic oscillations also emerge (see Fig.~\ref{fig:business_cycles}-c).

We stress that such persistent oscillations, observed in the rather large portions of the phase diagram, are not due to external perturbations, absent in these simulations (compare with section \ref{sec:excess_vol} where small external shocks are amplified by the proximity of an instability). Rather, this is a region of the phase diagram where the volatility of the economy is  purely endogenous (see \cite{bonart2014instabilities} for similar observations). 

This provides yet another scenario to explain the ``small shock, large business cycle'' puzzle described by~\cite{bernanke1994financial}, different from the proximity of an unstable point, as in section \ref{sec:excess_vol}. Volatility may be high because of the existence of self-sustained oscillations/chaos, as reported here and in many previous work in which a dynamical systems approach to economics was advocated, see e.g. \cite{Goodwin1982,grandmont_paper,keen1997, Flaschel2008TheMO, rosser} and also \cite{delligatti2005, chiarella, gualdi2015tipping,pangallo_synchronization_2020} in the context of ABMs.

\subsubsection{Intermittent Crises}

This additional dynamical class is represented in Fig.~\ref{fig:crises}. Here, a fast relaxation to equilibrium is followed by spontaneous destabilisation. The system enters a cycle of price inflation and plummeting  production. This is most likely due to a switch between different cones, characterised by different stability matrices, as discussed in section \ref{sec:cones}. {The first matrix is stable, whereas the second has at least one eigenvalue out of the unit circle, and therefore an unstable direction.}

Non-linear saturation effects then take over and quell the dynamics, and the system flows back towards equilibrium before the next crisis appears. These acute endogenous crises are one of the most interesting aspects of our model; they also appear in the Agent Based Models of \cite{gualdi2015tipping} and \cite{Sharma2020} where they result from a generic synchronisation mechanism, as made explicit by~\cite{Gualdi2015}.
\begin{figure}[t]
    \centering
    \includegraphics{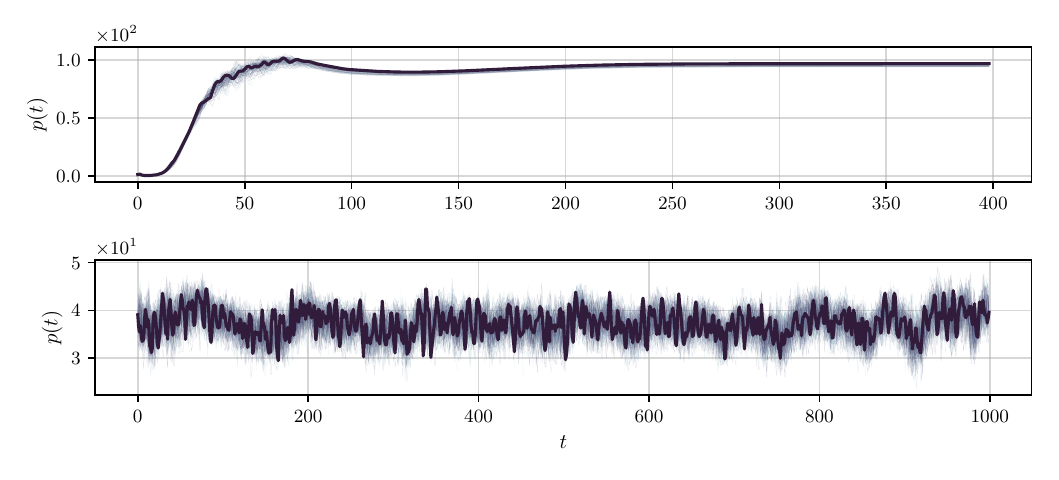}
    \caption{Different possible price (or production) dynamics in the unstable phase $\varepsilon=-5$, for initial conditions for prices and productions randomly chosen between 1 and 2 times the equilibrium values. Top: Rapid oscillations for  $\omega=\omegap=0.01$, $\alpha=\alphap=\beta=\betap=0.45$, $\sigma=0.2$. Bottom: deflationary equilibrium for $\omega=\omegap=0.02$, $\alpha=\alphap=\beta\in[0.4,0.45]$, $\sigma\in[0.2,0.8]$. The dark lines correspond to one randomly picked firm.}
    \label{fig:eps_neg}
\end{figure}
\subsection{The Low-Productivity Phase \texorpdfstring{$\varepsilon<0$}{Lg}}

A weakness of the naive model of section~\ref{sec:naive} was that it can only produce divergent trajectories whenever $\varepsilon<0$, i.e. in the low productivity phase. As illustrated in Fig. \ref{fig:eps_neg}, our full model produces instead a wide range of interesting behaviour in this case, from deflationary equilibria to oscillations. Of course, since there is no well-defined equilibrium, the convergent phase is now proscribed. However, an important message is that a viable economy can exist {\it even if the Hawkins-Simon condition is violated}, but at the expense of either substantial stationary imbalances or oscillatory/chaotic behaviour. 

\begin{figure}[t]
    \centering
    \includegraphics{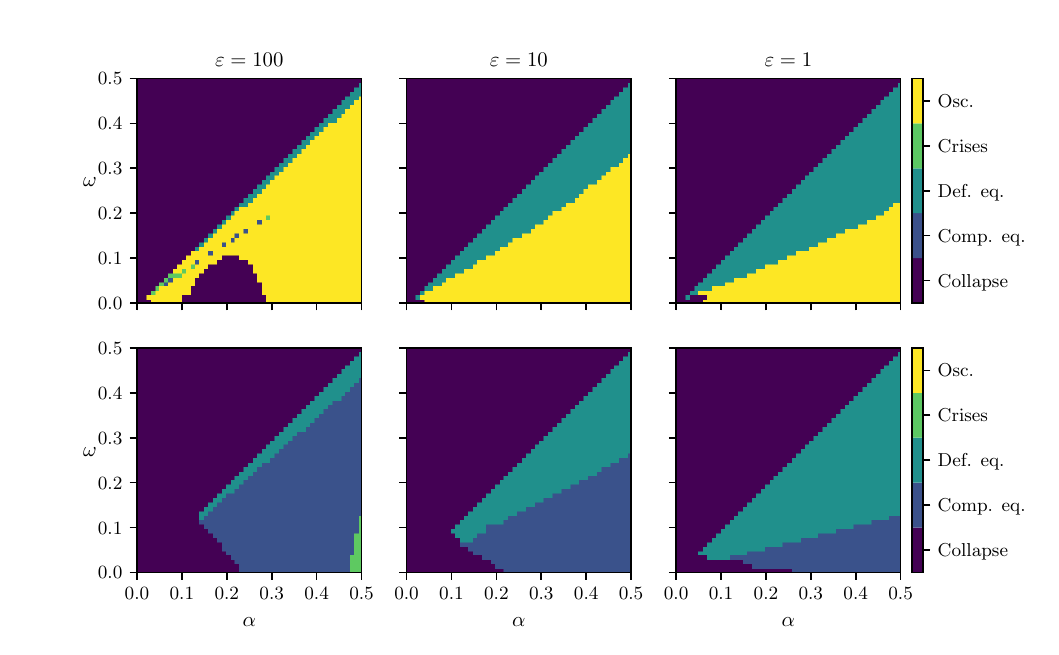}
    \caption{Phase diagrams for non-perishable ($\sigma=0$, top row) and immediately perishable ($\sigma=\infty$, bottom row) goods in the $\alpha,\omega$ plane, for different values of $\varepsilon$ (left: $\varepsilon=100$, middle: $\varepsilon=10$, right: $\varepsilon=1$).}
    \label{fig:phase_diagrams_alpha_omega}
\end{figure}

\subsection{The Role of Perishability}

Finally, we illustrate here the crucial role of inventories in determining the type of dynamics we observe.
As shown in the phase diagrams of Fig.~\ref{fig:phase_diagrams_alpha_omega} in the $(\alpha, \omega)$ plane at fixed $\sigma$, goods that perish immediately ($\sigma=\infty$) lead to simple relaxation towards equilibrium (deflationary/competitive) or to a collapse. In a sense, this limit is as close as possible to the ``naive'' model of section \ref{sec:naive} where all inventory effects were overlooked.

On the other hand, non-perishable goods lead to oscillating, highly volatile economies. Intuitively, if firm $i$ has a stock $I_{ik}$ of good $k$, it will decrease its demand to firm $k$, leading to a decrease of its production. This lasts until all stocks are exhausted. A phase of booming demands and increase in production follows, firms' stocks begin to pile up again and the economy enters another cycle. This is similar to the well-known ``bull-whip effect'' suggested by \cite{Bullwhip}, where inventories are known to lead to instability effects. These instabilities indeed  disappear completely when $\sigma = \infty$ (right column of Fig. \ref{fig:phase_diagrams_alpha_omega}).

\subsection{Sensitivity to Initial Conditions}

{The phase diagrams of Figs~\ref{fig:phase_diagrams} and \ref{fig:phase_diagrams_alpha_omega} have been established by classifying the behaviour of the system after an initial perturbation around equilibrium of magnitude $10^{-3}$. But  our system is non-linear, larger perturbations may lead to different outcomes for the same set of parameters. In this section, we study of the impact of initial conditions on the dynamics.}

\subsubsection{Basin of Attraction of Equilibrium}

\begin{figure}
    \centering
    \includegraphics[trim=0cm 5cm 0cm 5cm, clip]{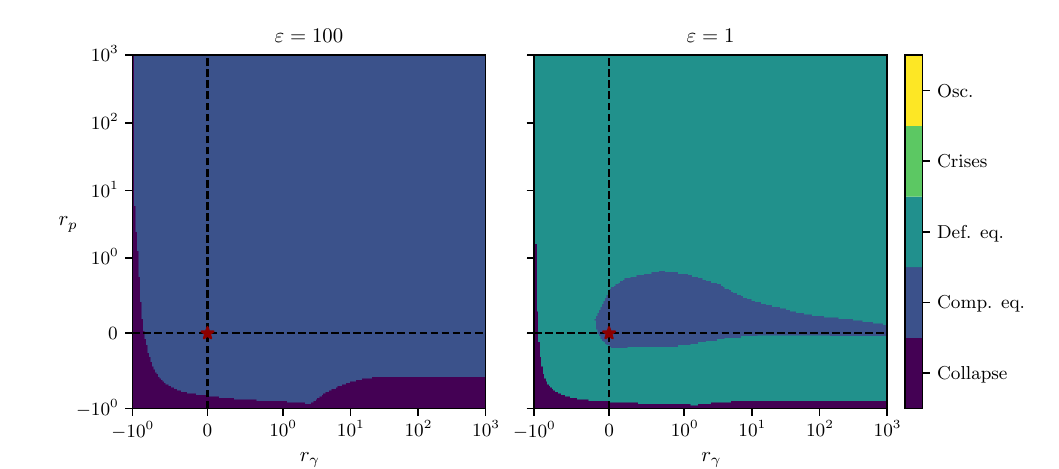}
    \caption{Approximate basins of attraction of the dynamics for $\alpha=\beta=\betap=\alphap=0.45$, $\omega=0.1$, $\sigma=\infty$. Left: $\varepsilon=100$, $\alpha\gg\alpha_d(100)\approx0.115$ which allows very large perturbations of equilibrium values. Right: $\varepsilon=1$, $\alpha\gsim\alpha_d(1)\approx0.395$. Large perturbations leads to the system reaching a deflationary equilibrium. Also note that downwards perturbation can lead to deflationary equilibrium. In this case, the system overreacts and blows up to reach a deflationary equilibrium.  Dashed black lines separate the regions of positive and negative perturbation on prices or productions. The red star corresponds to no-perturbation.}
    \label{fig:basin_attraction}
\end{figure}

{

As for any non-linear dynamical system, the basin of attraction of a given fixed point is defined as the set of initial conditions that will allow the system to reach it. The analytical determination of basins of attraction is a notoriously difficult question, especially for high dimensional systems.

As pointed out in section \ref{sec:cones}, it is possible to linearise the dynamics around equilibrium. The subsequent dynamics is piece-wise linear and described by the evolution of a $N^2+4N+1$-dimensional vector. As it would be unrealistic to explore separately the effects of a perturbation on each and every component of this vector, we will restrict our  study to uniform perturbations on current productions, prices and production targets. We thus parametrise the perturbations as
\begin{equation}
p_i(0)=\peq{i}(1+r_p),\;\gamma_i(0)=\preq{i}(1+r_\gamma),\;\widehat{\gamma}_i(1)=\preq{i}(1+r_\gamma),
\end{equation}
where $r_p$ and $r_\gamma$ are the  perturbation radii ranging from $-1$ (initial values at $0$) and $+\infty$. For a given $r_p$, we scan all values of $r_\gamma$ and find the largest upward and downward possible perturbation allowing the system to revert back to equilibrium. Beyond this domain, the dynamics may drive the system to another phase. 

Fig. \ref{fig:basin_attraction} shows the approximate regions for which the dynamics reach equilibrium after a perturbation of size $(r_p,r_\gamma)$. For large $\varepsilon$, the system is able to sustain very large perturbations when $r_p,r_\gamma>0$. However, whenever either $r_p$ or $r_\gamma$ is negative, the system can end up in the collapse region. We will discuss this point in the next section. 

Finally, as one expects, the basin of attraction drastically shrinks as $\varepsilon$ is reduced (Fig. \ref{fig:basin_attraction}-right). One can see that the system is still able to cope with large perturbations on production provided that prices are not too far from equilibrium. 

The shrinking of the basin of attraction of the competitive equilibrium state as $\varepsilon \to 0$ again reveals how network effects are crucial to understand the fragility of the economy, since the value of $\varepsilon$ is, we recall, determined by productivity on the one hand, and the structure of the input-output network on the other. 
}

\subsubsection{Direction of Perturbations}\label{sec:directions_pert}
{
On top of the importance of the magnitude of perturbation, the direction of the perturbation matters as well. This is a consequence of the separation of state space in different cones. For a small perturbation around equilibrium, if the system is initialised in the no-shortage cone, the dynamics will behave differently than if it were in the full shortage cone. As an illustration, Fig. \ref{fig:short_noshort} shows the phase diagrams in the plane $(\alpha,\omega)$ for the same parameters but for different initial perturbations. On the left, a small upward perturbation is applied on equilibrium prices and productions but initial targets are set to $\preqv$. This prepares the system in the no-shortage cone since production is higher than in equilibrium and household's demand lower. We see that the collapse region is very well described by the stability of the matrix $\mathbb{D}_0$ defined in section \ref{sec:cones}, which in this case keeps the trajectory inside the no-shortage cone. 

On the other hand, the right-hand plot shows that initialising the system in a mixture of no-shortages/shortages adds an additional wedge of collapsing dynamics (note that the same wedge is present on the diagrams of Fig. \ref{fig:phase_diagrams_alpha_omega}). Above this line, the matrix $\mathbb{D}_S$ drives the dynamics outside the partial-shortage cones. The system finally reaches the no-shortage cone, which is preserved by $\mathbb{D}_0$ and for which equilibrium is stable. Below this line, the dynamics is thrown into the full-shortage cone, which is preserved by $\mathbb{D}_N$ but where, on the other hand, equilibrium is unstable.

\begin{figure}
    \centering
    \includegraphics{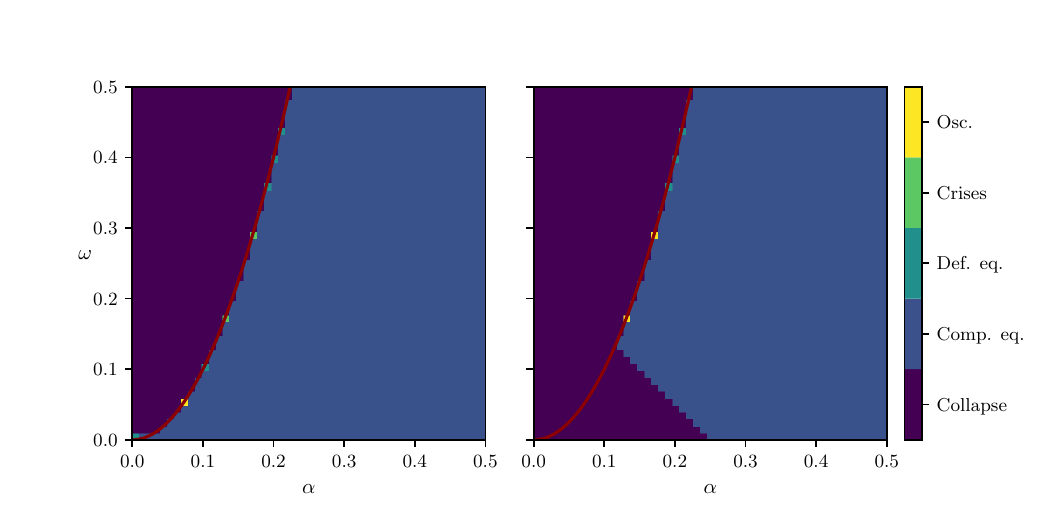}
    \caption{Phase diagrams $(\alpha, \omega)$ for $\varepsilon=10^4$, $\betap=0.1$, $\alphap=0.25$, $\sigma=\infty$ and $\alpha=\beta$. Left: The system is initialised in the no-shortage region by applying a small upward perturbation on equilibrium prices and productions of magnitude $10^{-4}$. Right: The system is initialised in a mixture of no-shortages and shortages ($50\%/50\%$) by applying a perturbation around equilibrium prices and productions. The red line corresponds to the prediction $\alpha_c=\sqrt{\betap\omega}$ (here $\alpha=\beta$) for the stability of the matrix $\mathbb{D}_0$. Note here that we fixed values of $\alphap$ and $\betap$ to illustrate another possible shape of the transition around $\alpha_c$. If we had chosen, as previously, $\alpha=\beta=\alphap=\betap$, the transition line would have been the line $\alpha=\omega$, as in Fig. \ref{fig:phase_diagrams_alpha_omega}.}
    \label{fig:short_noshort}
\end{figure}

To further illustrate this effect, we ran the dynamics of our model with mixed initial conditions
\begin{equation}
p_i(0)=\peq{i}(1\pm r_p),\;\gamma_i(0)=\preq{i}(1\pm r_\gamma),\;\widehat{\gamma}_i(1)=\preq{i}(1\pm r_\gamma),
\end{equation}
where we choose $+$ for $50\%$ of the firms and $-$ for the others. This prepares the system in a state where $50\%$ of the firms cannot fulfill demands. In Fig. \ref{fig:basin_attraction_5050}, we show the basin of attraction of equilibrium for perturbations $r_p,r_\gamma$ ranging from 0 to 1. As we see, large enough shortages can destabilise the dynamics even at large $\varepsilon$.
%One can notice that these basins look very much like the region of negative $r_p$ and $r_\gamma$ in Fig.\ref{fig:basin_attraction} meaning that destabilisation can occur even when some firms can cope with demands.

\begin{figure}
    \centering
    \includegraphics[trim=0cm 5cm 0cm 5cm, clip]{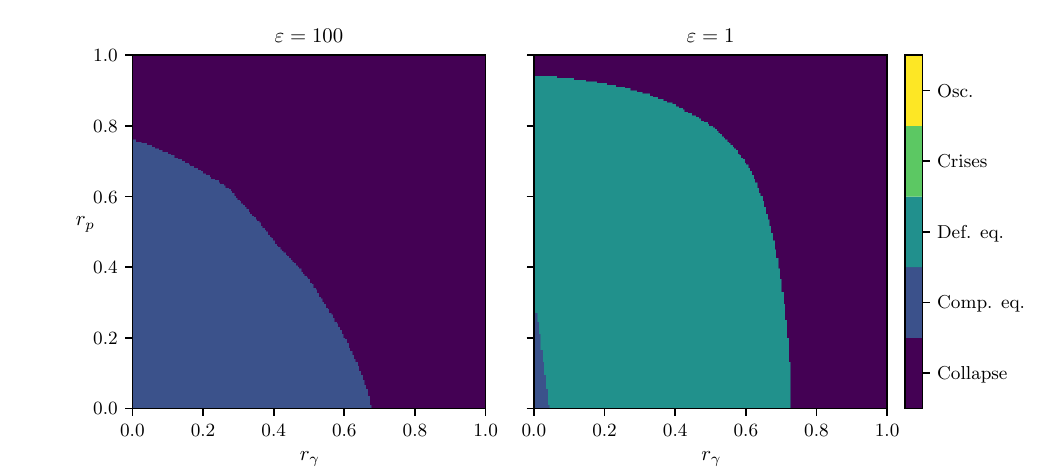}
    \caption{Approximate basins of attraction of the dynamics for mixed initial conditions (50 \% shortages, 50 \% mark-ups) for $\varepsilon=100$ (left) and $\varepsilon=1$ (right), and  $\alpha=\beta=\betap=\alphap=0.45$, $\omega=0.1$, $\sigma=\infty$.}
    \label{fig:basin_attraction_5050}
\end{figure}

}

\section{Summary \& Conclusion}

Let us first summarise the main messages of this paper. This work started from the observation made in \cite{moranb} that generic input-output network models cannot reach a competitive equilibrium state when productivity is too low, connectivity too high, or substitutability too low. This begs the question: what happens to the economy in such cases? 

We argued that the answer to such a question is necessarily of dynamical nature, and demands an extension of the standard equilibrium framework to out-of-equilibrium equations of motion, that aim to describe {\it how} imbalances regress in time and  {\it how fast} equilibrium is reached -- if it is reached at all. 

We first proposed what we called a ``naive'' model, based on the idea that forces driving the economy back to equilibrium are linear in the imbalances (profits and supply/demand imbalances). This leads to interesting non-linear (Lotka-Volterra type) differential equations for prices and productions which predict, among other things, that the equilibration time diverges as the network economy approaches the instability point at which competitive equilibrium is no longer admissible. We argued that this long time scale also leads to excess volatility, as the impact of past exogenous shocks cannot quickly dissipate. 

We then pointed out that the naive model does not correctly factor in physical constraints: excess demand cannot be satisfied, excess supply must be stored, consumption can only start after goods are produced, wages can only be spent after being paid, etc. Accounting for all these constraints within a consistent model considerably complexifies the resulting equations, but leads to a model which displays a much larger variety of possible dynamical behaviour, some very far from the competitive equilibrium. In fact, the dynamics of the model can remain well-behaved even in the region of parameters where equilibrium is inadmissible because some prices and/or productions would be negative, unless some firms are removed from the network.

A numerical investigation of the full model leads to rich phase diagrams, from which we extract the following salient features, with clear economic implications: 
\begin{itemize}
    \item The competitive equilibrium attracts the dynamics {\it only in a restricted range of parameters}: the speed at which firms adapt to imbalances must neither be too slow nor too fast, and the rate at which goods spoil must be high enough. Diminishing returns to scale also help convergence towards equilibrium. 
    \item When the adaptation speed is too large, or the perishability of goods too low, coordination breaks down and the economy enters a phase with periodic or chaotic business cycles of purely endogenous origin, as was also reported in \cite{bonart2014instabilities}.
    \item Close to the boundaries between the competitive equilibrium phase and the oscillating phase, one observes a regime of intermittent crises, with long periods of quasi-equilibrium interrupted by bursts of inflation.
    \item Another class of equilibria exists, with a negative inflation but with stationary real prices and production different from those pertaining to the competitive equilibrium. In particular, markets -- including the job market -- do not clear in such situations: labour supply is always larger than labour demand. These equilibria are however characterised by persistent discrepancies between forecasts and realized quantities, which presumably make them unstable against simple learning rules.
    \item For {\it inflationary} equilibria to exist, where labour demand is larger than labour supply, one needs to introduce precautionary savings and interest rates in the model. %(this will be reported in a separate publication).
    \item Finally, we have checked that the overall shape of the phase diagram is robust to changes of the structure of the network (although see Appendix \ref{ap:real_net} for additional information about the dynamics on real input-output networks) and of the specific form of the CES production function that one uses. This means that our results are generic and should hold in realistic situations as well.
\end{itemize}

Our model therefore suggests {\it two} distinct out-of-equilibrium routes to excess volatility (or ``large business cycles''): (a) purely {\it endogenous} cycles, resulting from over-reactions and non-linearities, or (b) persistence and amplification of {\it exogenous shocks}, governed by the proximity of a boundary in parameter space where the competitive equilibrium becomes unstable. While scenario (a) may appear at first sight to be more generic, the self-organized criticality scenario proposed long ago by \cite{bak1993aggregate} could make (b) plausible as well. Specific empirical work is needed to distinguish between these two scenarios.

It should however be borne in mind that many relevant features of the real economy are left out of the present version of the model. In particular, whereas firms are allowed to make losses, we have not accounted to the cost of credit that this would entail, and the impact of monetary policy, increasing or decreasing the interest rate in the face of inflation/deflation.

Nor have we introduced a bankruptcy mechanism when firms go too deep into debt, removing non-competitive firms along the lines of, e.g. \cite{Sharma2021}. But this would require moving from a static network of firms, as considered throughout this work, to a dynamically evolving network that rewires as some firms go bankrupt and others are created. In fact, another motivation for moving from such a static framework to a rewiring model is to be able to describe possible {\it cascades of bankruptcies} mediated by the input-output network, much as cascades of defaults can occur in banking networks. We leave this for further investigations.

The household sector also needs to be better described, moving away from the representative household assumption and introducing wage inequalities, confidence effects (as, for example, in \cite{morelli_pnas}) and debt. 
In fact, our dynamical model can be seen as a hybrid between traditional economic models (describing equilibrium) and Agent Based Models, where extra reasonable but {\it ad hoc} rules are implemented to account for out-of-equilibrium, dynamical aspects. As we have shown, in some swath of parameters, the classical competitive equilibrium is reached. If reached fast enough, the ``adiabatic'' assumption used in most standard descriptions will hold, whereas when the equilibration time is long (or even infinite) new phenomena appear. 

We hope that the possibility of recovering standard results in some limiting cases will make the ABM approach more palatable to economists, and at the same time elicit the inherent limits of general equilibrium ideas. Conversely, including firm network effects in ABMs like Mark-0 \citep{gualdi2015tipping, Gualdi2016} along the lines of the present model is certainly worthwhile. 

Finally, an appealing feature of our approach is the possibility to use highly dis-aggregated data on individual firms and prices (for example through the ``Billion Price Project'' \cite{Cavallo2016}) to calibrate the model and, hopefully, use it as a powerful descriptive and predictive tool. We look forward to working in that direction in the near future. 

\section*{Acknowledgments}

We thank Giulio Biroli, Doyne Farmer, Xavier Gabaix, Stanislao Gualdi,  Alan Kirman, Johannes Lumma and Francesco Zamponi for multiple feedbacks on our research program, Vu Do Chi Toai for helping setting up the web app, and Camille Boissel for her help in understanding this model and her numerous insights about economics. We also thank Antoine Mandel and Ben Moll for very useful comments on the first version of this manuscript. JPB benefited from numerous conversations with the members of the ``Rebuilding Macroeconomics'' project, including Angus Amstrong and Roger Farmer. JM would like to acknowledge countless discussions with Dhruv Sharma about Agent-Based Modelling. This research was conducted within the Econophysics \& Complex Systems Research Chair, under the aegis of the Fondation du Risque, the Fondation de l’Ecole polytechnique, the Ecole polytechnique and Capital Fund Management. 

\section*{Notations}

In this section, we summarise all the key notations that are used throughout the paper. 
\subsection*{Production function and network}

\begin{enumerate}[label=$\star$] \setlength\itemsep{0em}
    \item $q$ is the elasticity of substitution between inputs. The case $q=0$ corresponds to a Leontief production function where inputs are not substitutable to one another whereas $q=\infty$ corresponds to a Cobb-Douglas production function where inputs are fully substitutable. Furthermore, we call $\zeta=1/(q+1)$.
    \item $b$ is the return-to-scale parameter.
    \item $\mathbf{J}\in\mnm{N}{N+1}$ is the input-output matrix. Its entries $J_{ij}$ denote the amount of inputs made by $j$ needed by $i$ to produce one unit of its good, and therefore defines a weighed adjacency matrix and an interaction network. Conventionally, the input $j=0$ corresponds to labour and we use the notation $J_{i0}=V_i$.
    \item $\mathbf{a}\in\mnm{N}{N+1}$ is the substitution matrix. Its entries $a_{ij}$ and $a_{ik}$ indicate the ease with which firm $i$ can replace an input $k$ with another input $j$. For example, a large value of $a_{il}$ with respect to the other $a_{ik}$s means that input $l$ can easily substitute any other input.
    \item $\boldsymbol{\Lambda}\in\mnm{N}{N+1}=\mathbf{a}^{q\zeta}\circ\mathbf{J}^{\zeta}$ is the aggregate matrix for the Constant-Elasticity of Substitution production function.
    \item $\boldsymbol{\mathcal{M}}=\diag{z_i^\zeta}-\boldsymbol{\Lambda}$ is the network matrix with the productivity factors of the firms on the diagonal. We implicitly cross out the first column of $\boldsymbol{\Lambda}$. When $q=0$ (i.e Leontief production function) $\boldsymbol{\mathcal{M}}=\diag{z_i}-\mathbf{J}$.
    \item $\varepsilon$ is the smallest eigenvalue of the network matrix.
\end{enumerate}

\subsection*{Firms}
\begin{enumerate}[label=$\star$]\setlength\itemsep{0em}
    \item $N$ is the number of firms. 
    \item $z_i$ is the productivity factor of firm $i$.
    \item $\alpha$ is the log-elasticity of price growth rates against production surplus.
    \item $\alphap$ is the log-elasticity of price growth rates against profits.
    \item $\beta$ is the log-elasticity of production' growth rates against profits.
    \item $\betap$ is the log-elasticity of production growth rates against production surplus.
    \item $\omega$ is the log-elasticity of wage growth rate against labour market tensions.
    \item $\sigma_i$ is the depreciation parameter of good $i$.
    \item $p_i(t)\in\R^N$ is the price of good $i$ time $t$.
    \item $p_0(t)$ is the common wage used to pay the household at time $t$.
    \item $y_i(t):=z_i\gamma_i(t)$ is the production of firm $i$ at time $t$ along with the corresponding production levels $\gamma_i(t)$ at time $t$.
    \item $\widehat{y}_i(t):=z_i\widehat{\gamma}_i(t)$ is the targeted productions by firm $i$ at time $t$ along with the corresponding targeted production level $\widehat{\gamma}_i(t)$ at time $t$.
    \item $I_{ij}(t)\in\mn{N}$ is the inventory of good $j$ possessed by firm $i$. In particular, the diagonal terms $I_{ii}(t)$ corresponds to the stock of its own production.
    \item $\mathscr{G}_i(t)$, $\mathscr{L}_i(t)$, $\mathscr{S}_i(t)$ and $\mathscr{D}_i(t)$ correspond respectively to the proceeds of sales (``gains''), the production costs (``losses''), the supply and the demand for each firm at time $t$.
    \item $\pi_i(t):=\mathscr{G}_i(t)-\mathscr{L}_i(t)$ is firm $i$'s realised profits at time $t$.
    \item $\mathscr{E}_i(t):=\mathscr{S}_i(t)-\mathscr{D}_i(t)$ is firm $i$'s production surplus at time $t$.
    \item $\widehat{x}_{ij}(t)$ is the quantity of good $j$ that minimises the costs for firm $i$ given a certain production target and for a given production function. $\widehat{x}_{i0}(t):=\widehat{\ell}_i(t)$ corresponds to the optimal amount of work required.
    \item $x^{\text{d}}_{ij}(t)$ is the quantity of input $j$ that is demanded by firm $i$ to firm $j$. $x^{\text{d}}_{i0}(t):=\ell^{\text{d}}_i(t)$ corresponds to the demanded amount of work.
    \item $x_{ij}(t)$ is the quantity of input $j$ that is effectively exchanged. $x_{i0}(t):=\ell_i(t)$ corresponds to the amount of work the household is hired to do for firm $i$.
    \item $x^{\text{a}}_{ij}(t)$ is the quantity of input $j$ that is available for production. $x^{\text{a}}_{i0}(t):=\ell^{\text{a}}_i(t)$ corresponds to the available workforce for production.
    \item $x^{\text{u}}_{ij}(t)$ is the quantity of input $j$ that is effectively used for production. $x^{\text{u}}_{i0}(t):=\ell^{\text{u}}_i(t)$ corresponds to the available workforce for production.
    \item $\lambda$ is a behavioural parameter determining how firms forecast their future exchanges. 
\end{enumerate}

\subsection*{Household}

\begin{enumerate}[label=$\star$]\setlength\itemsep{0em}
    \item $\theta_i$ is the consumption preference of the household for good $i$.
    \item $\bar{\theta}=\sum_i\theta_i$.
    \item $L_0$ is the nominal number of hours that the household is willing to work.
    \item $\Gamma$ is the aversion to work parameter.
    \item $\varphi$ is the convexity-to-work parameter.
    \item $\omegap$ is a consumption confidence parameter.
    \item $\mathcal{U}(t)$ is the utility of the household at time $t$.
    \item $L^{\text{s}}(t)$ is the available supply of work at time $t$.
    \item $L^{\text{d}}(t)$ is the total demand for work at time $t$. 
    \item $L(t)$ is the actual amount of work done at time $t$. 
    \item $C^{\text{d}}_i(t)$ is the demanded consumption at time $t$.
    \item $C^{\text{r}}_i(t)$ is the realised consumption at time $t$.
    \item $B(t)$ is the budget at time $t$.
    \item $S(t)$ are the savings at time  $t$. 
\end{enumerate}

\bibliographystyle{apalike}
\bibliography{biblio.bib}

\newpage
\appendix

%!TEX root = ./draft_theo.tex
\section{General Equilibrium Conditions\label{ap:eq}}

In this appendix, we show the computations that lead to the equilibrium equations on prices and production levels in the case of a general CES production function and a non-constant return-to-scale $b$.

\subsection{Computation of equilibrium relations}

\subsubsection{Case \texorpdfstring{$q<+\infty$}{Lg}: Leontief and general CES}

We first enforce the market clearing condition
\begin{equation}
    z_i\preq{i}=\sum_{j=1}^N x_{\text{eq},ji}+C_{\text{eq},i},
\end{equation}
and inject it into the zero-profit condition using \eqref{eq:optimal_quantities}. We can deduce a nicer expression for the quantity $p^{\text{net}}_{\text{eq},i}=\sum_{j=0}^N\Lambda_{ij}\peq{j}^\zeta$ at equilibrium:
\begin{align*}
    z_{i}\peq{i}\preq{i}=\sum_{j=1}^N\Lambda_{ij}^{a}\peq{j}^{\zeta}\left(p^{\text{net}}_{\text{eq},i}\right)^{q}\preq{i}^{1/b}&\Longleftrightarrow z_{i}\peq{i}\preq{i}^{\frac{b-1}{b}}=\left(p^{\text{net}}_{\text{eq},i}\right)^{q}\sum_{j}\Lambda_{ij}\peq{j}^{\zeta}\\
    &\Longleftrightarrow \left(p^{\text{net}}_{\text{eq},i}\right)^{q+1}=z_{i}\peq{i}\preq{i}^{\frac{b-1}{b}}\\
    &\Longleftrightarrow p^{\text{net}}_{\text{eq},i}=\left(z_{i}\peq{i}\preq{i}^{\frac{b-1}{b}}\right)^{\zeta},
\end{align*}
and therefore a nicer expression for the exchanged quantities
\begin{equation}
    x_{ij}^{eq}=\Lambda_{ij}\peq{j}^{-q\zeta}z_{i}^{q\zeta}\peq{i}^{q\zeta}\preq{i}^{\frac{\zeta(b-1)+1}{b}}=z_{i}^{q\zeta}\Lambda_{ij}^{a}\left(\frac{\peq{i}}{\peq{j}}\right)^{q\zeta}\preq{i}^{\zeta\frac{bq+1}{b}}.
    \label{eq:quantities_eq}
\end{equation}

Using the null budget condition we can retrieve the equilibrium consumption
\begin{equation}
    C^{eq}_{i}=\frac{\theta_{i}}{\bar{\theta}^{\frac{\varphi}{1+\varphi}}\Gamma^{\frac{1}{1+\varphi}}}\frac{L_0}{\peq{i}}:=\frac{\kappa_i}{\peq{i}},\label{eq:cons_eq}
\end{equation}
so that we have every ingredients to get closed form equations on prices and production levels. We express \eqref{eq:quantities_eq} and \eqref{eq:cons_eq} back into the the zero profit condition to retrieve the first equilibrium equation:
\begin{align*}
    &\forall i,\;z_{i}\peq{i}\preq{i}-\sum_{j=1}^{N}\peq{j}z_{i}^{q\zeta}\Lambda_{ij}\left(\frac{\peq{i}}{\peq{j}}\right)^{q\zeta}\preq{i}^{\zeta\frac{bq+1}{b}}=z_{i}^{q\zeta}\Lambda_{i0}^{a}\peq{i}^{q\zeta}\preq{i}^{\zeta\frac{bq+1}{b}}\\
    \Longleftrightarrow \;\;&\forall i,\; z_{i}^{\zeta}\peq{i}^{\zeta}\preq{i}^{\zeta\frac{b-1}{b}}-\sum_{j=1}^{N}\Lambda_{ij}\peq{j}^{\zeta}=\Lambda_{i0}\\
    \Longleftrightarrow \;\;&\forall i,\; z_{i}^{\zeta}\peq{i}^{\zeta}-\sum_{j=1}^{N}\Lambda_{ij}\peq{j}^{\zeta}=\Lambda_{i0}+z_{i}^{\zeta}\peq{i}^{\zeta}\left(1-\preq{i}^{\zeta\frac{b-1}{b}}\right)\\
    \Longleftrightarrow \;\;& \boldsymbol{\mathcal{M}}{\peqv}^{\zeta}=\mathbf{V} +\mathbf{z}^{\zeta}\circ{\peqv}^{\zeta}\circ\left(1-{\preqv}^{\zeta\frac{b-1}{b}}\right),
\end{align*}

and then in the market clearing condition to retrieve the second equilibrium equation:
\begin{align*}
    &\forall i,\;z_{i}\preq{i}-\sum_{j=1}^{N}z_{j}^{q\zeta}\Lambda_{ji}\left(\frac{\peq{j}}{\peq{i}}\right)^{q\zeta}\preq{j}^{\zeta\frac{bq+1}{b}}=\frac{\kappa_i}{\peq{i}}\\
    \Longleftrightarrow\;\; &\forall i,\; z_{i}\preq{i}\peq{i}^{q\zeta}-\sum_{j=1}^{N}z_{j}^{q\zeta}\Lambda_{ji}\peq{j}^{q\zeta}\preq{j}^{\zeta\frac{bq+1}{b}}=\frac{\kappa_i}{\peq{i}^{\zeta}}\\
    \Longleftrightarrow\;\; &\forall i,\; z_{i}^{\zeta}\preq{i}z_{i}^{q\zeta}\peq{i}^{q\zeta}-\sum_{j=1}^{N}\Lambda_{ji}z_{j}^{q\zeta}\peq{j}^{q\zeta}\preq{j}^{\zeta\frac{bq+1}{b}}=\frac{\kappa_i}{\peq{i}^{\zeta}}\\
    \Longleftrightarrow\;\; &\forall i,\; z_{i}^{\zeta}\preq{i}^{\zeta\frac{bq+1}{b}}z_{i}^{q\zeta}\peq{i}^{q\zeta}-\sum_{j=1}^{N}\Lambda_{ji}z_{j}^{q\zeta}\peq{j}^{q\zeta}\preq{j}^{\zeta\frac{bq+1}{b}}=\frac{\kappa_i}{\peq{i}^{\zeta}}+z_{i}\peq{i}^{q\zeta}\preq{i}^{\zeta\frac{bq+1}{b}}\left(1-\preqv^{\zeta\frac{b-1}{b}}\right)\\
    \Longleftrightarrow\;\; &\boldsymbol{\mathcal{M}}^{\top}\diag{\mathbf{z}^{q\zeta}{\peqv}^{q\zeta}}{\preqv}^{\zeta\frac{bq+1}{b}}=\frac{\boldsymbol{\kappa}}{{\peqv}^{\zeta}}+\mathbf{z}\circ{\peqv}^{q\zeta}\circ{\preqv}^{\zeta\frac{bq+1}{b}}\left(1-{\preqv}^{\zeta\frac{b-1}{b}}\right).
\end{align*}

In the case where $q\to 0^+$ and $b=1$, one can check that \eqref{eq:eq_CRS} is retrieved.

\subsubsection{Case \texorpdfstring{$q=+\infty$}{Lg}: Cobb-Douglas}

To retrieve the equations in the case $q=+\infty$, we need to take this limit in \eqref{eq:optimal_quantities}. It yields
\begin{align*}
    \widehat{x}_{il}&=a_{il}^{q\zeta}J_{il}^{\zeta}p_l^{-q\zeta}\left(\sum_{j=0}^{N}a_{ij}^{q\zeta}J_{ij}^{\zeta}p_j^{\zeta}\right)^q \hat{\gamma}_i^{1/b}\\
    &=a_{il}^{q\zeta}J_{il}^{\zeta}p_l^{-q\zeta}\left(\sum_{\substack{j=0\\J_{ij}\neq0}}^{N}a_{ij}^{q\zeta}J_{ij}^{\zeta}p_j^{\zeta}\right)^q \hat{\gamma}_i^{1/b}\\
    &\hspace{-0.15cm}\underset{q\rightarrow+\infty}{\approx}a_{il}p_l\hat{\gamma}_i^{1/b}\exp\left\{q\log\left(\sum_{\substack{j=0\\J_{ij}\neq0}}^{N}a_{ij}\exp\zeta\log\left[\frac{J_{ij}}{a_{ij}}p_{j}\right]\right)\right\}\\
    &\hspace{-0.15cm}\underset{q\rightarrow+\infty}{\approx}a_{il}p_l^{-1}\hat{\gamma}_i^{1/b}\exp\left\{q\log\left(\sum_{\substack{j=0\\J_{ij}\neq0}}^{N}a_{ij} +\zeta \sum_{\substack{j=0\\J_{ij}\neq0}}^{N}\log\left[\frac{J_{ij}}{a_{ij}}p_{j}\right]\right)\right\}\\
    &\hspace{-0.15cm}\underset{q\rightarrow+\infty}{\approx}a_{il}p_l^{-1}\hat{\gamma}_i^{1/b}\exp\left\{q\log\left(1 +\zeta \sum_{\substack{j=0\\J_{ij}\neq0}}^{N}\log\left[\frac{J_{ij}}{a_{ij}}p_{j}\right]\right)\right\}\\
    &\hspace{-0.15cm}\underset{q\rightarrow+\infty}{\approx}a_{il}p_l^{-1}\hat{\gamma}_i^{1/b}\exp\left\{q\zeta \sum_{\substack{j=0\\J_{ij}\neq0}}^{N}\log\left[\frac{J_{ij}}{a_{ij}}p_{j}\right]\right\}\\
    &\hspace{-0.15cm}\underset{q\rightarrow+\infty}{\approx}a_{il}p_l^{-1}\hat{\gamma}_i^{1/b} \prod_{\substack{j=0\\J_{ij}\neq0}}^{N}\left(\frac{J_{ij}}{a_{ij}}p_{j}\right).
\end{align*}

We can then express the quantity $z_i\preq{i}^{\frac{b-1}{b}}\peq{i}$ through the zero profit condition as
\begin{equation}
    z_i\preq{i}^{\frac{b-1}{b}}\peq{i}=\prod_{\substack{j=0\\J_{ij}\neq0}}^{N}\left(\frac{J_{ij}}{a_{ij}}\peq{j}\right).
    \label{eq:p_net_cobb_douglas}
\end{equation}
Using the market clearing condition, we can get the first equilibrium equation in the Cobb-Douglas case:
\begin{align*}
    &\forall i,\;z_i\preq{i}=\frac{\kappa_i}{\peq{i}}+\sum_j a_{ji}\peq{i}^{-1}\preq{j}^{1/b}\prod_{\substack{j=0\\J_{ij}\neq0}}^{N}\left(\frac{J_{ij}}{a_{ij}}\peq{j}\right)\\
    \Longleftrightarrow\;\; &\forall i,\;
    z_i\preq{i}\peq{i}=\kappa_i+\sum_j a_{ji}z_j\preq{j}\peq{j}\\
    \Longleftrightarrow\;\; &\left(\mathbf{I}_{N}-\mathbf{a}^{\top}\right)\mathbf{z}\circ\preqv\circ\peqv=\boldsymbol{\kappa}\\
    \Longleftrightarrow\;\; &\mathbf{z}\circ\preqv\circ\peqv=\left(\mathbf{I}_{N}-\mathbf{a}^{\top}\right)^{-1}\boldsymbol{\kappa}.
\end{align*}

To get the second equation, we inject the previous into \eqref{eq:p_net_cobb_douglas} and take the logarithm. It reads

\begin{align*}
    &\forall i,\; \log{z_i\preq{i}\peq{i}}-\frac{1}{b}\log\preq{i}=\sum_{\substack{l=1\\J_{il}\neq0}}^{N}a_{il}\log\frac{J_{il}}{a_{il}}+\sum_{\substack{l=1\\J_{il}\neq0}}^{N}a_{il}\log\peq{i}\\
    \Longleftrightarrow\;\; &\forall i,\;\frac{b-1}{b}\log\left[\left(\mathbf{I}_{N}-\mathbf{a}^{\top}\right)^{-1}\boldsymbol{\kappa}\right]_i+\frac{1}{b}\log\peq{i}+\frac{1}{b}\log z_i=\sum_{\substack{l=1\\J_{il}\neq0}}^{N}a_{il}\log\frac{J_{il}}{a_{il}}+\sum_{\substack{l=1\\J_{il}\neq0}}^{N}a_{il}\log\peq{i}\\
    \Longleftrightarrow\;\; &\left(\frac{1}{b}\mathbf{I}_{N}-\mathbf{a}\right)\log\peqv=\frac{1-b}{b}\log\left(\mathbf{I}_{N}-\mathbf{a}^{\top}\right)^{-1}\boldsymbol{\kappa}-\frac{1}{b}\log\mathbf{z}+\mathbf{h}.
\end{align*}

where $h_i=\sum_{\substack{l=1\\J_{il}\neq0}}^{N}a_{il}\log\frac{J_{il}}{a_{il}}$. 

In the Cobb-Douglas case, a positive equilibrium for prices and productions always exists. Indeed, looking at the second equation, one sees that a solution generically exists (except in the very specific case where $b^{-1}$ is an eigenvalue of $\mathbf{a}$) for $\log{\peqv}$. Exponentiating this solutions shows that $\peqv$ will always be positive. For the first equation, the matrix $\mathbf{I}_N-\mathbf{a}$ is always invertible since the eigenvalues $\lambda$ of $\mathbf{a}$ are such that $|\lambda|\leq\sum_{j=1}^Na_{ij}=1-a_{i0}<1$ thanks to Gershgorin's theorem. This also proves that $\mathbf{I}_N-\mathbf{a}$ is in fact an M-matrix, which makes the solution of this equation positive, implying in turn that $\preqv$ is also positive.

\subsection{Positive equilibrium}

The question of positive solutions to non-linear algebraic systems is a complicated one. For now, no general recipe exists to prove that positive solutions exist except in some very specific cases. The previous set of equations for $q<+\infty$  and generic $b$ is no exception. However, first order approximations are possible. 

\begin{figure}
    \centering
    \includegraphics{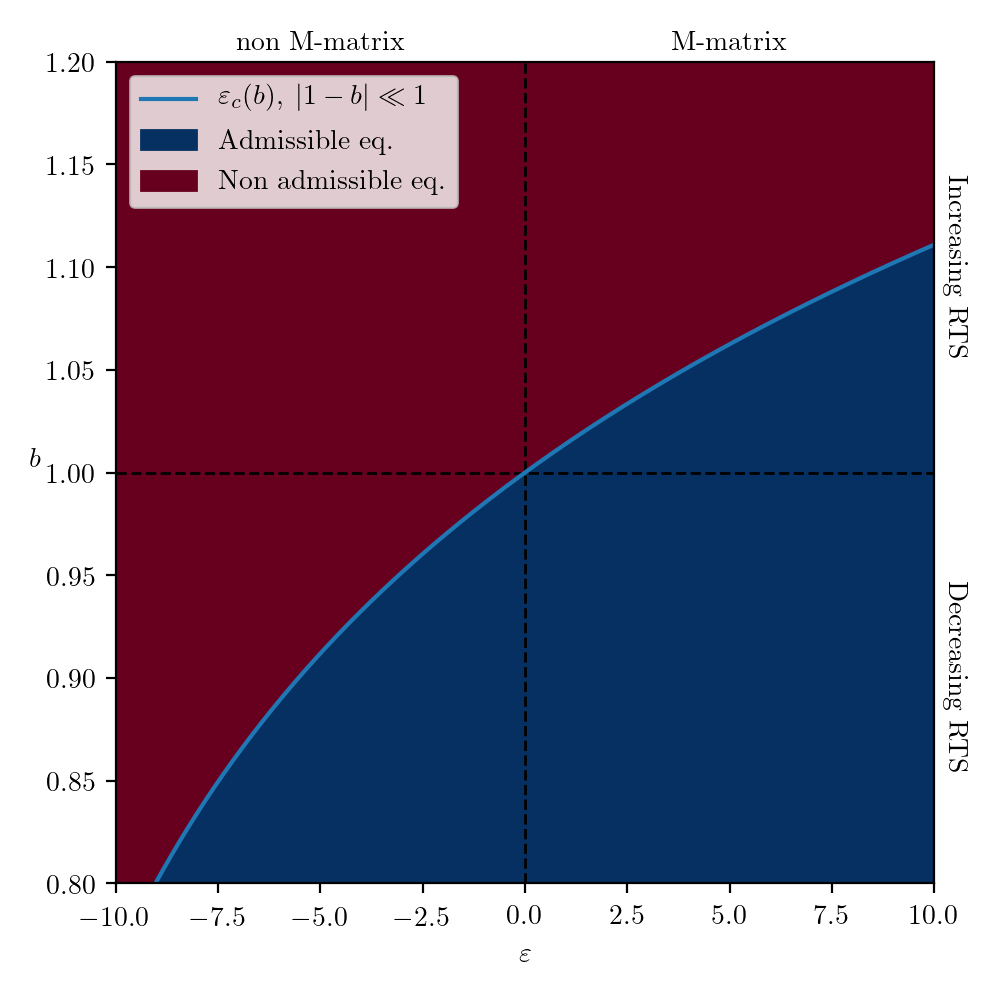}
    \caption{Region of existence of a positive equilibrium for a $d$-regular network on $n=100$ firms. We see that we retrieve the standard Hawkin-Simons transition $\varepsilon_c(0)=0$ for $\delta=0$ i.e $b=1$. The first order solution \ref{eq:eps_c_d_reg} is plotted in blue and fits very well for this range of values of $b$ and $\varepsilon$. However, upon zooming on the frontier on the right side of the plot, one begins to see a divergence between the blue line and the numerical frontier.}
    \label{fig:positive_eq}
\end{figure}

Setting $q=0$ (Leontief production function), the equations read
\begin{subequations}
\begin{eqnarray}
\boldsymbol{\mathcal{M}}{\peqv}&=&\mathbf{V} +\mathbf{z}{\peqv}\left(1-{\preqv}^{\frac{b-1}{b}}\right)\\
    \boldsymbol{\mathcal{M}}^{\top}\preqv^{1/b}&=&\frac{\boldsymbol{\kappa}}{{\peqv}}+\mathbf{z}\preqv^{1/b}\left(1-{\preqv}^{\frac{b-1}{b}}\right).
\end{eqnarray}
\end{subequations}
We know the solutions $\peqv^{(0)}, \preqv^{(0)}$ for $b=1$ and we set $b=1-\delta$. Assuming that equilibrium prices and productions are of the form $\peqv=\peqv^{(0)}+\peqv^{(1)}(\delta), \preqv=\preqv^{(0)}+\preqv^{(1)}(\delta)$ such that at least $\|\peqv^{(1)}(\delta)\|_\infty\ll\|\peqv^{(0)}(\delta)\|$ (and same for $\preqv$), we can write an equation for $\peqv^{(1)}$ yielding
\begin{equation}
    \boldsymbol{\mathcal{M}}{\peqv^{(1)}}=\mathbf{z}\peqv^{(0)}\left(1-\exp{\left(-\delta \log\preqv^{(0)}\right)}\right).
\end{equation}
Developing the solution at leading order in $\varepsilon$ (we use results and notations from Appendix \ref{ap:block_stability_matrix}) we get
\begin{equation}
    \peqv^{(1)}=\ket{r_N}\frac{\rho_n}{\varepsilon^2}\ps{\ell_N}{V}\bra{\ell_N}\diag{1-\exp{\left(-\delta \log\frac{\ket{\ell_N}}{\ps{\ell_N}{V}}\right)}}\ket{r_N},
\end{equation}
where any function of a vector is understood as component-wise. We can write $\peqv$ at leading order in $\varepsilon$
\begin{equation}
    \peqv=\ket{r_N}\frac{\ps{\ell_N}{V}}{\varepsilon}\left(1+\frac{\rho_n}{\varepsilon}\bra{\ell_N}\diag{1-\exp{\left(-\delta \log\frac{\ket{\ell_N}}{\ps{\ell_N}{V}}\right)}}\ket{r_N}\right),
\end{equation}
and get a first approximation of $\varepsilon_c(\delta)$ under which equilibrium prices are negative
\begin{equation}
    \varepsilon_c(\delta)\sim-\rho_n\left(1-\exp\left(-\delta\sum_j\ell_{N,j}r_{N,j}\log\frac{\ell_{N,j}}{\ps{\ell_N}{V}}\right)\right).
\end{equation}
In the case of an undirected d-regular network, we have
\begin{equation}
    \varepsilon_c(\delta)\sim-d\left(1-\exp\left(\delta\log\sum_j V_j\right)\right).
    \label{eq:eps_c_d_reg}
\end{equation}
Fig. \ref{fig:positive_eq} shows the region where an admissible equilibrium exists.

\section{Relaxation Time for the Naive Model\label{ap:relax_time}}

The non-linear dynamics of the naive model are given by Eqs. \eqref{eq:sys_tox_model}. In this Appendix, we derive the relaxation time of the system in various limits.

\subsection{Linearisation of the dynamics}
To linearise the system ,we write $p_i(t)=\peq{i} + \delta p_i(t)$ and $\gamma_i(t)=\preq{i} + \delta \gamma_i(t)$ and inject these expressions into \eqref{eq:sys_tox_model}. After a few computations, we establish the following linear equation in the variable $\mathbf{U}(t)=(\delta\mathbf{p}(t),\delta\boldsymbol{\gamma}(t))^\top$ to first order:

\begin{equation}
    \frac{d\mathbf{U}}{dt}=\begin{pmatrix}
    \mathbf{D}_1 & \mathbf{D}_2\\
    \mathbf{D}_3 & \mathbf{D}_4
    \end{pmatrix}\mathbf{U}(t):=\mathbb{D}\mathbf{U}(t),
\end{equation}
where the different blocks of the matrix are
\begin{equation}
    \begin{alignedat}{2}
    	\mathbf{D}_1&=-\alpha\mu\diag{\frac{\theta_i}{z_i\preq{i}\peq{i}}}-\alpha'\diag{z_i^{-1}}\boldsymbol{\mathcal{M}}
    	&&\qquad \mathbf{D}_2=-\alpha\diag{\frac{\peq{i}}{z_i\preq{i}}}\boldsymbol{\mathcal{M}}^\top\\
    	\mathbf{D}_3&=\beta\diag{\frac{\preq{i}}{z_i\peq{i}}}\boldsymbol{\mathcal{M}}-\beta'\mu\diag{\frac{\theta_i}{z_i\peq{i}^2}}
    	&&\qquad \mathbf{D}_4=-\beta'\diag{z_i^{-1}}\boldsymbol{\mathcal{M}}^\top.
    \end{alignedat}
\end{equation}

\subsection{Relaxation time in the high productivity regime}

In this section, we assume that the productivity factors are large enough to ignore interactions between firms. In this regime, firms are efficient enough so that the actual amount of inputs does not matter in the final production. In this limit, we can give approximate expressions for the equilibrium prices and productions 
\begin{subequations}
\begin{eqnarray}
    \peq{i}&=&\frac{V_i}{z_i}\\
\preq{i}&=&\frac{\mu\theta_i}{V_i}.
\end{eqnarray}
    \label{eq:appendix_eq_large_z}
\end{subequations}

Similarly, we  approximate each block of the stability matrix:
\begin{equation}
    \begin{alignedat}{2}
    	\mathbf{D}_1&\underset{z_i\to\infty}{\approx}-(\alpha+\alpha') \mathbf{I}_{N}
    	&&\qquad \mathbf{D}_2\underset{z_i\to\infty}{\approx}-\alpha\diag{\frac{V_i^2}{z_i\mu\theta_i}}\\
    	\mathbf{D}_3&\underset{z_i\to\infty}{\approx}(\beta-\beta')\diag{\frac{z_i\mu\theta_i}{V_i^2}}
    	&&\qquad \mathbf{D}_4\underset{z_i\to\infty}{\approx}-\beta'\mathbf{I}_{N},
    \end{alignedat}
\end{equation}
and deduce the spectrum of the $\mathbf{D}$ by computing its characteristic polynomial and setting it to $0$:
\begin{align*}
    \det{\left(\sigma\mathbf{I}_{2N}-\mathbb{D}\right)}&=
    \begin{vmatrix}
    \sigma\mathbf{I}_{N}-\mathbf{D}_1 & -\mathbf{D}_2\\
    -\mathbf{D}_3 & \sigma\mathbf{I}_{N}-\mathbf{D}_4
    \end{vmatrix}\\
    &\hspace{-0.25cm}\underset{z_i\to\infty}{\approx}\det\left(\left(\sigma+\alpha+\alpha'\right)\left(\sigma+\beta'\right)\mathbf{I}_{N}+\alpha(\beta-\beta')\mathbf{I}_{N}\right)\\
    &=\left(\sigma^2+\sigma(\alpha+\alpha'+\beta')+\alpha\beta+\alpha'\beta'\right)^{N}\\
    &=0.
\end{align*}

Solving this equation yields two eigenvalues $\sigma_\pm$, both with degeneracy $N$, that read
\begin{equation}\label{eq:eigenvalues_toy_large_z}
\sigma_\pm=\frac{1}{2}\times
\left\{
\begin{matrix}
-\alpha^{\prime}-\beta^{\prime}-\alpha\pm\sqrt{(\alpha^{\prime}+\beta^{\prime}+\alpha)^{2}-4(\alpha\beta+\alpha^{\prime}\beta^{\prime})} &\text{ if }& (\alpha^{\prime}+\beta^{\prime}+\alpha)^{2}>4(\alpha\beta+\alpha^{\prime}\beta^{\prime}) \\
 -\alpha^{\prime}-\beta^{\prime}-\alpha\pm i\sqrt{4(\alpha\beta+\alpha^{\prime}\beta^{\prime})-(\alpha^{\prime}+\beta^{\prime}+\alpha)^{2}} &\text{ if }& (\alpha^{\prime}+\beta^{\prime}+\alpha)^{2}<4(\alpha\beta+\alpha^{\prime}\beta^{\prime})
\end{matrix}
\right..
\end{equation}

This in turn lets us deduce the relaxation time:
\begin{equation}\label{eq:relax_time_toy_large_z}
\tau_{relax}=2\times
\left\{
\begin{matrix}
(\alpha^{\prime}+\beta^{\prime}+\alpha-\sqrt{(\alpha^{\prime}+\beta^{\prime}+\alpha)^{2}-4(\alpha\beta+\alpha^{\prime}\beta^{\prime})})^{-1} &\text{ if }& (\alpha^{\prime}+\beta^{\prime}+\alpha)^{2}>4(\alpha\beta+\alpha^{\prime}\beta^{\prime}) \\
 (\alpha^{\prime}+\beta^{\prime}+\alpha)^{-1} &\text{ if }& (\alpha^{\prime}+\beta^{\prime}+\alpha)^{2}\geq4(\alpha\beta+\alpha^{\prime}\beta^{\prime})
\end{matrix}
\right..
\end{equation}

\subsection{Perturbation expansion in \texorpdfstring{$\varepsilon$}{Lg} for \texorpdfstring{$\mathbb{D}$}{Lg}}

Studying the behaviour of $\mathbb{D}$ as $\varepsilon\to 0^+$ requires understanding the behaviour of $\left(\boldsymbol{\mathcal{M}}, \mathbf{p}_{eq}, \boldsymbol{\gamma}_{eq}\right)$ in that limit. We now introduce the matrix $\widetilde{\mathbf{J}} = \diag{z_{\max}-z_i} + \mathbf{J}$ and denote by $\rho_\nu$ (resp. $\ket{r_\nu}$, $\bra{\ell_\nu}$)\footnote{We use here Dirac bra-ket notation, where $\ket{v}$ represents a column vector and $\bra{v}$ a row vector.} its eigenvalues (resp. right/left eigenvectors) ordered by their real parts. The  Perron-Froebenius theorem implies that the top eigenvalue $\rho_{N}$ is real, simple and associated to a full and positive eigenvector. We next use the following spectral representation of the matrix $\boldsymbol{\mathcal{M}}$:

\begin{align}
\boldsymbol{\mathcal{M}}&= \left(\rho_{N}\mathbf{I}_{N} - \widetilde{\mathbf{J}}\right)+\varepsilon \mathbf{I}_{N}\\
\begin{split}
\boldsymbol{\mathcal{M}}^{-1}&=\frac{1}{\varepsilon}\ket{r_{N}}\bra{\ell_{N}}+\sum_{\nu=1}^{N-1}\frac{1}{\rho_{N}-\rho_{\nu}+\varepsilon}\ket{r_{\nu}}\bra{\ell_{\nu}}\\
&=\frac{1}{\varepsilon}\ket{r_{N}}\bra{\ell_{N}}+\sum_{k=0}^{\infty}(-\varepsilon)^{k}\sum_{\nu=1}^{N-1}\frac{1}{(\rho_{N}-\rho_{\nu})^{k+1}}\ket{r_{\nu}}\bra{\ell_{\nu}},
\end{split}
\label{eq:left_right_decomp}
\end{align}
which lets us express the equilibrium prices and outputs as well as $\mathbb{D}$. We also use the notation $\boldsymbol{\mathcal{M}}_0$ to refer to the network matrix when $\varepsilon=0$. This matrix is singular and verifies
\begin{equation}
    \boldsymbol{\mathcal{M}}_0\ket{r_N}=0\quad,\quad\boldsymbol{\mathcal{M}}^\top_0\ket{\ell_N}=0.
\end{equation}

Expanding in $\varepsilon$ and neglecting factors of order $\varepsilon^4$ and higher gives the following for the blocks of the stability matrix:
\begin{equation}
    \begin{alignedat}{2}
    	\mathbf{D}_1&=\mathbf{D}_1^{(0)}+\varepsilon \mathbf{D}_1^{(1)}+\varepsilon^{2}\mathbf{D}_1^{(2)}+\varepsilon^{3}\mathbf{D}_1^{(3)}
    	&&\qquad \mathbf{D}_2=\frac{1}{\varepsilon}\mathbf{D}_2^{(-1)}+\mathbf{D}_2^{(0)}+\varepsilon \mathbf{D}_2^{(1)}+\varepsilon^{2}\mathbf{D}_2^{(2)}+\varepsilon^{3}\mathbf{D}_2^{(3)}\\
    	\mathbf{D}_3&=\varepsilon \mathbf{D}_3^{(1)}+\varepsilon^{2}\mathbf{D}_3^{(2)}+\varepsilon^{3}\mathbf{D}_3^{(3)}
    	&&\qquad \mathbf{D}_{4}=\mathbf{D}_{4}^{(0)}+\varepsilon \mathbf{D}_{4}^{(1)}+\varepsilon^{2}\mathbf{D}_{4}^{(2)}+\varepsilon^{3}\mathbf{D}_{4}^{(3)},
    \end{alignedat}
\end{equation}
where the exact definition of the perturbation terms $\mathbf{D}^{(l)}_i$ is given in the Appendix \ref{ap:block_stability_matrix}. To ease computations and give closed-form results, we consider an undirected network (symmetric $\boldsymbol{\mathcal{M}}$) with homogeneous productivity factors. The qualitative results are however unchanged when considering more general networks. In this setting, the eigenvectors of $\boldsymbol{\mathcal{M}}$ are denoted by $\ket{e_\nu}$.

\subsection{Marginal stability for \texorpdfstring{$\varepsilon=0$}{Lg}}

Interestingly enough, although the upper-right block of $\mathbb{D}$ diverges as $\varepsilon\to0$, its spectrum converges to a finite limit. To see this, we use the block determinant formula 
\[\begin{vmatrix}\mathbf{A} & \mathbf{B}\\ \mathbf{C} & \mathbf{D}\end{vmatrix}=\det\left(\mathbf{A}\mathbf{D}-\mathbf{B}\mathbf{C}\right),\]
for same-size matrices, where the commutator $\left[\mathbf{C},\mathbf{D}\right]=\mathbf{C}\mathbf{D}-\mathbf{D}\mathbf{C}=0$. In our case, we need $\left[\mathbf{D}_3,\mathbf{D}_4\right]=0$ which is true only in the limit $\varepsilon=0$. We can then write:\footnote{We do not need to consider terms of order one in the commutator because $\left[\mathbf{D}_3^{(1)},\mathbf{D}_4^{(0)}\right]=0$, see Appendix \ref{ap:block_stability_matrix}.}
\begin{align*}
    \det\left(\sigma \mathbf{I}_{2N}-\mathbb{D}\right)&\hspace{-0.15cm}\underset{\varepsilon\to0}{\approx}\det\left(\left(\sigma \mathbf{I}_{N}-\mathbf{D}_{1}^{(0}\right)\left(\sigma \mathbf{I}_{N}-\mathbf{D}_{4}^{(0}\right)-\mathbf{D}_{2}^{(-1)}\mathbf{D}_3^{(1)}\right)\\
    &=\det\left(\left(\sigma \mathbf{I}_{N}+\frac{\alpha'}{\rho_{N}}\boldsymbol{\mathcal{M}}_0\right)\left(\sigma \mathbf{I}_{N}+\frac{\beta'}{\rho_{N}}\boldsymbol{\mathcal{M}}_0\right)+\frac{\alpha\beta}{\rho_{N}^2}\boldsymbol{\mathcal{M}}_0^2\right)\\
    &=\det\left(\sigma^2\mathbf{I}_{N}+\sigma\frac{\alpha'+\beta'}{\rho_{N}}\boldsymbol{\mathcal{M}}_0+\frac{\alpha\beta+\alpha'\beta'}{\rho_{N}^2}\boldsymbol{\mathcal{M}}_0^2\right)\\
    &=\prod_{\nu=1}^{N}\left(\sigma^2+\sigma\frac{\alpha'+\beta'}{\rho_{N}}(\rho_{N}-\rho_\nu)+\frac{\alpha\beta+\alpha'\beta'}{\rho_{N}^2}(\rho_{N}-\rho_\nu)^2\right)\\
    &=\sigma^2\prod_{\nu\neq N}\left(\sigma^2+\sigma(\alpha'+\beta')\left(1-\frac{\rho_\nu}{\rho_{N}}\right)+(\alpha\beta+\alpha'\beta')\left(1-\frac{\rho_\nu}{\rho_{N}}\right)^2\right).
\end{align*}

Each factor in this product yields two eigenvalues:
\begin{itemize}
    \item If $(\alpha'-\beta')^2>4\alpha\beta$ then
    \begin{equation}
        \sigma_\pm^\nu=\frac{1}{2}\left(-\alpha'-\beta'\pm\sqrt{(\alpha'+\beta')^2-4(\alpha\beta+\alpha'\beta')}\right)\left(1-\frac{\rho_\nu}{\rho_{N}}\right),
        \label{eq:sup_eigenvalues_D}
    \end{equation}
    \item If $(\alpha'-\beta')^2<4\alpha\beta$ then
    \begin{equation}
        \sigma_\pm^\nu=\frac{1}{2}\left(-\alpha'-\beta'\pm i\sqrt{4(\alpha\beta+\alpha'\beta')-(\alpha'+\beta')^2}\right)\left(1-\frac{\rho_\nu}{\rho_{N}}\right),
        \label{eq:inf_eigenvalues_D}
    \end{equation}
    \item If $(\alpha'-\beta')^2=4\alpha\beta$ then
    \begin{equation}
        \sigma_0^\nu=-\frac{\alpha'+\beta'}{2}\left(1-\frac{\rho_\nu}{\rho_{N}}\right).
        \label{eq:eq_eigenvalues_D}
    \end{equation}
\end{itemize}

\begin{figure}[ht!]
\includegraphics[width=\textwidth]{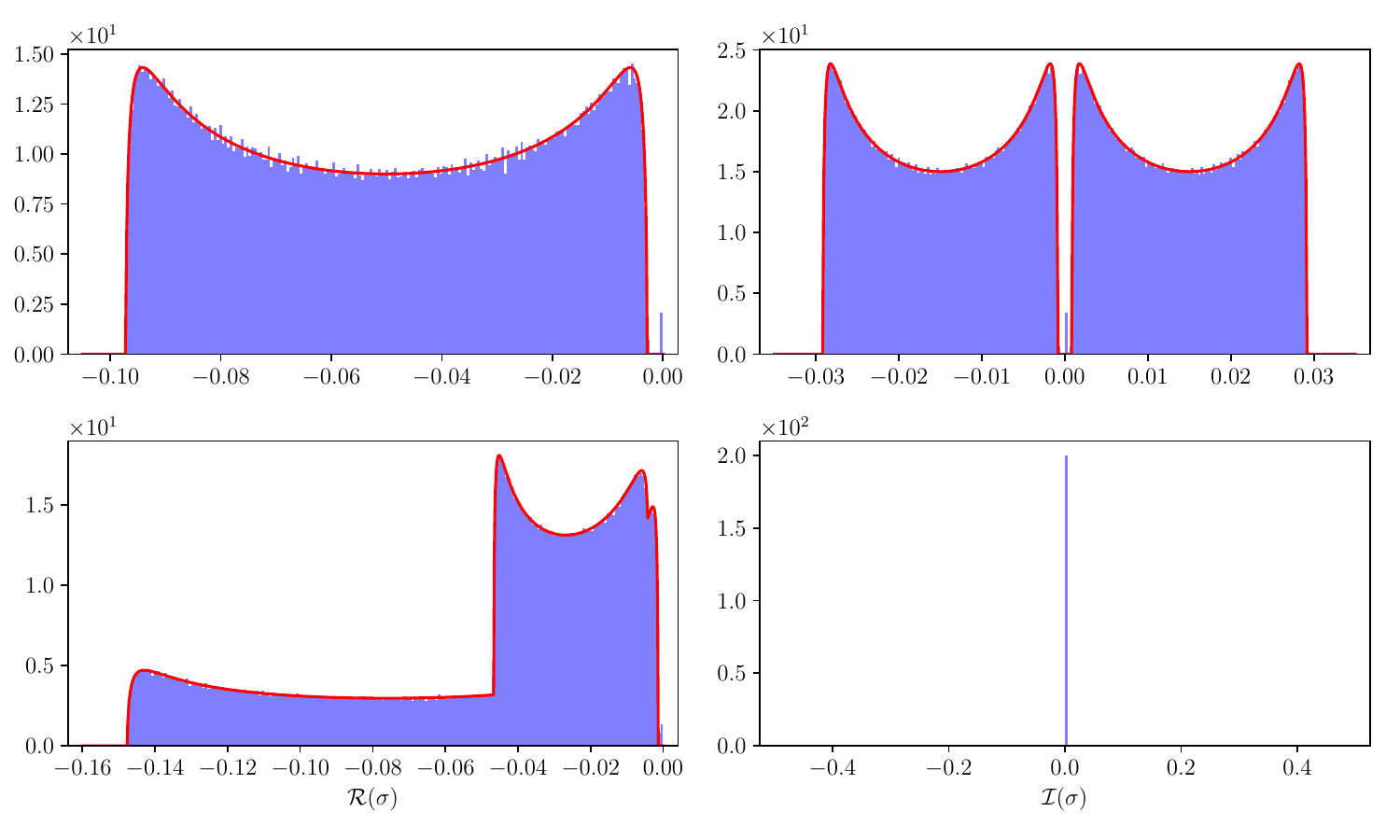}
\caption{\label{fig:spectrum_D} Histograms of the real parts (left column) and imaginary parts (right columns) of the spectrum of the stability matrix $\mathbb{D}$ (for $N=1000$ firms on a $d$-regular undirected network with $d=3$) obtained through numerical diagonalization. Top: Case $(\alpha'-\beta')^2<4\alpha\beta$. Bottom: $(\alpha'-\beta')^2>4\alpha\beta$. The red line is the thermodynamic computation accounting for \eqref{eq:inf_eigenvalues_D} and \eqref{eq:sup_eigenvalues_D} using the McKay density for the eigenvalues of a random $3$-regular graph~\cite{MCKAY1981203}. One can notice spikes at $0$ accounting for the case $\nu=N$ in \eqref{eq:inf_eigenvalues_D},\eqref{eq:sup_eigenvalues_D}.}
\end{figure}

The trailing factor shows that 0 is an eigenvalue of $\mathbb{D}$ (for $\nu=N$), twice degenerated as $\varepsilon\to0$. We deduce that the system exhibits marginal stability in this limit. Figure \ref{fig:spectrum_D} shows the empirical distribution of eigenvalues of $\mathbb{D}$ and the corresponding theoretical predictions.

\subsection{Relaxation time in the limit \texorpdfstring{$\varepsilon\to0$}{Lg}}

We have thus far shown that our system exhibits marginal stability at $\varepsilon=0$. We now prove that the relaxation time of the system behaves as $\tau_{relax}\sim\varepsilon^{-1}$. To this end, we use analytical perturbation theory as described in \cite{an_pert}, which in our setting reduces to the $\varepsilon$-perturbation of the characteristic polynomial of $\mathbb{D}(0)$\footnote{We have done a slight abuse of notation, since $\mathbb{D}(0)$ is not formally defined because of the diverging upper right block.} as $\varepsilon$ goes away from 0. This characteristic polynomial is given by 
\begin{equation}
    \chi(\sigma,0)=\sigma^2\prod_{\nu=1}^{N}\left(\sigma-\sigma_+^\nu\right)\left(\sigma-\sigma_-^\nu\right),
\end{equation}
with $\sigma_\pm^\nu$ given in the previous section.

We now try to find a perturbation of the $\sigma^2$ term to retrieve the perturbation on $\sigma_\pm^{N}=0$. Using analytical perturbation theory, we see that $(\varepsilon,\sigma)=(0,0)$ is a splitting point under the perturbation $\mathbb{D}(\varepsilon)$ ($\varepsilon=0$ is a multiple point -- since $\mathbb{D}$ has at least one multiple root for $\varepsilon=0$ -- and $\sigma_\pm^{N}=0$ is a multiple root.) In this setting, $\sigma_\pm^{N}=0$ splits under the perturbation $\mathbb{D}(\varepsilon)$ to give 2 perturbed eigenvalues. Henceforth, for small enough $\varepsilon$,  the prime factor $\sigma^{2}$ of $\chi(\sigma,0)$ is expressed as a second order polynomial whose coefficients depend on $\varepsilon$.

We may write
\[p_{0}(\sigma):=\sigma^{2}\overset{\mathbb{D}(\varepsilon)}{\longrightarrow}p_{0}(\sigma,\varepsilon):=\sigma^{2}(1+a_{2}^{(1)}\varepsilon+a_{2}^{(2)}\varepsilon^{2}+\cdots)+\sigma(a_{1}^{(1)}\varepsilon+a_{1}^{(2)}\varepsilon^{2}+\cdots)+a_{0}^{(1)}\varepsilon+a_{0}^{(2)}\varepsilon^{2}+\cdots.\]
This expansion makes sure that $p_{0}(\sigma,\varepsilon)\underset{\varepsilon\rightarrow0}{\longrightarrow}p_{0}(\sigma)$. Moreover, at least one of the $a_{0}^{(i)}$ is non-zero. Otherwise we would be able to factor out $\sigma$ in $p_{0}(\sigma,\varepsilon)$, meaning that for small enough (but non zero) $\varepsilon$, $0\in{\rm Sp}(\mathbb{D}(\varepsilon))$ which we know to be false because the system is stable for $\varepsilon>0$.

Furthermore, we know that the splitting behaviour of $\sigma_{\pm}^{N}=0$ is imposed, ensuring that the discriminant of $p_{0}(\sigma,\varepsilon)$ cannot vanish (leading to a multiple root), which yields another condition on the coefficients. Finally, since we are looking at complex roots in general, $p_{0}(\sigma,\varepsilon)$ will always factor into two irreducible and normalized polynomials of degree 1. This ensures that $\forall i\geq1,\;a_{2}^{(i)}=0$ and that $a_{0}^{(1)}=0$. 

This last point is not so straightforward and warrants an explanation. From \cite{an_pert}, the Puiseux series for the perturbed eigenvalues $\sigma_{N\alpha}(\varepsilon)$ can be written as 
\[\sigma_{N\alpha}(\varepsilon)=\sum_{x=1}^{\infty}b_{N\alpha x}\varepsilon^{x/g_{N\alpha}}\quad,\quad\alpha=1,2,\]
where $g_{N\alpha}$ is the degree of the polynomial from which the root $\sigma_{N\alpha}$ is extracted. In our setting $g_{N\alpha}=1$ meaning that the first perturbation to $\sigma_{N\alpha}$ is of order $\varepsilon$. Now, we also know that $\sigma_{N\alpha}$ is obtained by solving the second order equation $p_{0}(\sigma,\varepsilon)=0$. This means that both roots read
\[\sigma_{n\alpha}=o(\varepsilon)+\kappa_{\alpha}\sqrt{\Delta}.\]

We may now write $\Delta$ as
\[\Delta=o(\varepsilon^{2})-4a_{0}^{(1)}\varepsilon,\]
so that, if $a_{0}^{(1)}\neq0$, the dominant term of $\sigma_{N\alpha}$ will be of order $o(\sqrt{\varepsilon})$ which contradicts the previous analysis.

Finally, we can attempting looking for a perturbation resembling 
\[p_{0}(\sigma):=\sigma^{2}\overset{\mathbb{D}(\varepsilon)}{\longrightarrow}p_{0}(\sigma,\varepsilon):=\sigma^{2}+\sigma(a_{1}^{(1)}\varepsilon+a_{1}^{(2)}\varepsilon^{2}+\cdots)+a_{0}^{(2)}\varepsilon^{2}+\cdots.\]

To determine the different terms in this expansion, we  re-use the determinant computation that we carried in the previous section, but keeping now terms up to order $\varepsilon^2$. This yields:
\begin{align*}
    \det{\left(\sigma\mathbf{I}_{2N}-\mathbb{D}\right)}&=
    \begin{vmatrix}
    \sigma\mathbf{I}_{N}-\mathbf{D}_1 & -\mathbf{D}_2\\
    -\mathbf{D}_3 & \sigma\mathbf{I}_{N}-\mathbf{D}_4
    \end{vmatrix}\\
    &\hspace{-0.15cm}\underset{\varepsilon\to0}{\approx}\det\left((\sigma\mathbf{I}_{N}-\mathbf{D}_1)(\sigma\mathbf{I}_{N}-\mathbf{D}_4)-\mathbf{D}_2\mathbf{D}_3\right)\\
    &=\det\left[\underset{\boldsymbol{\Sigma}^{(0)}(\sigma)}{\underbrace{\sigma^2\mathbf{I}_{N}-\sigma\left(\mathbf{D}_1^{(0)}+\mathbf{D}_4^{(0)}\right)+\mathbf{D}_1^{(0)}\mathbf{D}_4^{(0)}-\mathbf{D}_2^{(-1)}\mathbf{D}_3^{(1)}}}\right.\\
    &\qquad\quad+ \varepsilon\left(\underset{\boldsymbol{\Sigma}^{(1)}(\sigma)}{\underbrace{-\sigma(\mathbf{D}_1^{(1)}+\mathbf{D}_4^{(1)})+\mathbf{D}_1^{(0)}\mathbf{D}_4^{(1)}+\mathbf{D}_1^{(1)}\mathbf{D}_4^{(0)}-\mathbf{D}_2^{(-1)}\mathbf{D}_3^{(2)}-\mathbf{D}_2^{(0)}\mathbf{D}_3^{(1)}}}\right)\\
    &\qquad\quad+ \varepsilon^2\left.\left(\underset{\boldsymbol{\Sigma}^{(2)}(\sigma)}{\underbrace{-\sigma(\mathbf{D}_1^{(2)}+\mathbf{D}_4^{(2)})+\mathbf{D}_1^{(0)}\mathbf{D}_4^{(2)}+\mathbf{D}_1^{(1)}\mathbf{D}_4^{(1)}+\mathbf{D}_1^{(0)}\mathbf{D}_4^{(1)}-\mathbf{D}_2^{(-1)}\mathbf{D}_3^{(3)}-\mathbf{D}_2^{(0)}\mathbf{D}_3^{(2)}-\mathbf{D}_2^{(1)}\mathbf{D}_3^{(1)}}}\right)\right]\\
    &\hspace{-0.15cm}\underset{\varepsilon\to0}{\approx}\det\boldsymbol{\Sigma}^{(0)}(\sigma)+\varepsilon\tr{\com{\boldsymbol{\Sigma}^{(0)}}^\top\boldsymbol{\Sigma}^{(1)}(\sigma)}+\varepsilon^2\tr{\com{\boldsymbol{\Sigma}^{(0)}}^\top(\sigma)\boldsymbol{\Sigma}^{(2)}(\sigma)}\\
    &\qquad\qquad+\varepsilon^2\frac{\left(\tr{\com{\boldsymbol{\Sigma}^{(0)}}^\top(\sigma)\boldsymbol{\Sigma}^{(1)}(\sigma)}\right)^2-\tr{\left(\com{\boldsymbol{\Sigma}^{(0)}}^\top(\sigma)\boldsymbol{\Sigma}^{(1)}(\sigma)\right)^2}}{2\det\boldsymbol{\Sigma}^{(0)}(\sigma)}.
\end{align*}

The constant term $\det\boldsymbol{\Sigma}^{(0)}(\sigma)$ is the characteristic polynomial of $\mathbb{D}$ for $\varepsilon=0$ so that $\det\boldsymbol{\Sigma}^{(0)}(\sigma)=\chi(\sigma,0)$. Similarly, it is easy to prove that, for a diagonalizable matrix $\mathbf{A}$ with eigenvalues $\lambda$ and associated eigenvector $\ket{\lambda}$, the matrix $\com{\mathbf{A}}$ can be diagonalized in the same basis and reads
\begin{equation}
    \com{\mathbf{A}}=\sum_\lambda\left(\prod_{\lambda'\neq\lambda}\lambda'\right)\ket{\lambda}\bra{\lambda}.
\end{equation}
Using this lemma, we can write 
\begin{equation}
    \com{\boldsymbol{\Sigma}^{(0)}(\sigma)}=\left(\prod_{\nu\neq N}\left(\sigma-\sigma_+^\nu\right)\left(\sigma-\sigma_-^\nu\right)\right)\ket{e_{N}}\bra{e_{N}}+\sum_{\nu\neq N}\left(\sigma^2\prod_{\mu\neq\nu,N}\left(\sigma-\sigma_+^{\mu}\right)\left(\sigma-\sigma_-^{\mu}\right)\right)\ket{e_\nu}\bra{e_\nu}.
\end{equation}
We now develop each trace term onto the eigenbasis of $\com{\boldsymbol{\Sigma}^{(0)}(\sigma)}$. From now on, we drop the $\sigma$ dependencies of the $\boldsymbol{\Sigma}$ matrices but bear in mind that these matrices are polynomials of order one in $\sigma$. The first trace reads
\begin{align*}
    \tr{\com{\boldsymbol{\Sigma}^{(0)}}^\top\boldsymbol{\Sigma}^{(1)}}=&\left(\prod_{\nu\neq N}\left(\sigma-\sigma_+^\nu\right)\left(\sigma-\sigma_-^\nu\right)\right)\bra{e_{N}}\boldsymbol{\Sigma}^{(1)}\ket{e_{N}}\\&+\sum_{\nu\neq N}\left(\sigma^2\prod_{\mu\neq\nu,N}\left(\sigma-\sigma_+^{\mu}\right)\left(\sigma-\sigma_-^{\mu}\right)\right)\bra{e_\nu}\boldsymbol{\Sigma}^{(1)}\ket{e_\nu}.
\end{align*}
Only the first term is of interest for us and we can use the explicit forms of the blocks of $\mathbb{D}$ to find
\begin{align*}
    \bra{e_{N}}\boldsymbol{\Sigma}^{(1)}\ket{e_{N}}&=\sigma\bra{e_{N}}\left(\mathbf{D}^{(1)}_1+\mathbf{D}^{(1)}_4\right)\ket{e_{N}}\\
    &=-\frac{\sigma}{\rho_{N}}(\alpha+\alpha'+\beta').
\end{align*}

The same computation can be carried out for the second trace term,
\begin{align*}
    \bra{e_{N}}\boldsymbol{\Sigma}^{(2)}\ket{e_{N}}&=\sigma\bra{e_{N}}\left(\mathbf{D}^{(2)}_1+\mathbf{D}^{(2)}_4\right)\ket{e_{N}}+\bra{e_{N}}\mathbf{D}^{(1)}_1\mathbf{D}^{(1)}_4\ket{e_{N}}-\bra{e_{N}}\mathbf{D}^{(0)}_2\mathbf{D}^{(2)}_3\ket{e_{N}}\\
    &=\frac{\sigma}{\rho_{N}^2}(\alpha'+\beta')-\frac{\alpha'\beta'+\alpha\beta}{\rho_{N}^2}+\sigma\kappa,
\end{align*}
with $\kappa=\bra{e_{N}}\mathbf{D}^{(2)}_1\ket{e_{N}}$ which we do not need to compute.

The square trace terms are very complicated, and we only sketch out their computation. The terms that could have entered in the perturbation of $p_0(\sigma)$ cancel out (these are sums of square terms). The terms that are rational fractions of polynomials (and could be pathological since we look for a polynomial perturbation) cancel out as well. The other terms do not enter the perturbation of $p_0(\sigma)$ and are non-pathological. \\

Finally the perturbation of $p_0(\sigma)$ resembles
\[p_{0}(\sigma):=\sigma^{2}\overset{\mathbb{D}(\varepsilon)}{\longrightarrow}p_{0}(\sigma,\varepsilon)\approx\sigma^{2}+\sigma\left(\varepsilon\frac{\alpha+\alpha'+\beta'}{\rho_{N}}-\varepsilon^2\frac{\alpha'+\beta'}{\rho_{N}^2}-\varepsilon^2\kappa\right)+\varepsilon^2\frac{\alpha'\beta'+\alpha\beta}{\rho_{N}^2}.\]

\begin{figure}[b]
	\centering
	\includegraphics[scale=0.8]{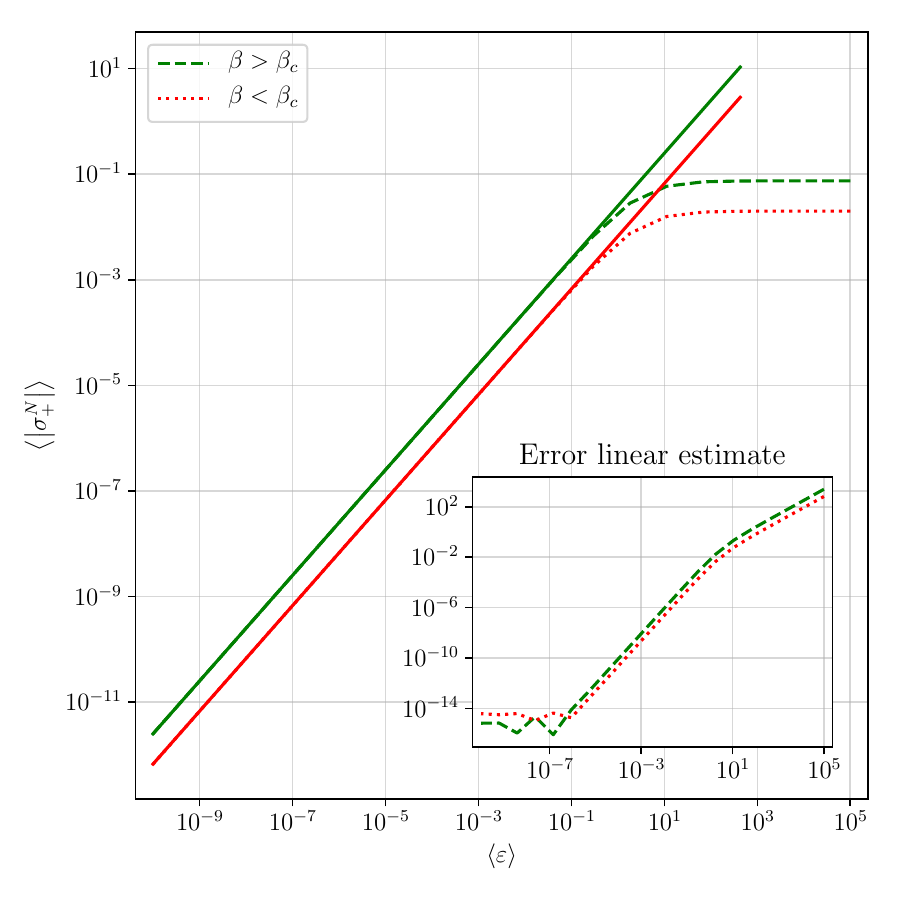}
	\caption{\label{fig:smallest_eig}Plain line: simulated smallest eigenvalue. Dashed: theoretical estimate for $\varepsilon\ll1$. Error plot give the error between the linear estimate from \eqref{eq:eigenvalues_toy_low_z} and the simulated eigenvalue. Simulations are made for an economy on $N=100$ firms on a $3$-regular undirected network with unit weights. We generate 50 such economies and average out the eigenvalue of $\mathbb{D}$ closest to 0 in real part.}
\end{figure}

We now write the discriminant of this polynomial at second order to get
\[\Delta(\varepsilon)=\frac{\varepsilon^2}{\rho_{N}^2}\left((\alpha+\alphap+\betap)^2-4(\alpha\beta+\alphap\betap)\right).\]
We retrieve the same separation as in the large $\varepsilon$ regime. Denoting by $\beta_c=\frac{(\alpha+\alphap+\betap)^2-4\alphap\betap}{4\alpha}$, we have at order one in  $\varepsilon$:
\begin{equation}\label{eq:eigenvalues_toy_low_z}
\sigma_\pm^N\underset{\varepsilon\to0}{\approx}\frac{\varepsilon}{2\rho_{N}}\times
\left\{
\begin{matrix}
-\alpha^{\prime}-\beta^{\prime}-\alpha\pm\sqrt{(\alpha^{\prime}+\beta^{\prime}+\alpha)^{2}-4(\alpha\beta+\alpha^{\prime}\beta^{\prime})} &\text{ if }& \beta<\beta_c \\
 -\alpha^{\prime}-\beta^{\prime}-\alpha\pm i\sqrt{4(\alpha\beta+\alpha^{\prime}\beta^{\prime})-(\alpha^{\prime}+\beta^{\prime}+\alpha)^{2}} &\text{ if }& \beta>\beta_c\\
 -\alpha^{\prime}-\beta^{\prime}-\alpha &\text{if} & \beta=\beta_c
\end{matrix}
\right..
\end{equation}
In the limit $\varepsilon\to0$, $\rho_{N}=z_{\max}$ and we retrieve the equations given in the text. Figure \ref{fig:smallest_eig} shows the adequacy between the theoretical estimate and the actual largest eigenvalue (obtained through numerical simulations of the matrix $\mathbb{D}$) as $\varepsilon\to0$.

\section{\label{ap:block_stability_matrix}Blocks of the stability matrix}

In this section, we give the values of the perturbation terms for the blocks of the stability matrix. We introduce several notations for quantities that simplify in the case of an undirected network with homogeneous productivity factors. Finally, we use the bra (resp. ket) notation to refer to a row (resp. column) vector $\ket{v}$ and we denote by $v_i$ its $i^{th}$ component.

\subsection{Perturbation of \texorpdfstring{$\peqv$ and $\preqv$}{Lg}}

\subsubsection{Prices}
	Equilibrium prices are easily obtained by applying $\boldsymbol{\mathcal{M}}^{-1}$ to the vector $\ket{V}$ yielding
	\begin{align*}
	p_{eq,j}&=\frac{1}{\varepsilon}\ps{\ell_N}{V}r_{N,j}+\sum_{\nu=1}^{N-1}\frac{\ps{l_{\nu}}{V}}{\rho_{N}-\rho_{\nu}}r_{\nu,j}-\varepsilon\sum_{\nu=1}^{N-1}\frac{\ps{l_{\nu}}{V}}{(\rho_{N}-\rho_{\nu})^{2}}r_{\nu,j}+\varepsilon^{2}\sum_{\nu=1}^{N-1}\frac{\ps{l_{\nu}}{V}}{(\rho_{N}-\rho_{\nu})^{3}}r_{\nu,j}\\
	&\quad\quad-\varepsilon^{3}\sum_{\nu=1}^{N-1}\frac{\ps{l_{\nu}}{V}}{(\rho_{N}-\rho_{\nu})^{4}}r_{\nu,j}\\
	&:=\frac{1}{\varepsilon}\pi_{-1}^{l}(V)_{j}+\pi_{0}^{l}(V)_{j}-\varepsilon\pi_{1}^{l}(V)_{j}+\varepsilon^{2}\pi_{2}^{l}(V)_{j}-\varepsilon^{3}\pi_{3}^{l}(V)_{j},
	\tag{1.a}
	\label{eq:1.a}
	\end{align*} 
	where we introduced for $i\geq0$
	\[\pi_{-1}^{l}(V)=\ps{\ell_N}{V}\ket{r_{N}}\;,\;\pi_{i}^{l}(V)=\sum_{\nu=1}^{N-1}\frac{\ps{l_{\nu}}{V}}{(\rho_{N}-\rho_{\nu})^{i+1}}\ket{r_{\nu}}.\]
\subsubsection{Productions}
	Equilibrium productions can be a little trickier to obtain. We first derive three useful identities to simplify calculations. For $s=1,\ldots,n$, we have
	\begin{align*}
	\frac{1}{\varepsilon p_{eq,s}}&=\frac{1}{\pi_{-1}^{l}(V)_{s}}-\varepsilon\frac{\pi_{0}^{l}(V)_{s}}{\left(\pi_{-1}^{l}(V)_{s}\right)^{2}}+\frac{\varepsilon^{2}}{\pi_{-1}^{l}(V)_{s}}\left(\frac{\pi_{1}^{l}(V)_{s}}{\pi_{-1}^{l}(V)_{s}}+\left(\frac{\pi_{0}^{l}(V)_{s}}{\pi_{-1}^{l}(V)_{s}}\right)^{2}\right)\\
	&\quad\quad\quad-\frac{\varepsilon^{3}}{\pi_{-1}^{l}(V)_{s}}\left(\frac{\pi_{2}^{l}(V)_{s}}{\pi_{-1}^{l}(V)_{s}}+2\frac{\pi_{0}^{l}(V)_{s}\pi_{1}^{l}(V)_{s}}{\pi_{-1}^{l}(V)_{s}}+\left(\frac{\pi_{0}^{l}(V)_{s}}{\pi_{-1}^{l}(V)_{s}}\right)^{3}\right)\tag{i}\label{eq:i}\\
	\frac{1}{p_{eq,s}}&=\frac{\varepsilon}{\pi_{-1}^{l}(V)_{s}}-\varepsilon^{2}\frac{\pi_{0}^{l}(V)_{s}}{\left(\pi_{-1}^{l}(V)_{s}\right)^{2}}+\frac{\varepsilon^{3}}{\pi_{-1}^{l}(V)_{s}}\left(\frac{\pi_{1}^{l}(V)_{s}}{\pi_{-1}^{l}(V)_{s}}+\left(\frac{\pi_{0}^{l}(V)_{s}}{\pi_{-1}^{l}(V)_{s}}\right)^{2}\right)\tag{ii}\label{eq:ii}\\
	\frac{\varepsilon}{p_{eq,s}}&=\frac{\varepsilon^{2}}{\pi_{-1}^{l}(V)_{s}}-\varepsilon^{3}\frac{\pi_{0}^{l}(V)_{s}}{\left(\pi_{-1}^{l}(V)_{s}\right)^{2}}\tag{iii}\label{eq:iii}\\
	\frac{\varepsilon^{2}}{peq_{s}}&=\frac{\varepsilon^{3}}{\pi_{-1}^{l}(V)_{s}}\tag{iv}\label{eq:iv}\\
	\frac{\varepsilon^{3}}{peq_{s}}&=o(\varepsilon^{3}).\tag{v}\label{eq:v}\\
	\end{align*}
	These results allow to write the equilibrium productions with $\psi_{eq,i}=\frac{\mu\theta_i}{\peq{i}}$
	\begin{align*}
	\gamma_{eq,j}&=\frac{\mu}{\varepsilon}\ps{r_{N}}{\psi_{eq}}l_{N,j}+\mu\sum_{\nu=1}^{N-1}\frac{\ps{r_{\nu}}{\psi_{eq}}}{\rho_{N}-\rho_{\nu}}l_{\nu,j}-\varepsilon\mu\sum_{\nu=1}^{N-1}\frac{\ps{r_{\nu}}{\psi_{eq}}}{(\rho_{N}-\rho_{\nu})^{2}}l_{\nu,j}+\varepsilon^{2}\mu\sum_{\nu=1}^{N-1}\frac{\ps{r_{\nu}}{\psi_{eq}}}{(\rho_{N}-\rho_{\nu})^{3}}l_{\nu,j}\\
	&\quad\quad\quad-\varepsilon^{3}\mu\sum_{\nu=1}^{N-1}\frac{\ps{r_{\nu}}{\psi_{eq}}}{(\rho_{N}-\rho_{\nu})^{4}}l_{\nu,j}\\
	&=\mu l_{N,j}\sum_{s=1}^{n}\frac{r_{n,s}\theta_{s}}{\pi_{-1}^{l}(V)_{s}}+\varepsilon\mu\sum_{s=1}^{n}\left\{\sum_{\nu=1}^{N-1}\frac{l_{\nu,j}r_{\nu,s}\theta_{s}}{(\rho_{N}-\rho_{\nu})\pi_{-1}^{l}(V)_{s}}-\frac{l_{N,j}r_{n,s}\theta_{s}\pi_{0}^{l}(V)_{s}}{\pi_{-1}^{l}(V)_{s}^{2}}\right\}\\
	&\quad+\varepsilon^{2}\mu\sum_{s=1}^{n}\left\{\frac{\pi_{1}^{l}(V)_{s}}{\pi_{-1}^{l}(V)_{s}}+\left(\frac{\pi_{0}^{l}(V)_{s}}{\pi_{-1}^{l}(V)_{s}}\right)^{2}-\sum_{\nu=1}^{N-1}\frac{l_{\nu,j}r_{\nu,s}\theta_{s}}{(\rho_{N}-\rho_{\nu})\pi_{-1}^{l}(V)_{s}}\left(\frac{1}{\rho_{N}-\rho_{\nu}}+\frac{\pi_{0}^{l}(V)_{s}}{\pi_{-1}^{l}(V)_{s}}\right)\right\}\\
	&\quad+\varepsilon^{3}\mu\sum_{s=1}^{n}\left\{\sum_{\nu=1}^{N-1}\frac{l_{\nu,j}r_{\nu,s}\theta_{s}}{(\rho_{N}-\rho_{\nu})\pi_{-1}^{l}(V)_{s}}\left(\frac{\pi_{1}^{l}(V)_{s}}{\pi_{-1}^{l}(V)_{s}}+\left(\frac{\pi_{0}^{l}(V)_{s}}{\pi_{-1}^{l}(V)_{s}}\right)^{2}+\frac{\pi_{0}^{l}(V)_{s}}{(\rho_{N}-\rho_{\nu})\pi_{-1}^{l}(V)_{s}}\right.\right.\\
	&\left.\left.\qquad\qquad\qquad\qquad\qquad\qquad\qquad\qquad\qquad+\left.\frac{1}{(\rho_{N}-\rho_{\nu})^{3}\pi_{-1}^{l}(V)_{s}}\right.^{}\right)^{}\right\}\\
	&:=\mu f_{0,j}+\varepsilon\mu f_{1,j}\tag{1.b}+\varepsilon^{2}\mu f_{2,j}+\varepsilon^{3}\mu f_{3,j}.\label{eq:1.b}
	\end{align*}

\subsection{Stability blocks}

The next step is to perturb the stability matrix itself. This yields no particular difficulty but calculations are a bit long so that we only give the results for the different blocks. We denote by $\tau g_k$ the coefficients of the expansion of $z\gamma$ where $\tau_i=\rho_{N}e_{N,i}/\ps{e_{N}}{V}$ for an undirected network. We have

\begin{align*}
(D^{(0)}_1)_{ij}&=-\alphap\frac{1}{\rho_{N}}M_{ij}\\
(D^{(1)}_1)_{ij}&=-\alpha\mu\frac{\theta_{i}}{\tau_{i}\pi_{-1}^{l}(V)_{i}}\delta_{ij}+\frac{\alpha'}{\rho_{N}^2}M_{ij}-\frac{\alphap}{\rho_{N}}\delta_{ij}\\
(D^{(2)}_1)_{ij}&=+\alpha\mu\frac{\theta_{i}}{\tau_{i}\pi_{-1}^{l}(V)_{i}}\left(g_{1,i}+\frac{\pi_{0}^{l}(V)_{i}}{\pi_{-1}^{l}(V)_{i}}\right)\delta_{ij}-\frac{\alpha'}{\rho_{N}^3}M_{ij}+\frac{\alphap}{\rho_{N}^2}\delta_{ij}\\
(D^{(3)}_1)_{ij}&=-\alpha\mu\varepsilon\frac{\theta_{i}}{\tau_{i}\pi_{-1}^{l}(V)_{i}}\left(g_{1,i}^{2}-g_{2,i}+\frac{\pi_{0}^{l}(V)_{i}}{\pi_{-1}^{l}(V)_{i}}g_{1,i}+\frac{\pi_{1}^{l}(V)_{i}}{\pi_{-1}^{l}(V)_{i}}+\left(\frac{\pi_{0}^{l}(V)_{i}}{\pi_{-1}^{l}(V)_{i}}\right)^{2}\right)\delta_{ij}+\frac{\alpha'}{\rho_{N}^4}M_{ij}-\frac{\alphap}{\rho_{N}^3}\delta_{ij},
\intertext{  }
(D^{(-1)}_2)_{ij}&=-\alpha\frac{\pi_{-1}^{l}(V)_{i}}{\tau_{i}}M_{ji}\\
(D^{(0)}_2)_{ij}&=\frac{\alpha}{\tau_{i}}\left[M_{ji}\left(\pi_{-1}^{l}(V)_{i}g_{1,i}-\pi_{0}^{l}(V)_{i}\right)-\pi_{-1}^{l}(V)_{i}\delta_{ij}\right]\\
(D^{(1)}_2)_{ij}&=\frac{\alpha}{\tau_{i}}\left[M_{ji}\left(\pi_{-1}^{l}(V)_{i}g_{1,i}^{2}+2\pi_{-1}^{l}(V)_{i}g_{1,i}g_{2,i}+\pi_{0}^{l}(V)_{i}g_{1,i}-\pi_{-1}^{l}(V)_{i}g_{2,i}-\pi_{1}^{l}(V)_{i}\right)\right.\\
&\quad+\left.\left(\pi_{-1}^{l}(V)_{i}g_{1,i}-\pi_{0}^{l}(V)_{i}\right)\delta_{ij}\right]\\
(D^{(2)}_2)_{ij}&=\frac{\alpha}{\tau_{i}}\left[M_{ji}\left(\pi_{1}^{l}(V)_{i}g_{1,i}+\pi_{2}^{l}(V)_{i}-\pi_{-1}^{l}(V)_{i}\left(g_{3,i}+g_{1,i}^{3}\right)+\pi_{0}^{l}(V)_{i}\left(g_{1,i}^{2}-g_{2,i}\right)\right)\right.\\
&\quad\quad\left.+\left(\pi_{-1}^{l}(V)_{i}\left(g_{1,i}^{2}-g_{2,i}\right)-\pi_{0}^{l}(V)_{i}g_{1,i}-\pi_{1}^{l}(V)_{i}\right)\delta_{ij}\right],
\intertext{  }
(D^{(1)}_3)_{ij}&=\frac{\mu\beta f_{0,i}}{\rho_{N}\pi_{-1}^{l}(V)_{i}}M_{ij}\\
(D^{(2)}_3)_{ij}&=\frac{\mu\beta f_{0,i}}{\rho_{N}}\left[\left(\frac{g_{1,i}}{\pi_{-1}^{l}(V)_{i}}-\frac{\pi_{0}^{l}(V)_{i}}{(\pi_{-1}^{l}(V)_{i})^{2}}+\frac{g_{1,i}}{\rho_{N}}-\frac{\pi_{0}^{l}(V)_{i}}{\rho_{N}\pi_{-1}^{l}(V)_{i}}\right)M_{ij}+\frac{1}{\pi_{-1}^{l}(V)_{i}}\delta_{ij}\right]-\betap\frac{\mu\theta_i}{\rho_{N}\pi_{-1}^{l}(V)_{i}^2}\delta_{ij}\\
(D^{(3)}_3)_{ij}&=\frac{\mu\beta f_{0,i}}{\rho_{N}}\left[M_{ij}\left(\frac{\pi_{1}^{l}(V)_{i}}{(\pi_{-1}^{l}(V)_{i})^{2}}+\frac{(\pi_{0}^{l}(V)_{i})^{2}}{(\pi_{-1}^{l}(V)_{i})^{3}}-\frac{\pi_{0}^{l}(V)_{i}}{(\pi_{-1}^{l}(V)_{i})^{2}}g_{1,i}+\frac{g_{1,i}}{\rho_{N}\pi_{-1}^{l}(V)_{i}}-\frac{\pi_{0}^{l}(V)_{i}}{\rho_{N}(\pi_{-1}^{l}(V)_{i})^{2}}\right)\right.\\
&\quad+\left.\left(\frac{g_{1,i}}{\pi_{-1}^{l}(V)_{i}}-\frac{\pi_{0}^{l}(V)_{i}}{(\pi_{-1}^{l}(V)_{i})^{2}}+\frac{1}{\rho_{N}}\left(g_{1,i}-\frac{\pi_{0}^{l}(V)_{i}}{\pi_{-1}^{l}(V)_{i}}\right)\right)\delta_{ij}\right]+\betap\frac{\mu\theta_i}{\rho_{N}\pi_{-1}^{l}(V)_{i}^2}\left(\frac{2\pi_{0}^{l}(V)_{i}}{\pi_{-1}^{l}(V)_{i}}+\frac{1}{\rho_{N}}\right)\delta_{ij},
\intertext{  }
(D^{(0)}_4)_{ij}&=-\betap\frac{1}{\rho_{N}}M_{ij}\\
(D^{(1)}_4)_{ij}&=\frac{\betap}{\rho_{N}^2}M_{ij}-\frac{\betap}{\rho_{N}}\delta_{ij}\\
(D^{(2)}_4)_{ij}&=-\frac{\betap}{\rho_{N}^3}M_{ij}+\frac{\betap}{\rho_{N}^2}\delta_{ij}\\
(D^{(3)}_4)_{ij}&=\frac{\betap}{\rho_{N}^4}M_{ij}-\frac{\betap}{\rho_{N}^3}\delta_{ij}.
\end{align*}

\section{\label{ap:fluctuation_sqrt_eps}Critical volatility of prices and outputs with fluctuations}

\subsection{General computation for marginally stable linear stochastic systems}

In this section, we consider a general evolution of a vector $\mathbf{U}(t)$ given by the linear stochastic equation
\begin{equation}\label{eq:appendix_dyn_fluctuations2}
    \frac{\mathrm{\text{d}}\mathbf{U}(t)}{\mathrm{\text{d}}t} = \mathbf{\mathbb{D}} \mathbf{U}(t) + \boldsymbol{\xi}(t),
\end{equation}
where $\mathbb{D}$ is a real $N\times N$ matrix and $\xi(t)$ is a Gaussian correlated noise such that 
\begin{align}
\left\langle\xi_i(t)\right\rangle&=0\\
\left\langle\xi_i(t)\xi_j(s)\right\rangle&=2\sigma^2\delta_{ij}G\left(|t-s|\right).
\end{align}
We assume the dynamical matrix $\mathbb{D}$ to be diagonalizable with real eigenvalues\footnote{The case with complex eigenvalues leads to the same conclusions. One must only take into account the fact that, since $\mathbb{D}$ is real, eigenvalues and eigenvectors will be conjugated so that their are two eigenvalues that are smallest in real parts. We make the same ordering of eigenvalues replacing the $\lambda$'s by their real parts.} such that 
\[\lambda_1\leq\lambda_2\leq\cdots\leq\lambda_{N-1}<\lambda_{N}:=-\varepsilon<0.\]
Negative eigenvalues means that the system is stable i.e $\left\langle\|\mathbf{U}(t)\|\right\rangle\to0$ as $t\to\infty$ for any initial condition. We assume that $\varepsilon\to0$ and show that the volatility of $\mathbf{U}(t)$ increases as $\varepsilon^{-1/2}$. We introduce the eigenvectors $\mathbf{e}_\nu$ associated to $\lambda_\nu$ and we express $\mathbf{U}$ into the diagonal basis
\begin{equation}
    \mathbf{U}(t)=\sum_{\nu=1}^{N}u_{\nu}\mathbf{e}_\nu.
\end{equation}
Injecting this expression into \eqref{eq:appendix_dyn_fluctuations2}, we get and evolution equation for the components of $\mathbf{U}(t)$ in the diagonal basis
\begin{equation}
    \frac{d}{dt}u_\nu=\lambda_\nu u_\nu+\boldsymbol{\xi}(t)\cdot \mathbf{e}_\nu.
\end{equation}
We can give an explicit solution for these components 
\begin{equation}
    u_\nu(t)=e^{\lambda_\nu t}\left[u_\nu(0)+\int_0^t ds\,e^{-\lambda_\nu s}\boldsymbol{\xi}(t)\cdot \mathbf{e}_\nu\right],
\end{equation}
and focus on $u_{N}$ since this is the component which yields the $\varepsilon^{-1/2}$-volatility. To do so, we compute the average value of $u_{N}(t)^2-\left\langle u_{N}(t)\right\rangle^2$
\begin{align*}
    \left\langle u_{N}(t)^2-\left\langle u_{N}(t)\right\rangle^2\right\rangle&=e^{-2\varepsilon t}\left\langle \left[u_{N}(0)+\int_0^t ds\,e^{\varepsilon s}\boldsymbol{\xi}(s)\cdot \mathbf{e}_{N}\right]^2\right\rangle-u_{N}(0)^2e^{-\varepsilon t}\\
    &=e^{-2\varepsilon t}\left[u_{N}(0)^2+2 u_{N}(0)\int_0^t ds\,e^{\varepsilon s}\left\langle \boldsymbol{\xi}(s)\cdot \mathbf{e}_{N}\right\rangle+\int dsds'\,e^{\varepsilon (s+s')}\left\langle(\boldsymbol{\xi}(s)\cdot \mathbf{e}_{N})( \boldsymbol{\xi}(s')\cdot \mathbf{e}_{N})\right\rangle\right]-u_{N}(0)^2e^{-\varepsilon t}\\
    &=e^{-\varepsilon t}\sum_{j,k}e_{N,j} e_{N,k}\int dsds'\,e^{\varepsilon (s+s')}\left\langle\xi_j(s)\xi_j(s') \right\rangle\\
    &=2\sigma^2\|\mathbf{e}_{N}\|^2e^{-\varepsilon t}\int dsds'\,e^{\varepsilon (s+s')}G(|s'-s|),\\
\intertext{we substitute $\tau=s'-s$ in the $s$ integral to get}
    &=2\sigma^2\|\mathbf{e}_{N}\|^2e^{-\varepsilon t}\int_0^t ds'\,e^{2\varepsilon s'}\int_0^{s'-t} d\tau e^{-\varepsilon\tau}G(\tau).\\
\intertext{Using the quick decay of the exponential term in the $\tau$ integral, we can expend the integration domain...}
&\approx2\sigma^2\|\mathbf{e}_{N}\|^2e^{-\varepsilon t}\int_0^t ds'\,e^{2\varepsilon s'}\int_0^{\infty} d\tau e^{-\varepsilon\tau}G(\tau),\\
\intertext{... and perform the integration over $s'$ with an approximately vanishing exponential remainder}
&\approx\frac{\sigma^2\|\mathbf{e}_{N}\|^2}{\varepsilon}\int_0^{\infty} d\tau e^{-\varepsilon\tau}G(\tau).\\
\end{align*}
Denoting be $\tau_\xi$ the typical correlation time of $G$, we see that
\begin{itemize}
\item if $\varepsilon\tau_\xi\ll1$ (meaning that $G$ correlates on short time-scales) then $G(\tau)\sim \delta(0)$ such that
    \[\int_0^{\infty} d\tau e^{-\varepsilon\tau}G(\tau)\approx 1,\]
    \item if $\varepsilon\tau_\xi\gg1$ (meaning that $G$ correlates on long time-scales) then $G(\tau)\sim G(0)$ on the decay time of the exponential such that
    \[\int_0^{\infty} d\tau e^{-\varepsilon\tau}G(\tau)\approx \frac{G(0)}{\varepsilon}.\]
\end{itemize}
Finally, the volatility of $\mathbf{U}(t)$ behaves as 
\begin{equation}\label{eq:volatility_U}
\sqrt{\left\langle u_{N}(t)^2-\left\langle u_{N}(t)\right\rangle^2\right\rangle}\propto
\left\{
\begin{matrix}
\varepsilon^{-1/2} &\text{ if }& \varepsilon\tau_\xi\ll1 \\
\varepsilon^{-1} &\text{ if }& \varepsilon\tau_\xi\gg1
\end{matrix}
\right..
\end{equation}

Note also that this result generalizes to discrete time processes (which is of interest in the case of the general ABM that we present)
\begin{equation}
    \mathbf{U}_{t+1}=\mathbb{D}\mathbf{U}_t+\boldsymbol{\xi}_t.
\end{equation}
The marginal stability condition can be written as $\lambda_{N}=1-\varepsilon$~\footnote{Or more generally for complex eigenvalues $\lambda_N=r_Ne^{i\theta_N}$ with $r_N=1-\varepsilon$.} with $\varepsilon\to0$. We can carry out the same kind of computation and derive the same result depending on the  behavior of the quantity $\sum_{\tau\geq0}(1-\varepsilon)^\tau G(\tau)$.

\subsection{Computation of the volatility induced by gaussian shocks on productivity factors}

If we consider shocks on productivity factors $z_i(t)=z_i+\xi_i(t)$ with $\xi(t)$ given as before, we can linearize the dynamics of the naive model in both small deviations from equilibrium and small shocks. The stochastic equation that we retrieve reads
\begin{equation}\label{eq:produtivity_shocks}
    \frac{\mathrm{\text{d}}\mathbf{U}(t)}{\mathrm{\text{d}}t} = \mathbf{\mathbb{D}} \mathbf{U}(t) + \boldsymbol{\Xi}(t),
\end{equation}
with a noise $\boldsymbol{\Xi}$ of the form
\begin{equation}
\boldsymbol{\Xi}(t)=\begin{pmatrix}
-\frac{\alpha+\alphap}{z_i}\peqv\circ\boldsymbol{\xi}(t)\\
\frac{\beta-\betap}{z_i}\preqv\circ\boldsymbol{\xi}(t)
\end{pmatrix}\underset{\varepsilon\to0}{\sim}\begin{pmatrix}
-\frac{\alpha+\alphap}{\varepsilon\rho_{N}}(\boldsymbol{\ell}_N\cdot\mathbf{V})\mathbf{r}_{N}\circ\boldsymbol{\xi}(t)\\
\frac{\beta-\betap}{(\boldsymbol{\ell}_N\cdot\mathbf{V})\rho_{N}}\mathbf{l}_{N}\circ\boldsymbol{\xi}(t)
\end{pmatrix},
    \label{eq:induced_{N}oise}
\end{equation}
with notations from \ref{ap:block_stability_matrix}. The correlations of this noise are slightly more complicated than before
\begin{equation}
\left\langle\Xi_i(t)\Xi_j(s)\right\rangle=\sigma^2G(|t-s|)\times
\left\{
\begin{matrix}
 \delta_{ij} \left(\frac{(\alpha+\alphap)(\boldsymbol{\ell}_N\cdot\mathbf{V})}{\rho_{N}}\right)^2r_{N,i}r_{N,j}\varepsilon^{-2}& \text{ if }&  i,j\leq n\\
 \delta_{ij} \left(\frac{\beta-\betap}{(\boldsymbol{\ell}_N\cdot\mathbf{V})\rho_{N}}\right)^2l_{N,i}l_{N,j}& \text{ if }&  i,j>n\\
 -\delta_{i,j-n} \frac{(\beta-\betap)(\alpha+\alphap)}{\rho_{N}^2}r_{N,i}l_{N,j}\varepsilon^{-1}& \text{ if }& i\leq n,\;j>n\\
 -\delta_{i-n,j} \frac{(\beta-\betap)(\alpha+\alphap)}{\rho_{N}^2}l_{N,i}r_{N,j}\varepsilon^{-1}& \text{ if }& i> n,\;j\leq n
\end{matrix}
\right..
\end{equation}

The dynamical matrix of the naive model with an undirected network two eigenvalues yields $\sigma_{N}^\pm=k^\pm\varepsilon\to 0$ with associated eigenvectors $\boldsymbol{\Sigma}_{N}^\pm=(\mathbf{e}_{N},\boldsymbol{\nu}^\pm\varepsilon)^\top$ for undirected networks. We assume $\beta<\beta_c$ so that the marginal eigenvalues are real as well as their eigenvectors.
It follows that, at leading order in $\varepsilon$, the volatility of the marginal components of $\mathbf{U}(t)$ behaves as $\varepsilon^{-3/2}$. Indeed
\begin{align*}
    \left\langle u_{N}^\pm(t)^2-\left\langle u_{N}^\pm(t)\right\rangle^2\right\rangle&=\frac{\sigma^2}{\left(k^\pm\right)^2\varepsilon^3}\left(\frac{(\alpha+\alphap)(\boldsymbol{e}_N\cdot\mathbf{V})}{\rho_{N}}\right)^2\mathscr{H}\left(\mathbf{e}_{N}\right)\int_0^{\infty} d\tau e^{-\varepsilon\tau}G\left(\frac{\tau}{\nu^\pm}\right),
\end{align*}
where $\mathscr{H}$ represents the inverse participation ratio. To retrieve the volatility as $\varepsilon^{-1/2}$ we may rescale $\delta p_i(t)$ (resp. $\delta\gamma_i(t)$) by $p_{eq,i}$ (resp. $\gamma_{eq,i}$). Denoting by $\mathbf{w}_i$ the $i^{th}$ canonical vector of $\R^{2N}$, we have
\begin{align*}
    {\rm Var}\left(\frac{\delta p_i(t)}{p_{eq,i}}\right)&=p_{eq,i}^{-2}{\rm Var}\left(\sum_{\substack{k=1\\\tau=\pm}}^N u_k^\tau(t)(\boldsymbol{\Sigma}_k^\pm\cdot\mathbf{w}_i)\right)\\
    &\hspace{-0.15cm}\underset{\substack{\varepsilon\to0\\\varepsilon t\ll1}}{\approx}\frac{\varepsilon^2}{e_{N,i}^2(\mathbf{e}_N\cdot\mathbf{V})^2}{\rm Var}\left(u_{N}^+(t)e_{N,i}+u_{N}^-(t)e_{N,i}\right)\\
    &=\frac{\varepsilon^2}{(\mathbf{e}_N\cdot\mathbf{V})^2}\left[{\rm Var}\left(u_{N}^+(t)\right)+{\rm Var}\left(u_{N}^-(t)\right)+2{\rm Cov}\left(u_{N}^+(t),u_{N}^-(t)\right)\right]\\
    &\propto\frac{1}{\varepsilon};\\
    {\rm Var}\left(\frac{\delta \gamma_i(t)}{\gamma_{eq,i}}\right)&=\gamma_{eq,i}^{-2}{\rm Var}\left(\sum_{\substack{k=1\\\tau=\pm}}^Nu_k^\tau(t)(\boldsymbol{\Sigma}_k^\pm\cdot\mathbf{w}_{i+N})\right)\\
    &\hspace{-0.15cm}\underset{\substack{\varepsilon\to0\\\varepsilon t\ll1}}{\approx}\frac{(\mathbf{e}_N\cdot\mathbf{V})^2}{e_{N,i}^2}{\rm Var}\left(u_{N}^+(t)\nu^+\varepsilon e_{N,i}+u_{N}^-(t)\nu^-\varepsilon e_{N,i}\right)\\
    &=(\mathbf{e}_N\cdot\mathbf{V})^2\varepsilon^2\left[(\nu^+)^2{\rm Var}\left(u_{N}^+(t)\right)+(\nu^-)^2{\rm Var}\left(u_{N}^-(t)\right)+2\nu^+\nu^-{\rm Cov}\left(u_{N}^+(t),u_{N}^-(t)\right)\right]\\
    &\propto\frac{1}{\varepsilon}.\\
\end{align*}

\section{Real World Networks}\label{ap:real_net}

\begin{figure}
    \centering
     \begin{subfigure}[b]{0.32\textwidth}
         \centering
         \includegraphics[width=\textwidth]{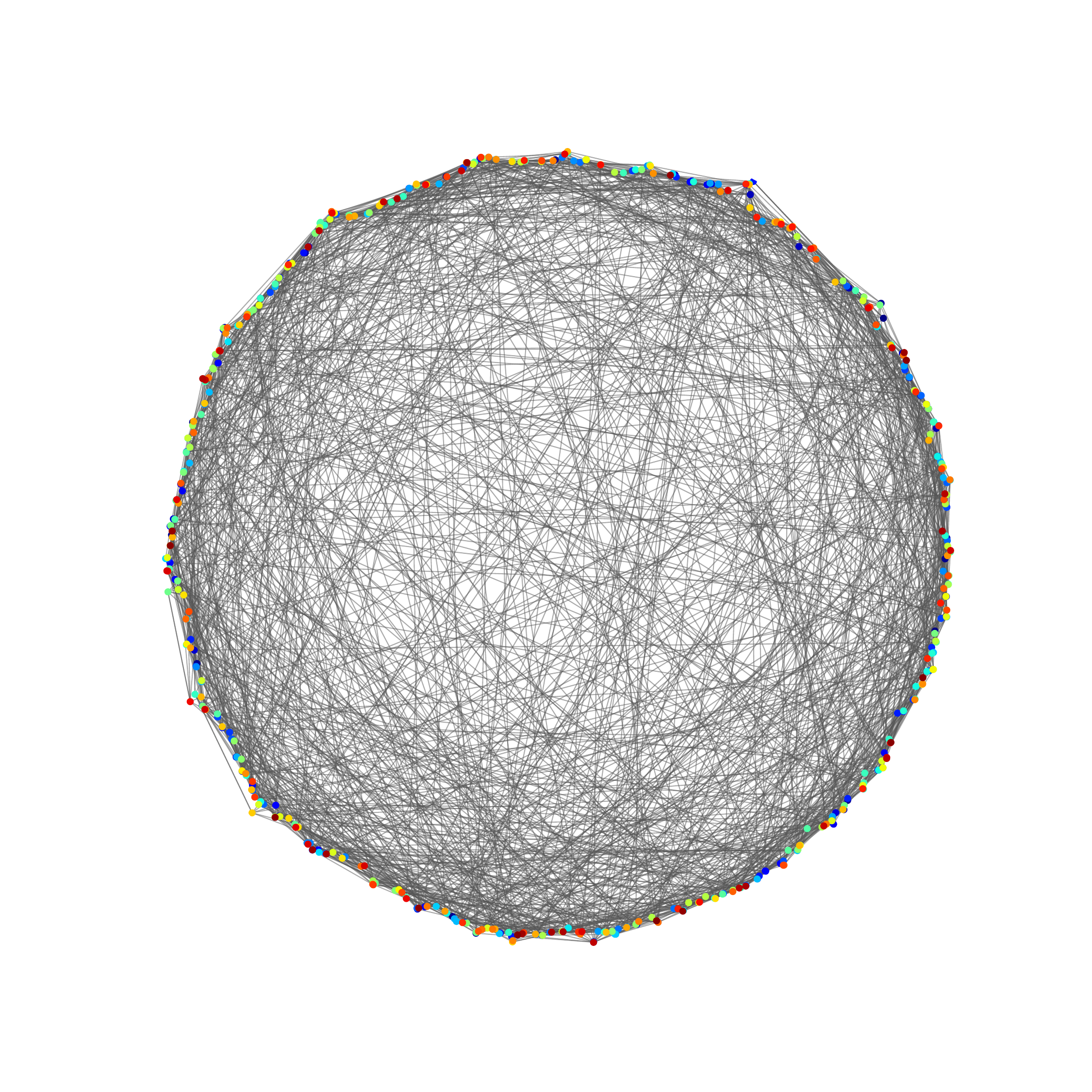}
         \caption{$d$-regular random network}
         \label{fig:rrg_net}
     \end{subfigure}
     \hfill
     \begin{subfigure}[b]{0.32\textwidth}
         \centering
         \includegraphics[width=\textwidth]{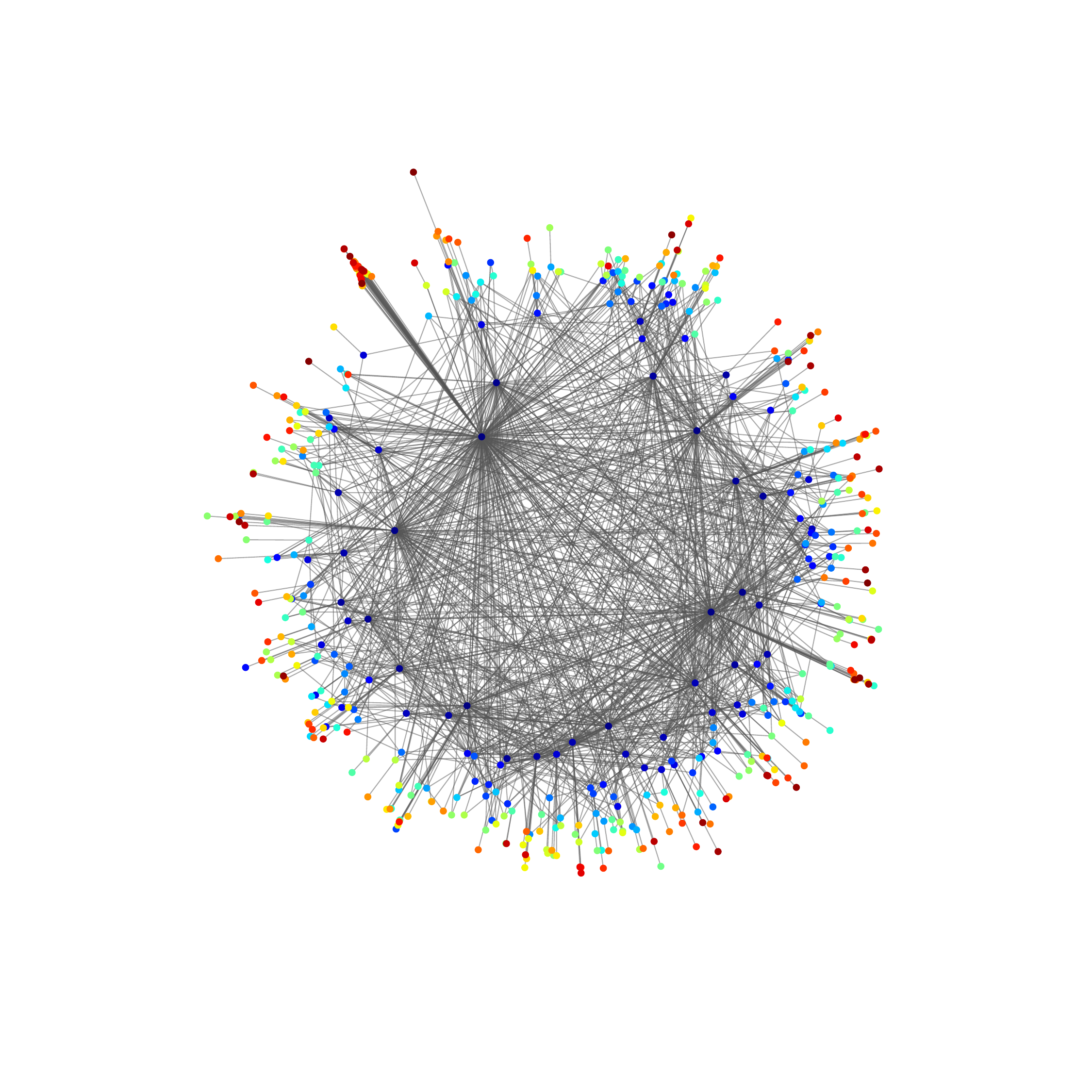}
         \caption{Scale-free network}
         \label{fig:scale_free_net}
     \end{subfigure}
     \hfill
     \begin{subfigure}[b]{0.32\textwidth}
         \centering
         \includegraphics[width=\textwidth]{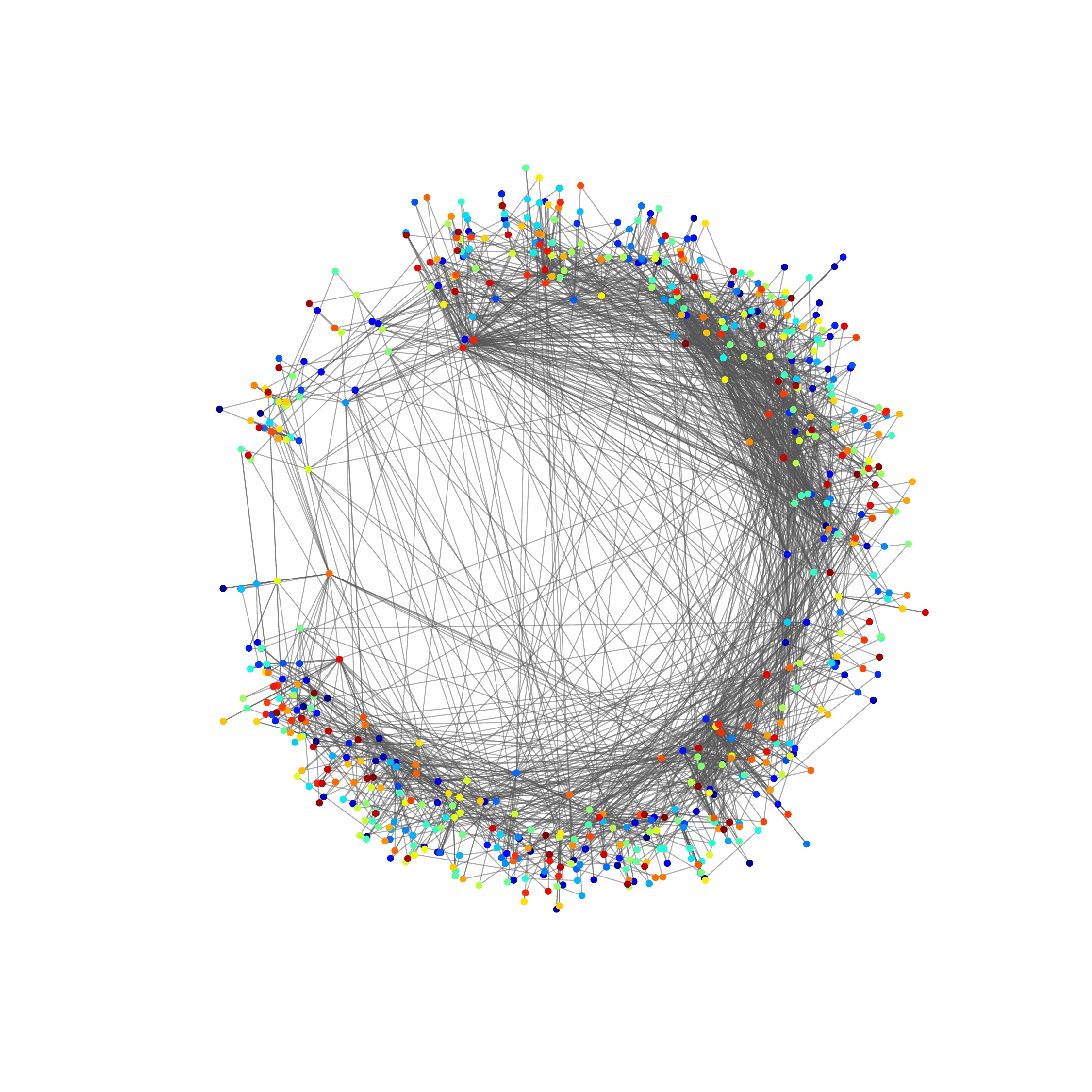}
         \caption{Real IO network}
         \label{fig:factset_net}
     \end{subfigure}
        \caption{Different types of networks. The original networks are directed but we omitted links' directions for clarity. Left: Random $4$-regular network on $n=500$ firms. The number of in and out links are fixed to $d=4$. Middle: Scale free network  on $n=500$ firms. The number of in and out links follows Pareto distributions with parameter $\mu_{in}=1.29$ and $\mu_{out}=1.25$ (from \cite{japIO}). Right: Input-output network over  inferred from FACTSET dataset. Links correspond to an existing supplier-buyer relationship between 2012 and 2015. These networks are represented with an embedding described in \cite{garciaperez2019mercator}.}
        \label{fig:networks}
\end{figure}

As mentioned at the beginning in \ref{sec:numerical_study}, random regular networks are a crude idealization of real interaction networks. Real networks have been studied extensively (see for instance \cite{atalay2011, japIO, NBERw24556}) and display well identified topological features such as power law distributed in and out vertex degrees. Fig.\ref{fig:networks} illustrates the topological discrepancies between regular and real world networks, and highlights similarities with scale-free networks. 

To build the network of \ref{fig:factset_net}, we use the FactSet Supply Chain Relationships database to build a supply chain network.
The FactSet dataset (\cite{factset}) contains a list of relational data between firms, stating if firms A and B have a client/supplier relation, if they are in competition or if they have a joint venture. It is built by collecting information from primary public sources such as SEC 10-K annual filings, investor presentations and press releases, and covers about 23, 000 publicly traded companies with over 325, 000 relationships. Since the relationships are inferred from data released to the public, we cannot be sure that it is an exhaustive database of all the relationships between firms, but the subset of relationships deemed important by the firm themselves. Such links between firms have a finite duration in time and have thus a beginning and end date. For our study, we have chosen the set of client/supplier relationships between the years 2012 and 2015. This allows us to build a graph G where a link $i\rightarrow j$ exists whenever i is reported to be a supplier of j or when j is reported to be a client of i. Furthermore, since this graph is not fully connected, we extracted its largest strongly connected component.

\begin{figure}
    \centering
    \begin{subfigure}[b]{\textwidth}
         \centering
         \includegraphics{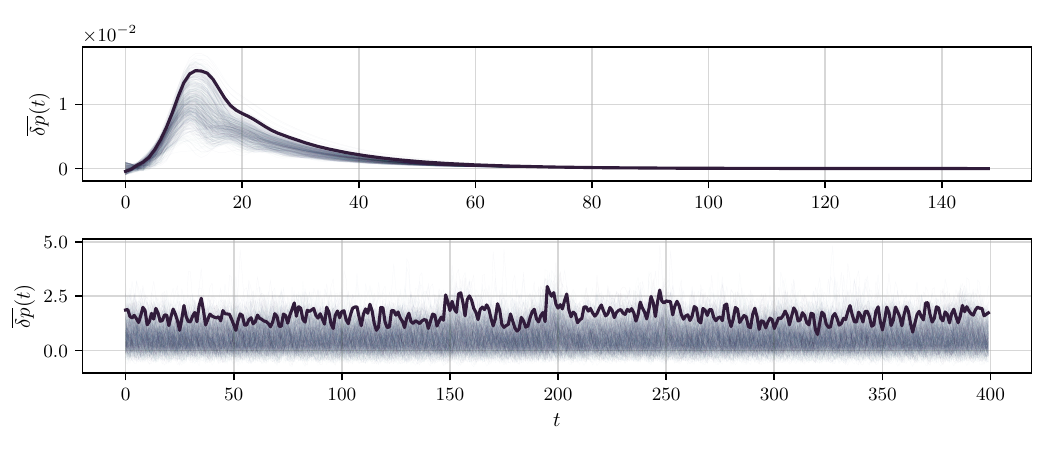}
         \caption{$\varepsilon=10$}
         %\label{fig:three sin x}
     \end{subfigure}
     \hfill
     \begin{subfigure}[b]{\textwidth}
         \centering
         \includegraphics{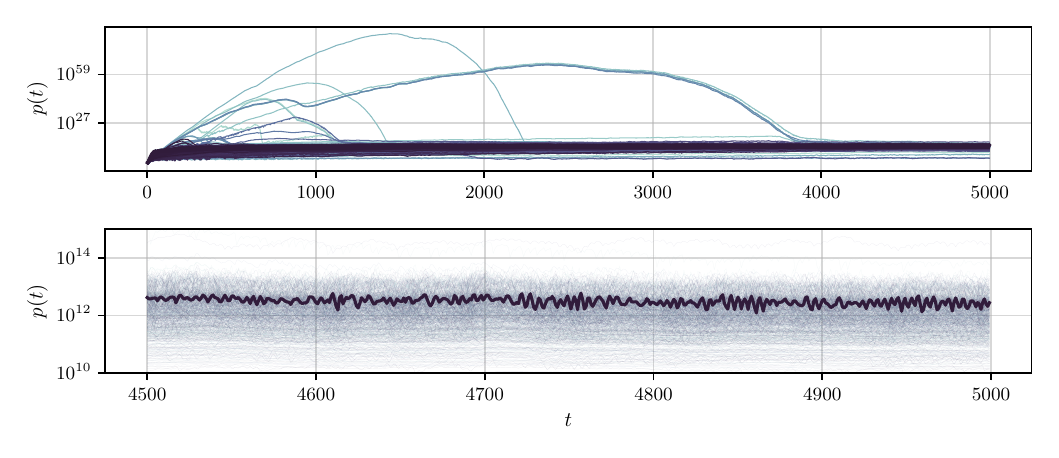}
         \caption{$\varepsilon=-3$}
         %\label{fig:three sin x}
     \end{subfigure}
    \caption{Example of dynamics on the network \ref{fig:factset_net}. (a) Dynamics for $\varepsilon=10$. Top: Relaxation towards equilibrium. Bottom: Oscillatory patterns. Oscillations are quite disordered relatively to random regular networks. (b) Dynamics for $\varepsilon=-3$. Top: Oscillations with quenched explosions for some firms. Bottom: Zoom on the last 500 time steps of the previous time-series. The dark lines correspond to one randomly picked firm.}
    \label{fig:real-world-network-dyn}
\end{figure}

Even though phase diagrams are not changed qualitatively if one changes the network, the features of the dynamics within one phase depends on the structure of interactions. As an illustration, we ran our model on the network represented on Fig. \ref{fig:factset_net} for $\varepsilon=10$ and $\varepsilon=-3$. Results are reported on Fig. \ref{fig:real-world-network-dyn}. While equilibria (whether competitive or deflationary) are reached in a somewhat similar manner as on a regular network, oscillatory patterns are much more disordered due to the inhomogeneity of in and out degrees.

{Finally, another effect closely related to network topology is worth mentioning. In the case of random regular networks, firms are always supplied by at least one firm. However it is possible for a firm to use labour as sole input. Upon simulating the dynamics on scale free networks with labour-supplied firms, we found that whenever deflationary equilibria occured, only a fraction of firms survived while the others saw their prices blow up exponentially and production plummet. Surviving firms are the ones for which going up the supplier network leads only to labor-supplied firms.}

\section{\label{ap:pseudo_code}Code for the simulation}

\subsection{Objects}

The simulation uses an object-oriented approach. Each object has attributes (parameters of the model, or other quantities that we can infer from the parameters) along with methods that carry out more complicated tasks. There are four objects in our simulation. The first two are the \texttt{firms} and \texttt{household} classes, and correspond to the smallest entities in our model (the agents). The \texttt{economy} class carries all of the static information -- essentially the different parameters describing the interactions between the agents -- of the model, along with instances of the \texttt{firms} and \texttt{household} classes. Finally, the \texttt{dynamics} class handles the evolution of the model by storing the time series of prices, productions and so on, along with the different methods that allow the model to move forward in time; the \texttt{dynamics} class contains an instance of the \texttt{economy} class that is used for the simulation. In our framework, classes need one another to function properly. For instance, firms need to know the input-output network to compute their optimal quantities as in \ref{eq:optimal_quantities}. As a consequence, some methods take instances of classes in their arguments, as is the case in e.g. the $\texttt{compute\_optimal\_quantities}$ method of the firm \texttt{firms} class.  We detail this last method as an example in Procedure \ref{alg:class_firms}.

\begin{algorithm}[H]
\begin{algorithmic}
\Class{Firms}
\Atr{}
\State $z$: \textsc{$1d$-array}\Comment{Productivity factors}
\State $\alpha$: \textsc{$1d$-array}\Comment{Log elasticity factors of prices to surplus}
\State $\alphap$: \textsc{$1d$-array}\Comment{Log elasticity factors of prices to profits}
\State $\beta$: \textsc{$1d$-array}\Comment{Log elasticity factors of productions to profits}
\State $\betap$: \textsc{$1d$-array}\Comment{Log elasticity factors of productions to surplus}
\State $\omega$: \textsc{float}\Comment{Log elasticity factors of wages to labor-market tensions}
\EndAtr
\Methods{}
\State \textsc{$1d$-array} $\texttt{update\_prices}(\mathbf{p}(t), \boldsymbol{\mathscr{S}}(t), \boldsymbol{\mathscr{D}}(t), \boldsymbol{\mathscr{G}}(t), \boldsymbol{\mathscr{L}}(t))$\Comment{Update prices according to \eqref{eq:final_price_update}}
\State \textsc{float} $\texttt{update\_wages}(L^s(t), L^d(t))$\Comment{Update wages according to \eqref{eq:final_wage_update}}
\State \textsc{$1d$-array} $\texttt{compute\_targets}(\mathbf{p}(t), \mathbb{E}_t[\mathbf{x}(t)], \boldsymbol{\mathscr{S}}(t), \boldsymbol{\gamma}(t))$\Comment{Compute targets according to \eqref{eq:final_target_update}}
\State \textsc{$4d$-array} $\texttt{compute\_forecasts}(\mathbf{p}(t), \mathbb{E}_t[\mathbf{x}(t)], \boldsymbol{\mathscr{S}}(t))$\Comment{Compute forecasts according to (\ref{eq:expected_balance},~\ref{eq:expected_profits})}
\State \textsc{$1d$-array} $\texttt{compute\_optimal\_quantities}(\boldsymbol{\hat{\gamma}}(t), \mathbf{p}(t), \texttt{economy})$\Comment{Compute optimal quantities according to \eqref{eq:optimal_quantities}}
\State \textsc{$4d$-array} $\texttt{compute\_profits\_balance}(\mathbf{p}(t), \mathbf{x}^{e}(t), \boldsymbol{\mathscr{S}}(t), \boldsymbol{\mathscr{D}}(t))$\Comment{Compute profits and balance according to (\ref{eq:realized_profits},~\ref{eq:realized_balance})}
\EndMethods

\EndClass
\end{algorithmic}
\caption{The \texttt{firms} class}
\label{alg:class_firms}
\end{algorithm}

\subsection{Pseudo-code to execute one step of the time-line}

In Procedure \ref{algo:time_step}, we present a pseudo-code to execute one time-step of the model. In order to get the full dynamics, one loops over this procedure during a time $T$, after a careful initialization.

To initialize, one needs to give the dynamics an initial value for prices $p_i(t=1)$, wages $p_0(t=1)$, production levels $\gamma_i(t=1)$, targets $\hat{\gamma}_i(t=2)$, stocks $I_{ij}(t=1)$ and savings $S(t=1)$. One also needs to carry out the initial planning by the household to have a value for $\mathbb{E}_1[C^d(t=1)]$ and $L^s(t=1)$. This entire process of initialization and loop over Procedure \ref{algo:time_step} in encapsulated into a class \texttt{dynamics}. This class stores the entire history of the most fundamental quantities (prices, demand matrix...) into array of the appropriate size, and uses reconstruction methods for all the inferable quantities (productions, targets, profits...).

This way, the algorithm is quicker and less memory-demanding. Finally, Procedure \ref{algo:time_step} is quite detailed compared to the actual implementation. Bearing in mind complexity issues, most of the loops of Procedure \ref{algo:time_step} are implemented in a single line through matrix multiplication. Using the result $(\Delta M)_{ij}=\Delta_{ii}M_{ij}$ and $(M\Delta)_{ij}=\Delta_{jj}M_{ij}$ with $\Delta$ a diagonal matrix, one can implement the procedure to go from demanded quantities to exchanged quantities as
\begin{equation*}
\begin{split}
\mathbf{x}(t) = &\diag{\left[\min\left(1, \frac{B(t)}{\sum_i p_i(t)C_i^d(t)\min\left(1,\mathscr{S}_i(t)/\mathscr{D}_i(t)\right)}\right),1,\ldots,1\right]}\\
&\mathbf{x}^d(t)\\
&\diag{\left[\min\left(1,\frac{L^s(t)}{L^d(t)}\right),\min\left(1,\frac{\mathscr{S}_1(t)}{\mathscr{D}_1(t)}\right),\ldots, \min\left(1,\frac{\mathscr{S}_N(t)}{\mathscr{D}_N(t)}\right)\right]}
\end{split}
\end{equation*}
where we use the convention $x_{00}^d=x_{00}^{e}=0$. Finally, we denote by $\partial_i^{in}$ (resp. $\partial_i^{out}$) the set of  suppliers (resp. buyers) of firm $i$.

\newpage
\begin{algorithm}[H]
\caption{Fundamental time-step}
\label{algo:time_step}
\begin{algorithmic}
\Algphase{Phase 1 - Planning}
\Input $L^s(t)$, $\boldsymbol{\gamma}(t)$, $\mathbf{p}(t)$, $\mathbf{I}(t)$, $\mathbf{x}(t)$
\ForAll {firms $i$}
    \State $\mathscr{S}_i(t) \leftarrow z_i\gamma_i(t)+I_{ii}(t)$
    \State $\hat{\gamma}_i(t+1) \leftarrow \texttt{compute\_targets}(\mathbf{p}(t),\mathbb{E}_t[\mathbf{x}(t)],\mathscr{S}_i(t),\gamma_i(t))$\Comment{Computation of targets according to forecasts}
\EndFor
\State $\boldsymbol{\widehat{x}}(t) \leftarrow \texttt{compute\_optimal\_quantities}(\boldsymbol{\hat{\gamma}}(t+1),\mathbf{p}(t),\texttt{economy})$
\ForAll {firms $i$}
    \State $x^d_{i0}(t):=\ell^d_i(t) \leftarrow \widehat{x}_{i0}(t)$
    \ForAll {firms $j\in\partial_i^{in}$}
        \State $x^d_{ij}(t) \leftarrow \max\left(0,\widehat{x}_{ij}-I_{ij}\right)$ 
    \EndFor
\EndFor
\Output $\mathscr{S}_i(t)$, $\boldsymbol{\hat{\gamma}}_i(t)$, $\mathbf{\widehat{x}}(t)$, $\mathbf{x}^d(t)$, $\ell^d(t)$, $S(t)$
\Algphase{Phase 2 - Exchanges \& Updates}
\Input $\mathscr{S}_i(t)$, $\mathbf{\widehat{x}}(t)$, $\mathbf{x}^d(t)$, $\ell^d(t)$, $C_i^d(t)$, $\mathbb{E}_t[B(t)]$
\ForAll{ firms $i$}
    \State $x_{i0}:=\ell_i\leftarrow\ell^{d}_i\min\left(1,\frac{L^s(t)}{L^d(t)}\right)$\Comment{Workers are hired}
\EndFor%
\State $B(t)\leftarrow S(t)+\sum_{i=1}^{n}\ell^{e}_i(t)$\Comment{Wages are paid}
\ForAll{firms $i$}
    \State $x^d_{0i}:=C_i^{d}(t)\leftarrow C_i^d(t)\left(\nu + (1-\nu)\min\left(1, \frac{B(t)}{\mathbb{E}_t[B(t)]}\right)\right)$\Comment{Household adjusts its consumption demands ($\nu=1$ in this paper)}
    \State $\mathscr{D}_i(t)\leftarrow\sum_{j\in\partial_i^{out}}x_{ji}^d(t)$\Comment{Firms compute their total demand}
    \ForAll{ firms $j\in\partial_i^{out}$}
        \State $x_{ji}\leftarrow x^d_{ji}\min\left(1,\frac{\mathscr{S}_i(t)}{\mathscr{D}_i(t)}\right)$\Comment{Exchanges of goods are carried out}
    \EndFor
    \State $C^{r}_i(t)\leftarrow C^d_{i}\min\left(1,\frac{\mathscr{S}_i(t)}{\mathscr{D}_i(t)}\right)\min\left(1,\frac{B(t)}{\mathbf{p}(t)\cdot\mathbf{C}^{e}(t)}\right)$\Comment{Household consumes according to its budget}
    
    \State $\mathscr{G}_i(t)$, $\mathscr{L}_i(t)\leftarrow  p_i(t)\sum_{j\in\partial_i^{out}}x_{ji}^{e}(t)$,  $\sum_{j\in\partial_i^{in}}x_{ij}^{e}(t)p_j(t)$
\EndFor%
\State $S(t+1)\leftarrow B(t)-\mathbf{p}(t)\cdot\mathbf{C}^{e}(t)$\Comment{The household saves unspent money}
\ForAll{ firms $i$}
    \State $x^{a}_{i0}(t)\leftarrow x_{i0}(t)$\Comment{Labor available for production is the hired workforce}
    \ForAll{ firms $j\in\partial_i^{in}$}
        \State $x^{a}_{ij}(t)\leftarrow x_{ij}(t)+\min\left(\widehat{x}_{ij}(t), I_{ij}(t)\right)$\Comment{Available goods depend on exchanges and current stocks}
    \EndFor
\EndFor
\State $p_0(t+1)\leftarrow \texttt{update\_wage}(L^s(t),L^d(t),\omega)$\Comment{Wage is updated}
\ForAll{ firms $i$}
    \State $p_i(t+1)\leftarrow\texttt{update\_price}(\mathscr{S}_i(t),\mathscr{D}_i(t),\mathscr{G}_i(t),\mathscr{L}_i(t),\alpha,\alphap,\beta,\betap)$\Comment{Prices are updated}
\EndFor
\Output $\mathbf{x}^e(t)$, $\mathbf{x}^p(t)$, $S(t+1)$, $B(t)$, $p_0(t+1)$, $\mathscr{G}_i(t)$,$\mathscr{G}_i(t)$
\Algphase{Phase 3 - Production}
\Input $S(t+1)$, $B(t)$, $p_0(t+1)$, $\mathscr{G}_i(t)$,$\mathscr{G}_i(t)$,$\mathbf{x}^e(t)$, $\mathbf{x}^p(t)$, $\mathbf{\widehat{x}}(t)$, $\mathbf{I}(t)$
\ForAll{ firms $i$}
    \State $\gamma_i(t+1)\leftarrow\texttt{production\_function}\left([x^a_{ij}]_{j\in\partial_i^{in}}\right)$\Comment{Production begins}
    \State $I_{ii}(t)\leftarrow e^{-\sigma_i}\left(\mathscr{S}_i(t)-\sum_{j\in\partial_i^{out}}x^{e}_{ji}\right)$\Comment{Firms update inventories for their own good}
    \If{ $q=0$}\Comment{If the economy is Leontief, firms need to stock other goods in addition to their own}
        \State $j^\star\leftarrow\underset{j}{\arg\min}\left([x^p_{ij}]_{j\in\partial_i^{in}}\right)$
        \ForAll{ firms $j\in\partial_i^{in}$}
            \State $x^{u}_{ij}(t)\leftarrow\frac{J_{ij}}{J_{ij^\star}}x^{p}_{ij^\star}(t)$
            \State $I_{ij}(t+1)=e^{-\sigma_j}\left[x^{a}_{ij}(t)-x_{ij}^{u}(t)\right]$
        \EndFor
    \EndIf
\EndFor
\ForAll{ firms $i$}
    \State $p_i(t+1)\leftarrow p_i(t+1)/p_0(t+1)$\Comment{Prices are updated}
\EndFor
\State $B(t)$, $S(t+1)$, $p_0(t+1)\leftarrow B(t)/p_0(t+1)$, $S(t+1)/p_0(t+1)$, $1$\Comment{Rescaling of monetary quantities}
\State $C^d_i(t+1)$, $L^s(t+1) \leftarrow \texttt{compute\_demands\_labor}(S(t), L^s(t), L^d(t), \mathbf{p}(t+1),\omega^\prime, \varphi)$\Comment{The household starts planning}
\Output $B(t)$, $S(t+1)$, $p_0(t+1)=1$, $p_i(t+1)$, $\gamma_i(t+1)$, $C^d_i(t+1)$, $L^s(t+1)$, $\mathbf{I}(t+1)$
\end{algorithmic}
\end{algorithm}

\end{document}